\documentclass[12pt]{article}

\usepackage{amssymb,amsmath}
\usepackage{cite}

\usepackage{graphicx}
\usepackage{subfigure}

\newcommand{\be}{\begin{equation}}
\newcommand{\ee}{\end{equation}}
\newcommand{\ra}{\rightarrow}
\newcommand{\prt}{\partial}
\newcommand{\al}{\alpha}
\newcommand{\bt}{\beta}
\newcommand{\dlt}{\delta}
\newcommand{\Dlt}{\Delta}
\newcommand{\Om}{\Omega}
\newcommand{\om}{\omega}
\newcommand{\gm}{\gamma}
\newcommand{\sgm}{\sigma}
\newcommand{\Gm}{\Gamma}
\newcommand{\lbd}{\lambda}

\newcommand{\dgr}{\dagger}
\newcommand{\ep}{\varepsilon}
\newcommand{\vp}{\varphi}

\newcommand{\bB}{{\bf B}}

\newcommand{\bM}{{\bf M}}
\newcommand{\bP}{{\bf P}}

\newcommand{\ba}{{\bf a}}

\newcommand{\br}{{\bf r}}
\newcommand{\bk}{{\bf k}}

\newcommand{\cF}{{\cal F}}

\font\tenmsb=msbm10 scaled\magstep 1
   \font\sevenmsb=msbm7 scaled \magstep 1
   \font\faivemsb=msbm5 scaled \magstep 1
\newfam\msbfam
      \textfont\msbfam=\tenmsb
      \scriptfont\msbfam=\sevenmsb
      \scriptscriptfont\msbfam=\faivemsb

\begin{document}

\begin{center}

{\Large\bf{Theory of Cold Atoms: Basics of Quantum Statistics} \\ [5mm]

V.I. Yukalov} \\ [3mm]

{\it Bogolubov Laboratory of Theoretical Physics, \\
Joint Institute for Nuclear Research, Dubna 141980, Russia} \\ [3mm]

\end{center}

\vskip 2cm

{\bf Abstract}

\vskip 2mm

The aim of this Tutorial is to present the basic mathematical techniques 
required for an accurate description of cold trapped atoms, both Bose and  
Fermi. The term {\it cold} implies that considered temperatures are low, such 
that quantum theory is necessary, even if temperatures are finite. And the term 
{\it atoms} means that the considered particles are structureless, being 
defined by their masses and mutual interactions. Atoms are {\it trapped} in the 
sense that they form a finite quantum system, though their number can be very 
large allowing for the use of the methods of statistical mechanics. This 
Tutorial is the first part of several tutorials, giving general mathematical 
techniques for both types of particle statistics. The following tutorials will 
be devoted separately to Bose atoms and Fermi atoms. The necessity of carefully 
explaining basic techniques is important for avoiding numerous misconceptions 
often propagating in literature.

\vskip 1cm

 
\vskip 1cm


\newpage

\section{Introduction}

Physics of cold atoms has recently been attracting high attention because of
the breakthrough in experimental techniques of cooling and trapping atoms, 
resulting in a series of experiments with cold trapped atoms, when quantum 
effects become principal. In these experiments, different atoms can be prepared 
and studied in a number of states, with varying density, temperature, atomic 
interactions, and external fields. This allows for a detailed comparison of
theory and experiment. However, in order that such a comparison would be 
meaningful, the used theory must be self-consistent and mathematically correct. 
In the other case, various misconceptions and paradoxes can arise. In theoretical
literature, one can meet numerous incorrect statements and unjustified 
approximations, which especially concerns systems with Bose-Einstein condensate.
It looks, therefore, that a Tutorial, explaining in detail the mathematical 
techniques that are necessary for a correct description of cold atoms is timely
and important. 

The present Tutorial is devoted to the general techniques of characterizing
atomic systems for which quantum theory is necessary. Applications to Bose and
Fermi systems are planned to be treated separately in the following tutorials. 
To make the presentation self-consistent, it is written with sufficient details,
in order that the reader would not be forced to jump to other numerous 
publications. The related literature is cited. But, since the Tutorial is not a 
review, the citations are mainly limited to books, review articles, and most 
important original papers. Some topics, relevant to the physics of cold atoms 
have been described in the books \cite{Lieb_1,Pitaevskii_2,Letokhov_3,Pethick_4} 
and review articles 
\cite{Courteille_5,Andersen_6,Yukalov_7,Bongs_8,Yukalov_9,Morsch_10,Posazhennikova_11,
Yukalov_12,Proukakis_13,Yurovsky_14,Moseley_15,Bloch_16,Ketterle_17,Giorgini_18,
Yukalov_19,Chevy_20,Yukalov_21,Yukalov_22}. Several results, presented in this 
Tutorial are new.  

Throughout the text, the system of units is employed, where the Planck constant 
$\hbar\ra 1$ and the Boltzmann constant $k_B\ra 1$.

\section{Phenomenological relations}

Before starting the exposition of a rigorous microscopic approach, it is useful 
to recall in brief the main phenomenological relations that often are common for 
quantum as well as classical systems and, when applied to quantum systems, 
describe the relations between their average characteristics.

\subsection{Thermodynamic quantities}

Thermodynamic relations are the same for either quantum or classical systems,
provided that the considered system is in equilibrium and is sufficiently 
large, with its number of atoms $N$ and volume $V$ such that $N\gg 1$ and
the volume being proportional to $N$. 

The transition to bulk matter corresponds to the {\it thermodynamic limit}
\be
\label{1}
N\ra\infty, \quad V\ra\infty, \quad \frac{N}{V}\ra const \; ,
\ee
so that the average atomic density $\rho\equiv N/V$ is finite,
$0 < \rho < \infty$. 

In the case of finite systems, it is not necessary to invoke the 
thermodynamic limit, though, for the validity of thermodynamic relations,
the system is to be sufficiently large, as is mentioned above. When atoms 
are trapped in an external confining potential, instead of limit (1), one
defines \cite{Yukalov_23} the {\it effective thermodynamic limit}
\be
\label{2}
N\ra\infty, \quad A_N\ra\infty, \quad \frac{A_N}{N}\ra const \; ,
\ee
where $A_N$ is an extensive observable quantity. If $A_N$ is proportional 
to $V$, then the effective limit (2) reduces to the standard limit (1).    

The notion of thermodynamic limit is important not only for bulk matter, 
but for finite systems as well, being closely related to system stability. 
Even dealing with finite systems, it is instructive to check the validity
of thermodynamic relations under the effective thermodynamic limit, which
makes it possible to understand whether the finite system is stable. 

Thermodynamic quantities are distinguished into intensive and 
extensive. {\it Intensive} quantities, being defined for a large system, 
do not depend on asymptotically large $N$, remaining finite for $N\ra\infty$. 
Examples of such intensive quantities are temperature $T$, pressure $P$,
chemical potential $\mu$, magnetic field ${\bf B}$, electric field 
${\bf E}$, specific heat $C_V$, isothermal compressibility $\kappa_T$,
and other susceptibilities. 

Contrary to this, {\it extensive} quantities are proportional to $N$ in 
some positive power, so that they tend to infinity together with $N$. For 
systems confined in a volume $V$, extensive quantities are proportional 
to $V$ as well as to $N$. When the atomic system is trapped in a confining 
potential, and the volume is not defined, then extensive quantities can be 
proportional to $N^\alpha$, with $\alpha \leq 1$. Examples of extensive 
quantities are energy $E$, entropy $S$, magnetic moment (magnetization) 
${\bf M}$, electric moment (polarization) ${\bf P}$, and thermodynamic 
potentials.

{\it Thermodynamic potentials} define the thermodynamic state of the 
system. Each thermodynamic potential is a function of those thermodynamic 
variables that are necessary and sufficient for a unique description of 
thermodynamics of an {\it equilibrium system}. The most often used 
thermodynamic potentials are as follows. 

{\it Internal energy} $E = E(S,V,N,{\bf M})$. In the presence of an 
electric field ${\bf E}$, the latter enters all equations similarly to 
the magnetic field ${\bf B}$, because of which we show below only the 
dependence on ${\bf B}$. In the same way, electric polarization ${\bf P}$ 
enters similarly to magnetization ${\bf M}$. The internal energy 
differential is
\be
\label{3}
dE = TdS - PdV +\mu dN +\bB \cdot d\bM \; .
\ee
Hence, one has
\be 
\label{4}
T = \frac{\prt E}{\prt S} \; , \quad P = -\; \frac{\prt E}{\prt V} \quad
\mu = \frac{\prt E}{\prt N} \; , \quad B^\al =\frac{\prt E}{\prt M^\al} \; ,
\ee
where $\al = x,y,z$. 

{\it Free energy} $F = F(T,V,N,{\bf M})$. It is related to the internal 
energy as 
\be
\label{5}
F = E-TS \; ,
\ee 
and its differential is
\be
\label{6}
dF = - SdT - PdV + \mu dN +\bB \cdot d\bM \; . 
\ee
Therefore, one has
$$
S = -\; \frac{\prt F}{\prt T} \; , \quad P = -\; \frac{\prt F}{\prt V}
\quad \mu = \frac{\prt F}{\prt N} \; , 
\quad B^\al = \frac{\prt F}{\prt M^\al} \; .
$$

{\it Gibbs potential} $G = G(T,P,N,{\bf B})$. It can be represented as
\be
\label{7}
G = F + PV -\bB \cdot \bM = \mu N \; .
\ee
And its differential is
\be
\label{8}
dG = -SdT + V dP + \mu dN -\bM \cdot d\bB
\ee
From here it follows:
$$
S = - \; \frac{\prt G}{\prt T} \; , \quad V = \frac{\prt G}{\prt P}
\quad \mu = \frac{\prt G}{\prt N} \; , \quad
M^\al = -\;\frac{\prt G}{\prt B^\al} \; . 
$$

{\it Grand potential} $\Om = \Om(T,V,\mu,{\bf M})$. It can be written as 
\be
\label{9}
\Om = F -\mu N = -PV + \bB \cdot \bM \; ,
\ee
with the differential 
\be
\label{10}
d\Om = - S dT - P dV - Nd\mu +\bB \cdot d\bM \: .
\ee
Respectively, 
$$
S = -\;\frac{\prt\Om}{\prt T} \; , \quad P = -\;\frac{\prt\Om}{\prt V}
\quad N = -\;\frac{\prt\Om}{\prt\mu} \; , \quad
B^\al = \frac{\prt\Om}{\prt M^\al} \; .
$$

Some other thermodynamic potentials can be found in Kubo \cite{Kubo_24}.

\subsection{Equilibrium and stability}

An equilibrium state of a system corresponds to an extremum of a 
thermodynamic potential, when 
\be
\label{11}
\dlt E = \dlt F = \dlt G = \dlt\Om = 0, 
\ee
with respect to the variation of parameters.

An equilibrium can be stable, metastable, or unstable. The absolute 
minimum of a thermodynamic potential corresponds to a thermodynamically 
absolutely stable system. A relative minimum represents a metastable 
system. And a maximum characterizes an unstable system. The sufficient 
conditions for the thermodynamic stability are
\be
\label{12}
\dlt^2 E > 0 \; , \quad \dlt^2 F > 0 \; , \quad \dlt^2 G > 0 \; ,
\quad \dlt^2 \Om > 0 \; .
\ee
These stability conditions can be expressed through second derivatives 
of thermodynamic potentials defining the related susceptibilities. 

The system response to temperature variation is described by 
{\it specific heat} (heat capacity)
\be
\label{13}
C_V \equiv \frac{T}{N} \left ( \frac{\prt S}{\prt T} \right )_V \; .
\ee
Since $S = - (\prt F/\prt T )_V$, we have
\be
\label{14}
C_V \equiv -\;\frac{T}{N} \left ( \frac{\prt^2 F}{\prt T^2} \right )_V \; .
\ee
Using the relations $F = E - TS$ and  
$$
T\left ( \frac{\prt S}{\prt T}\right )_V =
\left ( \frac{\prt E}{\prt T}\right )_V \; ,
$$
we get
\be
\label{15}
C_V = \frac{1}{N} \left ( \frac{\prt E}{\prt T}\right )_V \; .
\ee

The {\it isothermal compressibility} describes the system response to
pressure variation,
\be
\label{16}
\kappa_T \equiv -\; \frac{1}{V}\left ( \frac{\prt V}{\prt P}\right )_{TN} \; .
\ee
Substituting $V = (\prt G/\prt P)_{TN}$ gives
\be
\label{17}
\kappa_T = -\; \frac{1}{V}\left ( \frac{\prt^2 G}{\prt P^2}\right )_{TN} \; .
\ee
The isothermal compressibility can also be written as
\be
\label{18}
\kappa_T = -\; \frac{1}{V} \left ( \frac{\prt P}{\prt V} \right )^{-1}_{TN}
= \frac{1}{V} \left ( \frac{\prt^2 F}{\prt V^2}\right )^{-1}_{TN} \; .
\ee
And, with $\rho\equiv N/V$, one has
\be
\label{19}
\kappa_T = \frac{1}{\rho} \left ( \frac{\prt\rho}{\prt P} \right )_{TN}
= \frac{1}{\rho} \left ( \frac{\prt P}{\prt\rho}\right )^{-1}_{TN} \; .
\ee

The isothermal compressibility is a very important quantity characterizing
the system thermodynamic stability, because of which it is useful to have
its other representations. Let there be no magnetic and electric fields. 
Then the Gibbs potential differential is $dG = - SdT + V dP +\mu dN$. 
Since $G = \mu N$, it follows $N d\mu= - SdT + V dP$, from where
$$
\left (\frac{\prt\mu}{\prt N}\right )_{TV} = \frac{1}{\rho}
\left (\frac{\prt P}{\prt N}\right )_{TV} \; .
$$
Inverting this yields
$$
\left ( \frac{\prt N}{\prt\mu}\right )_{TV} = 
\rho\left (\frac{\prt N}{\prt P} \right )_{TV} \; .
$$
For the related variables $N,P,V$, one can write
$$
dN = \left (\frac{\prt N}{\prt P}\right )_V\; dP +
\left (\frac{\prt N}{\prt V}\right )_P \; dV \; .
$$
Under the fixed $N$, implying the zero total variation $dN=0$, we have
$$
\left (\frac{\prt N}{\prt P}\right )_V \left (
\frac{\prt P}{\prt V}\right )_N +
\left ( \frac{\prt N}{\prt V}\right )_P = 0 \; ,
$$
from where
$$
\left (\frac{\prt N}{\prt P}\right )_V \left (
\frac{\prt P}{\prt V}\right )_N
\left ( \frac{\prt V}{\prt N}\right )_P = -1 \; .
$$
Using the Maxwell relations
$$
\left ( \frac{\prt V}{\prt N}\right )_{TP} =
\left ( \frac{\prt\mu}{\prt P}\right )_{TN}
$$
and the equalities
$$
\left ( \frac{\prt\mu}{\prt P}\right )_{TN} = \frac{1}{N}\left (
\frac{\prt G}{\prt P}\right )_{TN} = \frac{1}{\rho} \; ,
$$
we get
$$
\left (\frac{\prt N}{\prt P} \right )_{TV} = -\rho\left (
\frac{\prt V}{\prt P}\right )_{TN} \; .
$$
Hence
$$
\left (\frac{\prt V}{\prt P}\right )_{TN} = -\; \frac{1}{\rho} \left (
\frac{\prt N}{\prt P}\right )_{TV} = -\; \frac{1}{\rho^2}
\left ( \frac{\prt N}{\prt\mu} \right )_{TV} \; .
$$
This gives
\be
\label{20}
\kappa_T \equiv \frac{1}{N}\left ( \frac{\prt N}{\prt P}\right )_{TV} =
\frac{1}{N\rho}\left ( \frac{\prt N}{\prt\mu}\right )_{TV} =
\frac{V}{\rho^2} \left ( \frac{\prt^2 F}{\prt\rho^2} \right )_{TV}^{-1} \; .
\ee

The isothermal compressibility is closely connected with the sound velocity
$s$, defined by the equation
\be
\label{21}
s^2 \equiv \frac{1}{m} \left ( \frac{\prt P}{\prt\rho}\right )_{TN} \; ,
\ee
where $m$ is atomic mass. As is clear,
\be
\label{22}
\kappa_T = \frac{1}{m\rho s^2} \; .
\ee

{\it Magnetic susceptibility} is defined through the susceptibility tensor
\be
\label{23}
\chi_{\al\bt} \equiv \frac{1}{N} \left ( \frac{\prt M^\al}{\prt B^\bt} \right ) \; .
\ee
Using here $M^\al = - \; \prt G/\prt B^\al$ gives
\be
\label{24}
\chi_{\al\bt} = -\; \frac{1}{N} \left (
\frac{\prt^2 G}{\prt B^\al\prt B^\bt}\right ) \; .
\ee
Diagonal parts $\chi_{\al\al}$ of tensor (23) are called magnetic 
susceptibilities. 

The thermodynamic characteristics, describing the system response to the
variation of thermodynamic variables and which are expressed through the
second derivatives of thermodynamic potentials, all can be termed 
susceptibilities. Thus, the specific heat $C_V$ is {\it thermal 
susceptibility} (susceptibility with respect to temperature variation), 
the isothermal compressibility $\kappa_T$ is {\it mechanical susceptibility}
(susceptibility with respect to pressure variation), and the magnetic 
susceptibility $\chi_{\al\al}$ defines the system response to magnetic 
field variation. According to the stability conditions (12), all these
susceptibilities are to be non-negative. Being intensive quantities, the
susceptibilities of stable equilibrium systems have to be finite, even
in thermodynamic limit. In this way, the necessary conditions for 
{\it thermodynamic stability} of an equilibrium system are:    
\be
\label{25}
0\leq C_V < \infty \; , \quad  0\leq \kappa_T < \infty \; , \quad
0\leq \chi_{\al\al} < \infty \; . 
\ee

These stability conditions are the direct consequence of the convexity 
properties (12) of thermodynamic potentials and of the fact that  
susceptibilities are intensive quantities. The compulsory finiteness
of the susceptibilities is easily understood. Take, for instance, specific 
heat $C_V$. If it would be infinite, then, according to Eq. (15), an 
infinitesimally weak temperature increase would yield infinite system 
energy, which is meaningless. If compressibility $\kappa_T$ would be 
infinite, this, by definition (16), would imply that an infinitesimally
small fluctuation of pressure would make the system volume either zero 
or infinity. Similar arguments are applicable to magnetic susceptibility.

\subsection{Phases and symmetries}

{\it Thermodynamic phase}, defined by a set of thermodynamic variables, 
is a system state qualitatively different from other states of the same 
system \cite{Kubo_24}. To make this vague definition more precise, one introduces 
{\it order parameters} such that to distinct phases there would correspond 
different orders. An order parameter is a quantity specifying a given 
thermodynamic phase and qualitatively distinguishing it from other phases.

For example, the distinction between ferromagnetic and paramagnetic phases
is done by means of magnetization ${\bf M}$, so that $M\equiv|\bM|$ is 
nonzero for ferromagnetic phase ($M > 0$, ferromagnet) and is zero for
paramagnetic phase ($M = 0$, paramagnet). Similarly, the nonzero or zero 
polarization ${\bf P}$ distinguishes ferroelectric from paraelectric phases. 

Crystalline phase is characterized by a set of spatial vectors $\{\ba\}$, 
called lattice vectors, with respect to which the particle density is 
periodic: $\rho(\br + \ba) = \rho(\br)$. While liquid-phase density is
uniform: $\rho(\br) = \rho$. The role of an order parameter, distinguishing 
between the crystalline and liquid phases can be played by the difference
$$
\Dlt\rho\equiv \max_{\br} \rho(\br) - \min_{\br} \rho(\br) \; .
$$
For crystalline phase, $\Dlt\rho > 0$, while for liquid phase, $\Dlt\rho = 0$.

Different types of crystalline lattices are described by different sets of
lattice vectors. For instance, one lattice, with the set $\{\ba\}$ can be 
simple cubic, with the particle density $\rho_1(\br + \ba) = \rho_1(\br)$,
while another lattice, with the set $\{ {\bf b}\}$, can be face-centered 
cubic, having the density periodicity $\rho_2(\br + {\bf b}) = \rho_2(\br)$.

Gaseous and liquid phases both are characterized by uniform density. Hence
the latter cannot be an order parameter for distinguishing between these 
phases. The difference between liquid and gas can be noticed on the 
microscopic level by studying the pair correlation function displaying
a kind of short-range order in the case of liquid. This type of short-range 
order can be captured by means of the order indices \cite{Coleman_25,Yukalov_26}. 

Systems with Bose-Einstein condensate and without it can be distinguished
by the related condensate density $\rho_0$. When $\rho_0 > 0$, the system 
is Bose-condensed, and when $\rho_0 = 0$, it is not Bose-condensed.

Superfluid and normal fluid are distinguished by superfluid density $\rho_s$.
The system is superfluid if $\rho_s > 0$, while it is normal if $\rho_s = 0$.

Dielectric and metallic phases can be distinguished by the single-electron 
energy gap $\Dlt E$, with $\Dlt E > 0$ representing dielectric phase and 
$\Dlt E = 0$  corresponding to metallic phase. 

Superconductor and normal conductor are characterized by the presence or 
absence of a gap $\Dlt$ in the spectrum of collective excitations. For
a superconductor, $\Dlt > 0$, while for a conductor, $\Dlt = 0$.  

A system can be described by several order parameters corresponding to 
different properties. For example, it can be ferromagntic and dielectric 
or ferromagnetic and metallic. 

Thermodynamic phases, having different order parameters (or order indices),
are usually related to different types of symmetry. Under symmetry one 
understands invariance under some transformations forming a group. A 
symmetry group is defined as follows. 

Let $\hat g_i$ be a transformation labelled by an index $i$. And let the 
family of such transformations satisfy the additivity property 
$\hat g_i \cdot \hat g_j = \hat g_{i+j}$. If the family includes a unit 
element $\hat g_0 = \hat 1$ and, for each $\hat g_i$ there exists an
inverse element, such that $\hat g^{-1}_i \hat g_i = \hat 1$, then the 
family $\{\hat g_i\}$ forms a group. 

As an example, consider the {\it inversion symmetry} that can be 
illustrated by inverting the magnetization vector $\bM$ of a ferromagnet. 
Denoting the inversion transformation by $\hat I$, we have 
$\hat I\bM = -\bM$. If the order parameter is given by $|-\bM|$, it is
symmetric with respect to inversion: ${\hat I} M = |-\bM| = |\bM| = M$.
Two elements, the inversion transformation and unity, form the inversion 
group $\{\hat I, \hat 1\}$. 

Another example is the {\it rotation symmetry}. Denote the spatial rotation
transformation by $\hat R$. The order parameter $M$ is invariant under the 
rotation group $\{\hat R\}$, since $\hat R M = |\hat R\bM| = |\bM| = M$. 

One more example is the {\it translation symmetry}. Let $\{\hat T(\br)\}$
be a translation transformation by a vector ${\bf r}$. If ${\bf a}$ is a 
vector of a crystalline lattice, then the crystal density is invariant 
under the transformation 
$$
\hat T(\ba)\rho(\br) = \rho(\br+\ba) = \rho(\br) \; .
$$
The family $\{\hat T(\ba)\}$ composes the lattice translation group. 

In the case of a liquid, one has $\hat T(\br)\rho =\rho$ for any vector 
${\bf r}$. Then the family $\{\hat T(\br)\}$ forms the total translation 
group.

\subsection{Phase transitions}

A phase transition is a transformation of one thermodynamic phase into 
another. The change happens under a varying thermodynamic parameter, say, 
temperature $T$. When $T$ reaches the stability boundary of one phase, 
the system becomes unstable and, under the following variation of $T$, 
transforms into a stable phase. The point of a phase transition is the 
point of instability. Therefore, at this point, some susceptibilities may 
diverge. Since thermodynamic phases can usually be characterized by order 
parameters, at the phase transition point, the order parameter qualitatively 
changes. Depending on the kind if this change, one distinguishes three types 
of phase transitions.      

Let an order parameter be denoted by $\eta$, being in particular cases 
either magnetization $M$, or particle density $\rho$, or condensate density 
$\rho_0$, or superfluid density $\rho_s$, or single-particle gap $\Dlt E$, 
or the gap in the spectrum of collective excitations $\Dlt$. And let 
temperature $T$ be varying, with $T_0$ being the point of a phase transition. 
The latter are classified into three types.

(i) {\it First-order phase transition}. This is a discontinuous transition,
with a finite jump of an order parameter $\eta$ at the transition point:  
$$
|\eta(T_0 + 0) - \eta(T_0 - 0)|>0 \; . 
$$
The values of $\eta(T_0 \pm 0)$ are not necessarily zero. An example of a 
first-order phase transition is the melting-crystallization transition 
between a crystal and a liquid, where there occurs a discontinuity of the 
density. 

(ii) {\it Second-order phase transition}. This is a continuous transition,
with a continuous variation of an order parameter between its non-zero and 
zero values:
$$
|\eta(T_c + 0) - \eta(T_c - 0)| = 0 \; , 
$$
where $\eta(T)$ becomes zero for one of the phases, say for $T\geq T_c$,
being non-zero for the other phase. The point of a second-order phase 
transition is also called the {\it critical point}, since some of the 
susceptibilities diverge there. Respectively, the transition temperature 
is denoted as $T_c$, to distinguish it from the first-order transition 
point $T_0$. Examples of second-order transitions are different magnetic 
transitions or Bose-Einstein condensation.  

(iii) {\it Crossover transition}. This is a fast but gradual variation of 
an order parameter that is non-zero from both sides of the crossover point
$T_c$. The crossover point can be defined as the {\it inflection point}
of the order parameter:
$$
\frac{\prt^2\eta}{\prt T^2} = 0 \qquad (T=T_c) \; .
$$
That is, the first derivative of the order parameter displays an extremum
at the inflection point. 

In the vicinity of a phase-transition point, the system properties may 
essentially change, because of which a region around a transition point, 
where such changes are strong, is termed the {\it critical region}. 

If thermodynamic phases are described by different symmetries, the latter
also change at the phase transition point. When the symmetry of one of the 
phases is lower than that of the other phase, the transition to the 
lower-symmetry phase is called {\it spontaneous symmetry breaking}. It is 
termed {\it spontaneous}, since the symmetry breaking occurs in a 
self-organized way, without action of external fields. The cause for the 
spontaneous symmetry breaking is that the phase with a lower symmetry 
happens to be stable, when the symmetric phase becomes unstable.     

More information on phase transitions can be found, e.g., in the books
\cite{Pathria_27,Landau_28,Huang_29,Bogolubov_30,Yukalov_31,Uzunov_32}.

\subsection{Heterophase mixtures}

In addition to pure thermodynamic phases, there can arise mixtures of 
different phases, where the latter are not spatially completely separated, 
but are intermixed in such as way that each of the phases occupies randomly
distributed in space regions of {\it mesoscopic} size. Such mesoscopic 
mixtures are also called {\it heterophase mixtures} \cite{Yukalov_33}. For 
example, we can imagine paramagnetic regions in a ferromagnet, ferromagnetic 
clusters in a paramagnet, disordered regions in a crystal, crystalline 
clusters in a liquid, gas bubbles in a liquid, and liquid droplets in a gas.

For illustration, let us keep in mind the case of a heterophase mixture of
two phases, characterized by different order parameters, $\eta_1$ and $\eta_2$. 
When the system is composed of two macroscopic thermodynamic phases, they are 
said to be in thermal equilibrium with each other if their temperatures 
coincide, $T_1 = T_2$. The phases are in mechanical equilibrium, if their
pressures are the same, $P_1 = P_2$. And they are in chemical equilibrium, 
if their chemical potentials are equal, $\mu_1 = \mu_2$. A mesoscopic 
heterophase mixture, strictly speaking, is quasi-equilibrium, because of 
which the above equilibrium conditions are valid only on average 
\cite{Yukalov_33}. 

Let $V_\nu$ be the volume occupied by a $\nu$-phase and $N_\nu$ be the number 
of particles in the $\nu$-phase. Mesoscopic phases can be separated by means
of an {\it equimolecular separating surface} \cite{Gibbs_34,Gibbs_35}. In that 
case, all extensive quantities are additive, e.g., 
$$
V_1 + V_2 = V \; , \quad N_1 + N_2 = N \; .
$$
The geometric weights of each phase are quantified by  
$$
w_\nu \equiv \frac{V_\nu}{V} \quad (\nu=1,2) \; .
$$
This quantity has the meaning of geometric phase probability and satisfies 
the conditions
$$
 0\leq w_\nu \leq 1 \; ,  \quad  w_1 + w_2 = 1 \; . 
$$

Equations for the phase probabilities are obtained from the minimization
of a thermodynamic potential under the above normalization condition. 
Defining, for convenience,
$$
w_1 \equiv w \; , \quad w_2 \equiv 1 - w \; , 
$$
one has to find the absolute minimum of a thermodynamic potential, say, of 
the free energy $F = F(w)$:
$$
F = {\rm abs}\;\min_w F(w) \; . 
$$
The minimization has to include the consideration of $w$ at the boundary, 
where either $w = 0$ or $w = 1$. That is, it is necessary to compare three 
quantities: $F(w)$, with $0 < w < 1$, and the boundary values $F(0)$, 
and $F(1)$. 

The condition of heterophase equilibrium reads as
$$
\frac{\prt F}{\prt w} = 0 \; .
$$
And the condition of heterophase stability is
$$
\frac{\prt^2 F}{\prt w^2} >  0\; .
$$
These equations define the geometric phase probabilities $w_\nu$ playing 
the role of additional order parameters. The second derivative 
$\prt^2 F/\prt w^2$ has the meaning of the {\it heterophase susceptibility}
that is required to be non-negative and finite for guaranteeing the system
stability, similarly to stability conditions (25).

\subsection{Landau potential}

If the considered phase transition is of second order or of weak first 
order, when the order parameter in the vicinity of the transition point 
is small, it is possible, assuming that the free energy is an analytical 
function of the order parameter, to expand the free energy in powers of 
the latter, as suggested by Landau \cite{Landau_36}. 

Let us consider the critical region of temperatures $T\approx T_c$, where
the order parameter $\eta = \eta(T)$ is small. The order parameter $\eta$
is assumed to be real-valued and such that $\eta(T)\neq 0$ for $T < T_c$,
but $\eta(T) = 0$ for $T > T_c$. The free energy $F = F(\eta)$ is expanded
in powers of the order parameter as
$$
F = F(0) +a_1\eta +a_2\eta^2+a_3\eta^3+a_4\eta^4 \; .
$$
To have a minimum at finite $\eta$, it is necessary that $a_4 > 0$. 

In equilibrium, one has
$$
\frac{\prt F}{\prt\eta} = a_1 + 2a_2\eta +3a_3\eta^2 + 4a_4\eta^3 = 0 \; ,
$$
which requires that $a_1 = 0$, when $\eta = 0$. The condition of stability
is
$$
\frac{\prt^2 F}{\prt \eta^2} = 2\left (a_2 + 3a_3\eta + 6a_4\eta^2\right ) \geq 0 \; .
$$
There can happen two cases, with and without inversion symmetry.

In the case of {\it inversion symmetry}, the free energy is such that 
$F(-\eta) = F(\eta)$. Then $a_3 = 0$ and the condition of equilibrium reads 
as
$$
\eta(a_2+2a_4\eta^2)=0 \; . 
$$
From here, we see that the order parameter at low temperatures can take one
of the two values
$$
\eta_\pm = \pm \sqrt{-\; \frac{a_2}{2a_4}} \qquad (T<T_c) \; .
$$
And above $T_c$, one has
$$
\eta=0 \quad (T>T_c) \; .
$$
The stability condition yields
$$
a_2 + 6a_4\eta^2 \geq 0 \; ,
$$
which implies that 
$$
a_2(T) < 0 \qquad (T<T_c) \; ,
$$
$$
a_2(T) \geq 0 \qquad (T>T_c) \; .
$$
This suggests that one can accept the form
$$
a_2(T)=\al(T-T_c) \; . 
$$

The free energy is symmetric with respect to the inversion of $\eta$
to $-\eta$. The two order parameters have the same magnitude, 
$|\eta_+| = |\eta_-|$. This situation corresponds to {\it macroscopic 
degeneracy}. When temperature is lowered, the system passes from the 
symmetric phase, with $\eta = 0$, to the phase with a non-zero value
of the order parameter, taking spontaneously either $\eta_+$ or $\eta_-$.
This is an example of spontaneous symmetry breaking. 

When there is {\it no inversion symmetry}, then $a_3 \neq 0$, and the 
condition of equilibrium defines the order parameter as
$$
\eta_\pm = \frac{-3a_3\pm\sqrt{9a_3^2-32a_2a_4}}{8a_4} \; .
$$
To have a real-valued order parameter, it is necessary that 
$$
9a_3^2 - 32 a_2a_4 \geq 0 \; .
$$
This situation corresponds to a first-order phase transition.

\subsection{Ginzburg-Landau functional}

In the Landau expansion, the order parameter is treated as a scalar. 
In general, it can be an order function $\eta = \eta(\br,t)$ of the 
real-space vector $\br$ and time $t$. The generalization to this case has 
been proposed by Ginzburg and Landau \cite{Ginzburg_37}. 

One defines an effective Hamiltonian density 
$H(\br) = H (\eta(\br,t),T,\ldots)$ depending on the order function and
thermodynamic variables, such as temperature $T$. The effective Hamiltonian
$$
H = \int H(\br)\; d\br
$$
depends on $T$, reminding rather free energy. This effective Hamiltonian is
the {\it Ginzburg-Landau functional} of the order function.

The dependence of the order function on the spatial and temporal variables
implies that the system is not uniform and, in general, not equilibrium. 
The evolution equation for the order function is
\be
\label{26}
i\; \frac{\prt\eta}{\prt t} = \frac{\dlt H}{\dlt\eta^*} \; . 
\ee
Stationary states can be represented as
$$
\eta(\br,t) = \psi(\br) e^{-i\ep t} \; ,
$$
where $\ep$ is a real-valued quantity. Then the evolution equation reduces 
to the stationary form  
$$
\ep \psi = \frac{\dlt H}{\dlt\psi^*} \; .
$$

Introducing the grand effective Hamiltonian, or grand potential
$$
\Om \equiv H - \ep \int |\psi|^2\; d\br \; ,
$$
the stationary equation can be rewritten as
$$
\frac{\dlt\Om}{\dlt\psi^*} = 0 \; ,
$$
which is a condition of equilibrium. 

The function $\psi$ can be normalized to the number of particles $N$ by
the normalization condition
$$
\int |\psi|^2\; d\br = N\; , \quad N = -\; \frac{\prt\Om}{\prt\ep} \; .
$$
This shows that $\ep$ plays the role of a chemical potential. 

The dynamic stability of the stationary solution is checked by studying
small deviations from this solution. For this purpose, the order function
is written as 
$$
\eta(\br,t) = [\psi(\br) +\dlt\psi(\br,t)] e^{-i\ep t} \; , 
$$
with the deviation
$$
\dlt\psi(\br,t) = u(\br) e^{-i\om t} + v^*(\br) e^{i\om t} \; 
$$
representing the wave function of collective excitations. Substituting
the above representation into the evolution equation yields an equation
for $\om$. The condition of {\it dynamic stability} requires that the
admissible $\om$ be positive (non-negative). This is because if
$\om = \om_1 + i\om_2$ is complex-valued, then 
$$
e^{-i\om t} = e^{-i\om_1 t} e^{\om_2 t} \; , \quad
e^{i\om t} = e^{i\om_1 t} e^{-\om_2 t} \; ,
$$
and there appears a divergence for any $\om_2\neq 0$, as time increases,
which means that the considered solution is not dynamically stable.

\subsection{Superconductor model}

Ginzburg and Landau \cite{Ginzburg_37} illustrated their approach by a model
of superconductor. In the effective Hamiltonian $H = \int H(\br)\; d\br$,
the Hamiltonian density is written as 
$$
H(\br) = \frac{1}{2m}\left | \left (\hat{\bf p} -\; \frac{e}{c}{\bf A}
\right )\eta \right |^2 + U|\eta|^2 + \frac{1}{2}\; \Phi_0|\eta|^4 +
\frac{\bB^2}{8\pi} \; ,
$$
where $m$ is the effective mass equal to two electron masses, $m = 2m_0$, 
and the effective charge equals two electron charges, $e = -2e_0$; momentum
operator $\hat{\bf p} \equiv -i\nabla$, and ${\bf A}$ is the vector potential.
The quantities $U$ and $\Phi_0$ can be functions of thermodynamic variables 
$T,P,\ldots$. The order parameter $\eta$ is such that $|\eta|^2$ gives the 
superfluid electron density.  

In the presence of an external magnetic field $\bB_0$, the effective 
Hamiltonian becomes
$$
H = \int \left [ H(\br) - \; \frac{\bB\cdot\bB_0}{4\pi}\right ]\; d\br \; .
$$
The vector potential defines the magnetic field $\bB = \nabla\times{\bf A}$
and satisfies the Coulomb calibration $\nabla\cdot{\bf A} = 0$. In equilibrium, 
one has  
$$
\frac{\dlt H}{\dlt\eta^*(\br)} = 0\; . 
$$
This leads to the {\it equation for the order function}
\be
\label{27}
\frac{1}{2m} \left (\hat{\bf p} -\; \frac{e}{c}\; {\bf A}\right )^2 \eta +
U\eta +\Phi_0|\eta|^2\eta = 0 \; .
\ee

The extremization with respect to the vector potential,
$$
\frac{\dlt H}{\dlt A^\al} = 0 \; , 
$$
can be rewritten as the Euler-Lagrange equations
$$
\left ( \frac{\dlt H}{\dlt A^\al}\right )_{A'} -
\sum_\bt \frac{\prt}{\prt r^\bt} \left ( \frac{\dlt H}{\dlt A'_{\al\bt}}
\right ) = 0 \; ,
$$
in which $A'_{\al\bt} \equiv \prt A^\al/\prt r^\bt$. From here, one finds
the equation
$$
\nabla^2 {\bf A} = - \frac{4\pi}{c} {\bf J} \; ,
$$
with the supercurrent
$$
{\bf J} = -\; \frac{ie}{2m} \left (\eta^*\nabla\eta -
\eta\nabla\eta^* \right ) - \; \frac{e^2}{mc}\; |\eta|^2{\bf A} \; . 
$$

The equation for the order function $\eta$ is a kind of the nonlinear 
Schr\"{o}dinger equation. It is worth emphasizing that the order-function 
equation, being a kind of the nonlinear Schr\"{o}dinger equation, has 
appeared, first, in the paper by Ginzburg and Landau \cite{Ginzburg_37}. 
Therefore such equations can be called the {\it Ginzburg-Landau order-function
equations}.

\subsection{Superfluid vortex}

For systems of neutral Bose atoms, such as superfluid helium, the effective 
Hamiltonian density takes the form
$$
H(\br) = \frac{1}{2m}\; |\nabla\eta|^2 + U|\eta|^2 +
\frac{1}{2}\; \Phi_0|\eta|^4 \; ,
$$
where $m$ is atomic mass and the quantities $U$ and $\Phi_0$ can be functions 
of thermodynamic variables $T,\ldots$. The order parameter, or order function 
$\eta = \eta(\br,t)$ gives the density $|\eta|^2$ of atoms in the ordered state.

With the nonlinear Hamiltonian
\be
\label{28}
\hat H[\eta] \equiv -\; \frac{\nabla^2}{2m} + U + \Phi_0|\eta|^2 \; ,
\ee
the evolution equation for the order function
\be
\label{29}
i\; \frac{\prt\eta}{\prt t}  = \hat H[\eta]\eta
\ee
is the nonlinear Schr\"odinger equation, being a particular case of the
Ginzburg-Landau equation (27). For an equilibrium system, one gets the 
stationary equation
\be
\label{30}
\hat H[\psi]\psi = \ep \psi \; . 
\ee

The latter equation possesses vortex solutions \cite{Ginzburg_38}. To show
this, one passes to cylindrical coordinates $\br=\{r_\perp,\vp,z\}$, where
$r_\perp\equiv\sqrt{x^2+y^2}$ is the transverse spatial variable. In this 
representation, 
$$
\nabla^2 = \frac{\prt^2}{\prt r^2_\perp} + \frac{1}{r_\perp}\;
\frac{\prt}{\prt r_\perp} + \frac{1}{r^2_\perp}\; \frac{\prt^2}{\prt\vp^2}
+\frac{\prt^2}{\prt z^2} \; .
$$
Looking for the solution in the form $\psi(\br)=\psi_\perp(r_\perp)e^{i\nu\vp}$,
with $\psi_\perp(r_\perp)$ real-valued, and imposing the periodicity constraint
$$
\psi(r_\perp,\vp+2\pi,z)=\psi(r_\perp,\vp,z) \; ,
$$
one finds that the quantum number $\nu$, called the {\it winding number}, has 
to be quantized, $\nu = 0,\pm 1,\pm 2,\ldots$. 

The stationary equation (30) yields
\be
\label{31}
\left [ \frac{d^2}{dr^2_\perp} + \frac{1}{r_\perp}\; \frac{d}{dr_\perp} -\;
\frac{\nu^2}{r^2_\perp} + 2m(\ep - U)\right ]\psi_\perp -
2m\Phi_0\psi^3_\perp = 0 \; .
\ee
Considering a cylindrical system of radius $R$, length $L$, and volume 
$V = \pi R^2L$, the solution to this equation is normalized as
\be
\label{32}
2\pi L \int_0^R |\psi_\perp(r_\perp)|^2 r_\perp\; dr_\perp = N \; .
\ee

One introduces the {\it healing length}
\be
\label{33}
\xi \equiv \frac{1}{\sqrt{2m(\ep - U)}} \; .
\ee
Using this scale, one defines the dimensionless radius and the dimensionless 
order function,
\be
\label{34}
r\equiv \frac{r_\perp}{\xi} \; , \quad 
f(r)\equiv\xi\sqrt{2m\Phi_0}\;\psi_\perp(r_\perp) \; .
\ee
Then Eq. (31) reduces to
\be
\label{35}
\left ( \frac{d^2}{dr^2} + \frac{1}{r}\; \frac{d}{dr} + 1 - \;
\frac{\nu^2}{r^2}\right ) f - f^3 = 0 \; .
\ee
And the normalization condition (32) becomes
\be
\label{36}
\frac{\pi L}{m\Phi_0} \; \int_0^{R/\xi} f^2(r)r\; dr = N \; .
\ee
Looking for a nontrivial solution $f\neq 0$, we see that it depends on the
winding number $\nu$. 

If $\nu=0$, there is no vortex, which corresponds to the {\it ground state},
when $f_0 = \rm const$. From Eq. (35), we have $f_0^2 = 1$, hence $f_0 = \pm 1$.
This degeneracy is not important, since the physical quantity $|f_0|^2$
does not depend on the sign of $f_0$. Without the loss of generality, we may 
choose $f_0 = 1$. Normalization (36) gives
\be
\label{37}
\xi =\frac{1}{\sqrt{2m\rho\Phi_0}} \; , \quad 
\ep = U + \rho\Phi_0 \quad (\nu = 0) \; .
\ee

The first vortex solution appears for $\nu = 1$, describing the 
{\it basic vortex}. Then Eq. (35) can be solved numerically. Close to the 
vortex axis, where $r\ll 1$, that is $r_\perp\ll\xi$, the solution has the
asymptotic expansion
$$
f(r)\simeq\sqrt{C}\; r(1+c_2r^2+c_4r^4+c_6r^6) \; ,
$$
with $C$ being a constant and 
$$
c_2=-\; \frac{1}{8}\; , \quad c_4 = \frac{1+8C}{192}\; , \quad
c_6 = -\; \frac{1+80C}{9216} \; . 
$$
At large distance from the vortex axis, where $r\gg 1$, that is $r_\perp\gg \xi$,
the solution has the form
$$
f(r) \simeq 1 -\; \frac{1}{2}\; r^{-2} -\; \frac{9}{8}\; r^{-4} -\;
\frac{161}{16}\; r^{-6} \; . 
$$

Solutions $f = f_\nu(r)$, representing the vortices with higher winding numbers, 
satisfy the equation
$$
\left ( \frac{d^2}{dr^2} + \frac{1}{r}\; \frac{d}{dr} + 1 -\;
\frac{\nu^2}{r^2}\right ) f_\nu - f_\nu^3 = 0 \; .
$$
The normalization condition (36) defines the related healing length $\xi_\nu$
that is different from (37). Notation (32) gives the spectrum 
\be
\label{38}
\ep_\nu = U + \frac{1}{2m\xi_\nu^2} \; ,
\ee
labelled by the winding number $\nu = 1,2,\ldots$.  

If the system radius is very large, then the dependence of the healing 
length and, respectively, of the spectrum on the winding number is weak, since, 
in that case, the main contribution to the normalization integral (36) is due 
to the large region of $r \gg 1$, where $f_\nu(r)\ra 1$. However for an 
elongated system, with a small radius, spectrum (38) can essentially depend 
on the winding number, if the vortex axis is along the cylinder axis.   

The solution $\psi(\br)=\psi_\perp(r_\perp)e^{i\nu\vp}$, for $\nu > 0$, 
represents a vortex, since it defines a nonzero tangential velocity
\be
\label{39}
{\bf v}\equiv\ \frac{{\bf j}}{\rho} = \frac{\nu}{mr}\;{\bf e}_\vp \; ,
\ee
where
$$ 
\rho\equiv|\psi|^2 \; , \quad  {\bf j} = -\; \frac{i}{2m}\left (\psi^*\nabla\psi -
\psi\nabla\psi^* \right ) \; .
$$
The vortex circulation corresponds to the integral
\be
\label{40}
\oint {\bf v}\cdot {\bf dl} = 2\pi\; \frac{\nu}{m} \; .
\ee

It is worth emphasizing again that the evolution equations for order functions
naturally appear in the Ginzburg-Landau approach in a similar way for both
superconducting and superfluid systems 
\cite{Ginzburg_37,Ginzburg_38,Ginzburg_39,Tilley_40}.

\subsection{Landau criterion}

In order to find out whether the system can exhibit superfluid motion, one
employs a simple criterion, suggested by Landau \cite{Landau_36}. 

Suppose that a flow moves with superfluid velocity ${\bf v}$. In the moving 
frame, where the liquid is at rest, the energy per particle is
$$
E = U + \frac{p^2}{2m} \; ,
$$
with $p\equiv|{\bf p}|$. Its energy in the laboratory frame reads as
$$
E_v = U + \frac{({\bf p}+m{\bf v})^2}{2m} =  U + \frac{p^2}{2m} +
{\bf p}\cdot{\bf v} + \frac{mv^2}{2} \; ,
$$
with $v\equiv|{\bf v}|$. If the energy of an elementary excitation 
in the moving frame is $\ep({\bf p})$, then its energy in the laboratory 
frame is $\ep_v({\bf p}) = \ep({\bf p}) + {\bf p}\cdot{\bf v}$.

The fluid motion is dynamically stable when the energy of the elementary 
excitations is positive,
$$
\ep_v({\bf p}) > 0\; , \quad \ep({\bf p}) + {\bf p}\cdot{\bf v} > 0 \; ,
$$
hence, when $- {\bf p}\cdot{\bf v} < \ep({\bf p})$ for all ${\bf p}$. 
Maximizing the left-hand side of the latter inequality with respect to the 
direction of ${\bf p}$, one gets $pv < \ep({\bf p})$. Therefore, the 
superfluid velocity should be such that 
$$
v < \min_p\; \frac{\ep({\bf p})}{p} \; .     
$$
That is, it has to be lower than the critical velocity
$$
v_c \equiv \min_p\; \frac{\ep({\bf p})}{p} \; .
$$
For $v < v_c$, the superfluid motion is stable, while if $v$ exceeds $v_c$, 
the superfluidity is destroyed. 

Strictly speaking, superfluid flow is not absolutely stable, but metastable.
The absolutely stable state corresponds to the system at rest. 

As an example, when there is no superfluidity, it is possible to mention the 
ideal Bose gas with the spectrum $ \ep({\bf p}) = p^2/2m$, for which 
$\min_p \ep({\bf p})/p = 0$, hence $v_c = 0$, and there can be no superfluid 
motion in the ideal gas, even as a metastable state.

\section{Field representation}

Now the basic mathematical notions will be given, providing the fundamentals
for the microscopic theory of quantum systems. The first such a notion is
the so-called second quantization. Below, we shall describe it by employing
the field representation, following mainly Berezin \cite{Berezin_41} and
Schweber \cite{Schweber_42}.

\subsection{Fock space}

Let us enumerate the number of particles by $n=0,1,2,\ldots$. Each particle 
is characterized by a set $x$ of variables from the variable space $\{ x\}$. 
For the general mathematical exposition, the nature of the particle variables 
is not important. In particular, this can be the real-space vector $\br$, or
momentum vector ${\bf p}$, or the combination of the real-space vector and 
a spin $\{\br, s\}$, etc. 

The system microscopic states are the vectors of a Hilbert space. A 
{\it Hilbert space} ${\cal H}$ is a complete vector space with a scalar
product. In this space ${\cal H}=\{ f\}$, for any two vectors 
$f,f'\in{\cal H}$ there exists a scalar product $(f,f')$. The latter 
generates the vector norm $||f||\equiv\sqrt{(f,f)}$. The space is complete,
if for a sequence $\{ f^{(k)}\in {\cal H}\; |\; k=1,2,\ldots\}$, the limit
$\lim_{k\ra\infty} f^{(k)} = f^{(\infty)}\in {\cal H}$. The convergence
is understood as the convergence by norm: 
$\lim_{k\ra\infty}||f^{(k)}||=||f^{(\infty)}||$. 

Let us consider the space ${\cal H}_n = \{ f_n\}$, which is an 
{\it n-particle Hilbert space} composed of functions 
$f_n = f_n(x_1,x_2,\ldots,x_n)$. Let $\hat P_{ij}$ be a permutation operator 
interchanging the indices $i$ and $j$, with $i\neq j$, so that its action 
is defined by the equality
$$
\hat P_{ij} f_n(\ldots x_i \ldots x_j \ldots) =
f_n(\ldots x_j\ldots x_i \ldots) \; .
$$
It is assumed that all particles are classified into two types, bosons and 
fermions. The first type is described by symmetric states and the second, by 
antisymmetric. An $n$ - particle function $f_n$ is symmetrized provided that 
$$
\hat P_{ij}f_n = \pm f_n \; .
$$
In what follows, the upper sign corresponds to bosons, while the lower sign,
to fermions. The scalar product in ${\cal H}_n$ is given by the integral  
$$
(f_n,f_n')\equiv
\int f_n^*(x_1,\ldots, x_n) f_n'(x_1,\ldots, x_n)\; dx_1\ldots dx_n \; . 
$$
The integration for each $x_i\in\{ x_i\}$ is over all $\{ x_i\}$, so that 
$\int dx\equiv \int_{ \{ x\} } dx$. Each space ${\cal H}_n$ is complete by
norm $||f_n||\equiv\sqrt{(f_n,f_n)}$. In the case of discrete variables, 
the integration is to be understood as summation. For instance, if $x\ra\br,s$,
where $s$ is a spin index, then $\int dx\equiv \sum_s \int d\br$. In that case,
the multi-dimensional delta function is 
$\dlt(x-x')\equiv\dlt(\br-\br')\dlt_{ss'}$.

Consider a set of $n$-particle spaces ${\cal H}_0, {\cal H}_1,\ldots$, where
the first space ${\cal H}_0$ is composed of constants $f_0 = const$. 

The {\it Fock space} is the direct sum
$$
{\cal F}\equiv \bigoplus_{n=0}^\infty {\cal H}_n \; .
$$
The vectors of the space ${\cal F}=\{\vp\}$ have the form of columns
\begin{eqnarray}
\vp\equiv \left [ \begin{array}{c}
\nonumber
f_0 \\
f_1 \\
.  \\
. \\
. \\
f_n \\
. \\
. \\
.
\end{array} \right ] , \quad f_n=f_n(x_1,\ldots,x_n) \; .
\end{eqnarray}

The conjugate of $\vp$ is the row 
$\vp^+\equiv[f_0^* f_1^*\ldots f_n^* \ldots ]$. For brevity, we shall denote
$\vp\equiv[f_n]$ and $\vp^+\equiv [f_n]^+$. The scalar product in the Fock 
space is
$$
(\vp,\vp')\equiv\vp^+\vp'\equiv \sum_{n=0}^\infty(f_n,f_n') 
$$
for any $\vp,\vp' \in{\cal F}$. Hence, the Fock space is a Hilbert space. 
The norm is generated by the scalar product, $||\vp||\equiv\sqrt{(\vp,\vp)}$.

Recall that functions $f_n$ are assumed to be symmetrized, which can be done
by means of the {\it symmetrization operator} ${\hat S}$ that makes of 
functions either symmetric or antisymmetric combinations, depending on whether 
bosons or fermions are considered. Thus, a function of two variables is 
symmetrized as  
$$
{\hat S} h(x_i,x_j)\equiv\frac{1}{2}[ h(x_i,x_j)\pm h(x_j,x_i)] \; .
$$
Similarly, functions of many variables are symmetrized.
 
Sometimes, one assumes that the symmetrization is done over the products of 
the type $f(x_1)\ldots f(x_n)$. This, however, is not necessary, and the 
functions $f_n(x_1,\ldots,x_n)$ can be of any type, provided they are 
symmetrized.

The Fock space ${\cal F}$, containing only symmetrized functions $f_n$
is called the {\it ideal Fock space}. 

The {\it vacuum} in the Fock space is the column
\begin{eqnarray}
|0> \; \equiv \left [ \begin{array}{c}
\nonumber
1 \\
0 \\
0  \\
. \\
. \\      
. \\
. \\
. \\
.
\end{array} \right ] \; .
\end{eqnarray}

The {\it restricted Fock space} $\cF_N$ for the fixed number of particles $N$ 
consists of $\vp_N\in\cF_N$ such that $\vp_N = [ \dlt_{nN} f_N ]$. The 
restricted Fock space is the number-conserving space. This restriction makes
operations in the Fock space more complicated, because of which one prefers
to deal with the non-restricted Fock space. In what follows, we shall deal with
the ideal Fock space that, for short, will be called just the Fock space.

\subsection{Field operators}

The field operators act on the Fock space. One terms $\psi^\dgr(x)$ the 
creation operator and $\psi(x)$, annihilation operator. The action of the
creation operator is given by the formula

\begin{eqnarray}
\psi^\dgr(x)\vp \equiv \left [ \begin{array}{l}
\nonumber
0 \\
f_0 \dlt(x_1-x) \\
\sqrt{2}\; {\hat S} f(x_1)\dlt(x_2-x) \\
. \\
. \\
. \\
\sqrt{n}\; {\hat S} f_{n-1}(x_1,\ldots,x_{n-1})\dlt(x_n-x) \\   \; .
. \\
. \\
.
\end{array} \right ] \; .
\end{eqnarray}

In what follows, the latter form will be represented as 
$$
\psi^\dgr(x)\vp= \left [\sqrt{n}\;
{\hat S} f_{n-1}(x_1,\ldots,x_{n-1})\dlt(x_n-x)\right ] \; . 
$$

The annihilation operator acts according to the rule

\begin{eqnarray}
\psi(x)\vp \equiv \left [ \begin{array}{l}
\nonumber
f_1(x) \\
\sqrt{2} \;f_2(x_1,x)  \\
\sqrt{3}\; f_3(x_1,x_2,x) \\
. \\
. \\
. \\
\sqrt{n+1}\;f_{n+1}(x_1,\ldots,x_n,x) \\   
. \\
. \\
.
\end{array} \right ] \; .
\end{eqnarray}

This will be denoted as 
$$
\psi(x)\vp=\left [\sqrt{n+1}\; f_{n+1}(x_1,\ldots,x_n,x)\right ] \; .
$$
The field operators act on the vectors of the Fock space, but 
$\psi^\dgr(x)$ and $\psi(x)$ are not defined in the Fock space, since
$\psi^\dgr(x)\vp$ lies outside ${\cal F}$. In that sense, the field 
operators are the operator distributions. 

In the Fock space, one can define the operators
$$
\psi_f^\dgr \equiv \int f(x)\psi^\dgr(x) \; dx \; , \quad
\psi_f \equiv \int f^*(x)\psi(x)\; dx \; ,
$$
where $||f|| < \infty$. Their action is given by the expressions
$$
\psi^\dgr_f\vp =\left [ \sqrt{n}\; {\hat S}
f_{n-1}(x_1,\ldots,x_{n-1})f(x_n)\right ]  \; ,
$$
$$
\psi_f\vp =\left [ \sqrt{n+1}\; \int
f_{n+1}(x_1,\ldots,x_n,x)f^*(x)\; dx \right ] \; ,
$$
which are in ${\cal F}$. 

The action of the annihilation operator on the vacuum is
$$
\psi(x)|\; 0>\; = 0 \; .
$$
And the creation operator acts on the vacuum as

\begin{eqnarray}
\psi^\dgr(x)|\; 0> \; = \left [ \begin{array}{l}
\nonumber
0 \\
\dlt(x_1-x) \\
0 \\
. \\                  
. \\
.
\end{array} \right ] \; .
\end{eqnarray}

A double action of the creation operators yields

\begin{eqnarray}
\psi^\dgr(x_1')\psi^\dgr(x_2')|\; 0> \; = \left [ \begin{array}{l}
\nonumber
0 \\
0 \\
\sqrt{2}\; \hat S \dlt(x_1-x_1')\dlt(x_2-x_2') \\   
0 \\
. \\
. \\
. \\
\end{array} \right ] \; .
\end{eqnarray}

Acting on the vacuum by $n$ creation operators gives

\begin{eqnarray}
\prod_{i=1}^n \psi^\dgr(x_i')|\; 0>\; = \left [ \begin{array}{l}
\nonumber
0 \\
0 \\
. \\
. \\
. \\
0 \\
\sqrt{n!}\; \hat S \prod_{i=1}^n \dlt(x_i-x_i') \\    
0 \\
. \\
. \\
.
\end{array} \right ] \; .
\end{eqnarray}

Therefore, any $\vp\in{\cal F}$ can be created by the creation operators
by the formula
$$
\vp = \sum_{n=0}^\infty\; \frac{1}{\sqrt{n!}}\;
\int f_n(x_1,\ldots,x_n)\prod_{i=1}^n \psi^\dgr(x_i)\; dx_i|\; 0> \; .
$$

The double action of the annihilation operators gives
$$
\psi(x)\psi(x')\vp=\left [ \sqrt{(n+1)(n+2)}\;
f_{n+2}(x_1,\ldots,x_n,x,x')\right ]   \; .
$$
This should be compared with the double action of the creation operators
$$
\psi^\dgr(x)\psi^\dgr(x')\vp= \left [ \sqrt{n(n-1)}\;
{\hat S} f_{n-2}(x_1,\ldots,x_{n-2}) \dlt(x_{n-1}-x)
\dlt(x_n-x') \right ]  \; .
$$
Interchanging the variables $x$ and $x'$, we come to the relations 
$$
\psi(x')\psi(x)\vp=\pm\psi(x)\psi(x')\vp \; ,
$$
$$
\psi^\dgr(x')\psi^\dgr(x)\vp=\pm\psi^\dgr(x)\psi^\dgr(x')\vp \; .
$$
Since these equations are valid for any $\vp\in{\cal F}$, we have
$$
[\psi(x),\; \psi(x')]_\mp =
\left [ \psi^\dgr(x),\; \psi^\dgr(x')\right ]_\mp = 0  \; . 
$$
Here the notation for the commutator and anticommutator of two operators 
is used, $[A,\; B]_\mp \equiv AB \mp BA$.

Similarly, we find
$$
\psi(x)\psi^\dgr(x')\vp =\left [ \pm (n+1)
{\hat S} f_n(x_1,\ldots,x_{n-1},x)\dlt(x_n-x')\right ] \; ,
$$
$$
\psi^\dgr(x')\psi(x)\vp =\left [ n {\hat S}
f_n(x_1,\ldots,x_{n-1},x)\dlt(x_n-x')\right ] \; .
$$
From here,
$$
\psi_f\psi_g^\dgr\vp = [(n+1)f_n](f,g) \; ,
$$
$$
\psi^\dgr_g\psi_f\vp = [\pm nf_n](f,g) \; .
$$
This can be rewritten as 
$$
\left [ \psi_f,\; \psi_g^\dgr\right ]_\mp \vp = (f,g)\vp \; , \quad
\left [ \psi_f,\psi_g^\dgr\right ]_\mp = (f,g) \; .
$$
Since these equations are valid for any $\vp\in{\cal F}$ and any 
$f,g\in {\cal H}_1$, we come to the commutation (anticommutation) relations
$$
[\psi(x),\; \psi^\dgr(x')]_\mp = \dlt(x-x') \; .
$$
 
A vector $\vp\in\cF$ is called a quantum state or microstate. The Fock space 
$\cF$ is a space of quantum states or a space of microstates. As is shown, the 
field operators $\psi(x)$ and $\psi^\dgr(x)$ act on the Fock space and can 
create any vector $[f_n(x_1,\ldots,x_n)]$ in this space. Therefore, one tells 
that the Fock space is generated by $\psi$ and denote this Fock space as 
${\cal F}(\psi)$. 

From the variables $x$, one can pass to any other required variables, for 
instance, changing the representation from $x$ to $k$. For this purpose, one 
defines a complete orthonormal basis $\{\vp_k(x)\}$ and expands the field 
operators over this basis,
$$
\psi(x) = \sum_k a_k\vp_k(x) \; , \quad
\psi^\dgr(x) =\sum_k a_k^\dgr\vp_k^*(x) \; .
$$
The operators $a_k$ and $a_k^\dgr$ generate the Fock space ${\cal F}(a_k)$.
The vectors of the latter space have the form 
$[h_n(k_1,\ldots,k_n)]\in {\cal F}(a_k)$.

\subsection{Operators of observables}

In quantum mechanics, observable quantities are represented by operators of 
observables. An $n$-particle operator is written as 
$$
A_n =\sum_{i=1}^n A_1(x_i) +
\frac{1}{2} \sum_{i\neq j}^n A_2(x_i,x_j) + \ldots \; .
$$
It is a Hermitian or self-adjoint operator, such that $A_n^+ = A_n$. 
The kernels $A_n(x_1,\ldots,x_n)$ are quantum-mechanical operators that are 
also self-adjoint. Examples of the quantum-mechanical operators are the 
operator of momentum
$$
{\bf P}_n = \sum_{i=1}^n(-i\nabla_i)  \; ,
$$
where $\nabla_i \equiv \prt/\prt\br_i$, the kinetic-energy operator 
$$
K_n =\sum_{i=1}^n\left ( -\; \frac{\nabla^2_i}{2m}\right ) \; ,
$$
and the potential-energy operator
$$
W_n = \sum_{i=1}^n U(x_i) + \frac{1}{2} \sum_{i\neq j}^n \Phi(x_i,x_j) \; ,
$$
in which the pair interaction potential is symmetric, 
$\Phi(x_i,x_j) = \Phi(x_j,x_i)$. The operator of the total energy is the 
quantum-mechanical {\it Hamiltonian} $H_n = K_n + W_n$. 

In the {\it field representation}, the general form of an operator acting 
on the Fock space is  
$$
\hat A = \sum_{n=0}^\infty \; \frac{1}{n!} \int
\psi^\dgr(x_n) \ldots \psi^\dgr(x_1)\; A_n(x_1,\ldots,x_n)\;
\psi(x_1)\ldots\psi(x_n)\; dx_1\ldots dx_n \; ,
$$
with the kernel $A_n(x_1,\ldots,x_n)$ being a quantum-mechanical operator.
The operators of observables are self-adjoint, $\hat A^+ = \hat A$,
since $A_n^+ = A_n$. 

Examples of the operators of observables in the field representation are 
the number-of-particle operator
$$
\hat N =\int \psi^\dgr(x)\psi(x)\; dx\; , 
$$
the momentum operator
$$
\hat{\bf P} = \int \psi^\dgr(x)(-i\nabla)\psi(x)\; dx \; ,
$$
the kinetic-energy operator
$$
\hat K =\int \psi^\dgr(x)\left ( -\; \frac{\nabla^2}{2m}\right )\psi(x)\; dx \; ,
$$
and the potential-energy operator
$$
\hat W = \int \psi^\dgr(x) U(x)\psi(x)\; dx + \frac{1}{2}\;
\int \psi^\dgr(x)\psi^\dgr(x')\Phi(x,x')\psi(x')\psi(x)\; dx dx' \; .
$$

The total-energy operator is the {\it Hamiltonian} $\hat H = \hat K + \hat W$
that reads as
$$
\hat H = \int\psi^\dgr(x)\left [ -\; \frac{\nabla^2}{2m} + U(x)\right ]
\psi(x)\; dx + \frac{1}{2}\; \int
\psi^\dgr(x)\psi^\dgr(x')\Phi(x,x')\psi(x')\psi(x)\; dx dx' \; . 
$$

The operators of observables act on the Fock space as
$$
\hat N\vp=[nf_n]\; , \quad \hat {\bf P}\vp=[{\bf P}_nf_n] \; ,
$$
$$
\hat K\vp=[K_nf_n]\; , \quad \hat W\vp=[W_nf_n]\; , \quad
\hat H\vp=[H_nf_n] \; .
$$
In the resulting vector columns, one has quantum-mechanical expressions for $n=0,1,2,\ldots$.

\subsection{Statistical operator}

The system statistical state is characterized by a statistical operator 
$\hat\rho(t)$, which is a function of time $t\geq 0$. This is a self-adjoint 
operator, $\hat\rho^+(t)=\hat\rho(t)$. It is semipositive and bounded, in 
the sense of the inequalities
$$
0\leq \frac{\vp^+\hat\rho(t)\vp}{||\vp||^2} < \infty
$$
for all $\vp\in{\cal F}$ and $\vp\neq 0$. It is normalized to one,
${\rm Tr}\hat\rho(t) = 1$, with the trace over the Fock space. 

The trace can be taken over any basis $\{\vp_k\}$ in ${\cal F}$,
which is orthonormal, $\vp_k^+\vp_p=\dlt_{kp}$, and complete, 
$\sum_k\vp_k\vp_k^+=\hat 1$, where ${\hat 1}$ is the identity operator 
in ${\cal F}$. The trace of an operator ${\hat A}$ is given by
${\rm Tr}\hat A \equiv \sum_k \vp_k^+\hat A\vp_k$.

{\it Observable quantities} are defined as the {\it statistical averages}
\be
\label{41}
\langle \hat A(t) \rangle \;\equiv{\rm Tr}\hat\rho(t)\hat A \; . 
\ee

One considers finite systems, with a finite support in real space.
Therefore, the operators of observables $\hat A$ are termed the operators 
of local observables. The {\it algebra of local observables} 
${\cal A}\equiv\{\hat A\}$ is the family of all operators of local 
observables describing the treated system. And one calls the 
{\it statistical state} the set of all observable quantities 
$<{\cal A}>\equiv\{<\hat A>\}$. 

Being defined for a finite system, with $\br$ in volume $V$, the operator 
${\hat A}$ could be written as  
$$
\hat A_V =\sum_{n=0}^\infty \; \frac{1}{n!} \int_V \;
\psi^\dgr(\br_n)\ldots \psi^\dgr(\br_1)\; A_n(\br_1,\ldots,\br_n)\;
\psi(\br_1)\ldots\psi(\br_n)\; d\br_1 \ldots d\br_n \; ,
$$
where the integration is over the finite system volume. Respectively,
the observable quantities are $<\hat A>_V$. The average number of particles 
in the system is $N\equiv<\hat N>_V$. 

When taking the thermodynamic limit
$$
N\ra\infty\; , \quad V\ra \infty\; , \quad
\frac{N}{V}\ra const \; 
$$
for extensive quantities that are proportional to $N$, one should 
consider the ratio
$$
\left | \lim_{N\ra\infty}\; \frac{1}{N}\; <\hat A>_V\right | < \infty \; .
$$
In what follows, the index $V$ in the integrals and averages is omitted,
though always keeping in mind that the integration is over the system 
volume $V$. 

One tells that two operators
$$
\hat A = \int \hat A(\br)\; d\br \; , \quad
\hat B = \int \hat B(\br)\; d\br \; ,
$$
with the densities $\hat A(\br)$ and $\hat B(\br)$, satisfy the 
{\it clustering property} if 
\be
\label{42}
<\hat A(\br)\hat B(\br')>\; \simeq \; <\hat A(\br)><\hat B(\br')>
\ee
for $|\br -\br'| \ra \infty$. This property is assumed to hold for the 
densities of any two operators of observables.

\subsection{Evolution operator}

In the Schr\"{o}dinger picture of quantum mechanics, $n$ particles are 
described by the wave function $\Psi_n(t)=\Psi_n(x_1,\ldots,x_n,t)$, where 
$t\geq 0$ is time. Particles are termed {\it identical}, provided that  
$$
|\hat P_{ij}\Psi_n(t)|^2 = |\Psi_n(t)|^2 \; ,
$$
where ${\hat P}_{ij}$ is the operator of particle exchange. The 
{\it symmetry principle}  
$$
\hat P_{ij}\Psi_n(t) = \pm \Psi_n(t)
$$
is a sufficient condition for the particles to be identical. The upper
sign corresponds to the Bose-Einstein and the lower, to the Fermi-Dirac 
statistics.

The scalar product is given by 
$$
(\Psi_n,\Psi_n') \equiv \int \Psi_n^*(x_1,\ldots,x_n,t)
\Psi_n'(x_1,\ldots,x_n,t)\; dx_1\ldots dx_n  \; ,
$$
with the norm generated by this scalar product, 
$||\Psi_n(t)||=\sqrt{(\Psi_n,\Psi_n)}$. 

The number of particles $n$ is conserved in time, which imposes the 
condition $||\Psi_n(t)||=||\Psi_n(0)||$. The wave function pertains
to a Hilbert space ${\cal H}_n$ and satisfies the Schr\"{o}dinger 
equation
$$
i\; \frac{\prt}{\prt t} \; \Psi_n(t) =  H_n\Psi_n(t) \; . 
$$

In the Fock space the wave function can be represented as a collection 
of the Schr\"{o}dinger wave functions

\begin{eqnarray}
\Psi(t)  = \left [ \begin{array}{l}
\nonumber
\Psi_0(t) \\
\Psi_1(t) \\
. \\
. \\
. \\
\Psi_n(t) \\
. \\
.\\
. \\
\end{array} \right ] = [\Psi_n(t) ] \; ,
\end{eqnarray}
with $||\Psi(t)||=const$. The Fock wave function $\Psi(t)\in {\cal F}$
satisfies the Schr\"{o}dinger equation 
$$
i\; \frac{d}{dt}\; \Psi(t) = \hat H\Psi(t)  \; , 
$$
where
$$
\hat H\Psi(t) = [ H_n\Psi_n(t)] = 
\left [ i\; \frac{\prt}{\prt t}\; \Psi_n(t)\right ] \; , 
$$
which is equivalent to the set of the Schr\"{o}dinger equations 
in ${\cal H}_n$. 

The time evolution of the wave function in the Fock space can be 
represented by means of an {\it evolution operator} $\hat U(t)$ defined 
so that
$$
\Psi(t)=\hat U(t)\Psi(0) \; , \quad \Psi^+(t)=\Psi^+(0)\hat U^+(t) \; .
$$
The evolution operator is additive, $\hat U(t+t')=\hat U(t)\hat U(t')$.
There exists the identity $\hat U(0) = \hat 1$ and the inverse 
$\hat U^{-1}(t)$. From the norm conservation $||\Psi(t)||^2=||\Psi(0)||^2$, 
we have
$$
\Psi^+(t)\Psi(t)=\Psi^+(0)\Psi(0) = \Psi^+(0)\hat U^+(t)\hat U(t)\Psi(0) \; , 
$$
which shows that $\hat U^+(t)\hat U(t) = \hat 1$. Hence the evolution 
operator is a unitary operator, such that $\hat U^+(t)= \hat U^{-1}(t)$,
with $\hat U^{-1}(t)\hat U(t) = \hat 1$. By these properties, the evolution 
operators, labeled by the time index, form the {\it evolution group}
$\{\hat U(t)\}$. 

The Schr\"odinger equation for the wave function $\Psi(t)$ is equivalent to
the Schr\"odinger equation for $\hat U(t)$, 
\be
\label{43}
i\; \frac{d}{dt}\; \hat U(t) = \hat H\hat U(t) \; , \quad
i\; \frac{d}{dt}\; \hat U^+(t) = -\hat U^+(t)\hat H \; ,
\ee
where the fact that the Hamiltonian is self-adjoint, $\hat H^+=\hat H$, is 
used. 

In general, the evolution operator cannot be represented in explicit form
as a function of the Hamiltonian. This is possible, only when the    
{\it Lyappo-Danilevsky condition} is valid:
\be
\label{44}
\left [ \hat H, \; \int_0^t \hat H\; dt \right ] = 0 \; .
\ee
If this is so, then the formal solution for the evolution operator is
$$
\hat U(t) = \exp\left \{ - i \int_0^t \hat H \; dt \right \} \; .
$$
This means that the commutator
$$
\left [ \hat H, \; \hat U(t) \right ] = 0 \; 
$$
has to be true. 

In particular, for $\hat H$ not explicitly depending on $t$,
$$
\hat U(t)=\exp\left ( -i\hat H t\right ) \; .
$$
But, in general, the evolution operator cannot be represented in such 
explicit forms.

\subsection{Liouville equation}

An $n$-particle wave function in the Fock space can be written as

\begin{eqnarray}
\vp_n(t)  = \left [ \begin{array}{l}
\nonumber
0 \\
. \\
. \\
. \\
0\\
\Psi_n(t) \\
0 \\
. \\
.\\
. \\
\end{array} \right ] \; .
\end{eqnarray}

A pure-state statistical operator is given by the product
$\hat\rho_n(t)=\vp_n(t)\vp_n^+(t)$. It is the normalized operator,
with ${\rm Tr}\hat\rho_n(t) = 1$, which follows from the wave-function 
normalizations $||\Psi_n(t)|| = 1$ and $||\vp_n(t)||=1$. The pure-state
operator is idempotent, $\hat\rho_n^2(t) = \hat\rho_n(t)$. 

A mixed-state statistical operator can be written as
$$
\hat\rho(t) =\sum_n w_n\vp_n(t)\vp_n^+(t) \; .
$$
Since the statistical operator has to be semipositive and normalized, 
${\rm Tr}\hat\rho(t) = 1$, it follows that the weights $w_n$ are to be such that
$$
0\leq w_n\leq 1 \;, \quad \sum_n w_n=1 \; .
$$

The time evolution reads as 
$$
\hat\rho(t) =\hat U(t) \sum_n w_n \vp_n(0)\vp_n^+(0)\hat U^+(t) \; .
$$
Denoting the initial statistical operator
$$
\hat\rho(0) = \sum_n w_n\vp_n(0)\vp_n^+(0) \; ,
$$
we come to the form
$$
\hat\rho(t) = \hat U(t)\hat\rho(0) \hat U^+(t)  
$$
defining the time evolution of the statistical operator. The latter form
leads to the {\it Liouville equation}
\be
\label{45}
i\; \frac{d}{dt}\; \hat\rho(t) = \left [ \hat H,\; \hat\rho(t)\right ] \; ,
\ee
in which $ [\hat A,\; \hat B] \equiv \hat A\hat B - \hat B\hat A$.

\subsection{Heisenberg picture}

The observable quantities are given by the averages 
$<\hat A(t)>\equiv{\rm Tr}\hat\rho(t)\hat A$, which, using the evolution 
operator, can be written as
$$
{\rm Tr}\hat\rho(t)\hat A ={\rm Tr}\hat\rho(0)\hat U^+(t)\hat A\hat U(t) \; .
$$
The {\it Heisenberg representation} for the operators of observables is
\be
\label{46}
\hat A(t)\equiv \hat U^+(t)\hat A(0)\hat U(t) \; ,
\ee
with the initial condition $\hat A(0) = \hat A$. Therefore, we have
$$
{\rm Tr}\hat\rho(0)\hat U^+(t)\hat A\hat U(t) ={\rm Tr}\hat\rho(0)\hat A(t) \; .
$$
Then the observable quantities can be written in two equivalent forms:
\be
\label{47}
<\hat A(t)>={\rm Tr}\hat\rho(t)\hat A = {\rm Tr}\hat\rho(0)\hat A(t) \; .
\ee
The relation (46) can also be written as the {\it Heisenberg equation of motion}
\be
\label{48}
i\; \frac{d}{dt}\; \hat A(t) = \left [ \hat A(t),\; \hat H(t)\right ] \; .
\ee
We should notice the difference of the Hamiltonian form in the Heisenberg 
equation from that in the Liouville equation. In the latter, we have ${\hat H}$,
while the former contains 
$$
\hat H(t)\equiv \hat U^+(t)\hat H\hat U(t) \; . 
$$
Generally, $\hat H(t)$ does not equal $\hat H$. This happens only when 
$\hat H$ commutes with the evolution operator $\hat U$, which holds true when 
the Hamiltonian $\hat H$ satisfies the Layppo-Danilevsky condition (44). 

The Heisenberg equation (48) leads to the time evolution of the observable
quantities
$$
i\; \frac{d}{dt}\; <\hat A(t)> \; = \; <\left [ \hat A(t),\; \hat H(t)
\right ] > \; .
$$

In this way, one can use two representations. In the Schr\"odinger picture,
one deals with $\Psi(t)$, $\hat\rho(t)$, and $\hat A(0)$. And in the 
Heisenberg picture, one employs $\Psi(0)$, $\hat\rho(0)$, and $\hat A(t)$.

The generic form of the operators of observables is
$$
\hat A = \sum_{n=0}^\infty \; \frac{1}{n!}
\int \psi^\dgr(x_n)\ldots\psi^\dgr(x_1) A_n(x_1,\ldots,x_n)
\psi(x_1)\ldots\psi(x_n)\; dx_1\ldots dx_n \; .
$$

The time evolution of the field operator reads as 
\be
\label{49}
\psi(x,t)\equiv \hat U^+(t)\psi(x,0)\hat U(t) \; ,
\ee
with the initial condition $\psi(x,0) = \psi(x)$. This is equivalent to
the equation of motion
\be
\label{50}
i\; \frac{\prt}{\prt t}\; \psi(x,t) = \left [ \psi(x,t),\; \hat H(t)
\right ] \; , 
\ee
where $\hat H(t) = \hat U^+(t)\hat H\hat U(t)$.

For the Heisenberg representation of an operator of observable 
$\hat A(t)=\hat U^+(t)\hat A\hat U(t)$, we find
$$
\hat A(t) = \sum_{n=0}^\infty \; \frac{1}{n!}
\int \psi^\dgr(x_n,t)\ldots\psi^\dgr(x_1,t) A_n(x_1,\ldots,x_n)
\psi(x_1,t)\ldots\psi(x_n,t)\; dx_1\ldots dx_n \; .
$$
If the kernel $A_n(x_1,\ldots,x_n,t)$ explicitly depends on time, then
the Heisenberg equation of motion reads as
\be
\label{51}
i\; \frac{d}{dt}\; \hat A(t) = \left [ \hat A(t), \; \hat H(t)\right ] +
\hat U^+(t)\; i\; \frac{\prt\hat A}{\prt t}\; \hat U(t) \; .
\ee
Again, one has to pay attention that, contrary to the Liouville equation, 
here we have $\hat H(t)$.

The commutation relations for the field operators from Sec. 2.2, can be
represented for the time-dependent field operators (49). To this end,
we see that   
$$
[\psi(x,t),\; \psi(x',t)]_\mp =\hat U^+(t) [\psi(x),\; \psi(x')]_\mp
\hat U(t) \; ,
$$
and similarly for other relations. Therefore, we obtain the equal-time 
commutation relations
\be
\label{52}
[\psi(x,t),\; \psi(x',t)]_\mp =0 \; , \quad 
[\psi^\dgr(x,t),\; \psi^\dgr(x',t)]_\mp =0 \; , \quad
[\psi(x,t),\; \psi^\dgr(x',t)]_\mp =\dlt(x-x') \; .
\ee

According to Sec. 2.3, the Hamiltonian reads as
$$
\hat H(t) = \int \psi^\dgr(x,t) \left [ -\; \frac{\nabla^2}{2m} +
U(x,t)\right ] \psi(x,t)\; dx +
$$
\be
\label{53}
+  \frac{1}{2} \int \psi^\dgr(x,t)
\psi^\dgr(x',t)\Phi(x,x')\psi(x',t)\psi(x,t)\; dx dx' \; , 
\ee
so that the equation of motion (50) becomes
\be
\label{54}
i\; \frac{\prt}{\prt t}\; \psi(x,t) = \left [ -\; \frac{\nabla^2}{2m}
+ U(x,t) \right ] \psi(x,t) + \int \Phi(x,x')\psi^\dgr(x',t)\psi(x',t)
\psi(x,t)\; dx \; .
\ee

It is important to stress that the Heisenberg equation of motion (50)
for the field operator is equivalent to the variational equation
\be
\label{55}
i\; \frac{\prt}{\prt t} \; \psi(x,t) =
\frac{\dlt\hat H(t)}{\dlt\psi^\dgr(x,t)} \; .
\ee
The proof of their equivalence is given in Ref. \cite{Yukalov_43}.

\subsection{Local dynamics}

Using the evolution equations in the Heisenberg representation, it is 
straightforward to derive the equations for different local density 
operators \cite{Bogolubov_44}. Let us consider the local density of 
particles 
\be
\label{56}
\hat n(x,t)\equiv \psi^\dgr(x,t)\psi(x,t) \; .
\ee
The density commutes with itself, $[\hat n(x,t),\hat n(x',t)] = 0$, 
for any statistics. The integral of the density gives the operator of the 
total number of particles 
$$
\hat N = \int \hat n(x,t)\; dx \; .
$$
And let us define the local current density of particles 
$\hat{\bf j}=\hat{\bf j}(x,t)$ by the equation
\be
\label{57}
\hat{\bf j} \equiv -\;\frac{i}{2m}\left [ \psi^\dgr(\nabla\psi) -
(\nabla\psi^\dgr)\psi\right ] \; ,
\ee
where, for short, we write $\psi=\psi(x,t)$ and $x=\{\br,\ldots\}$.

First, it is easy to derive the {\it continuity equation}
\be
\label{58}
\frac{\prt\hat n}{\prt t} + \nabla\cdot\hat{\bf j} = 0 \; .
\ee
The equation for the current density follows:
\be
\label{59}
\frac{\prt\hat{\bf j}}{\prt t} + \frac{1}{4m^2}\left \{
(\nabla^2\psi^\dgr)(\nabla\psi) - \psi^\dgr\nabla(\nabla^2\psi) +
(\nabla\psi^\dgr)(\nabla^2\psi) -\left [ \nabla(\nabla^2\psi^\dgr)
\right ] \psi \right \} =
\ee
$$
=  -\; \frac{1}{m}\; \psi^\dgr\left [\nabla( U + \hat \Phi)\right ] \psi \; ,
$$
where
$$
 \hat\Phi = \hat\Phi(x,t) \equiv \int \Phi(x,x')\; \hat n(x',t)\; dx' \; . 
$$

Analogously, we can get the equation for the time variation of the  
Hamiltonian density (density of energy)
\be
\label{60}
\hat H(x,t) \equiv \frac{1}{2m}\; \nabla\psi^\dgr(x,t) \cdot
\nabla\psi(x,t) + \psi(x,t)\left [ U(x) + \frac{1}{2}\; \hat\Phi(x,t)
\right ] \psi(x,t) \; ,
\ee
which can also be written in the form
\be
\label{61}
\hat H(x,t) \equiv \psi^\dgr(x,t)\left [ -\; \frac{\nabla^2}{2m} +
U(x) + \frac{1}{2}\; \hat\Phi(x,t) \right ] \psi(x,t) \; .
\ee
The total energy Hamiltonian is $\hat H = \int \hat H(x,t)\; dx$.

It is worth emphasizing that the variation in space bears some analogy
with the variation in time. To show this, let us treat $x$ as $\br$.
Consider the operator of momentum
$$
\hat{\bf P} = m \int \hat{\bf j}(\br)\; d\br = 
\int \psi^\dgr(\br)(-i\nabla) \psi(\br)\; d\br \; .
$$
We see that the commutator
$$
[\psi(\br),\; \hat{\bf P}] = -i\nabla\psi(\br) 
$$
reminds us the Heisenberg evolution equation (50). Introducing the 
{\it translation operator} 
\be
\label{62}
\hat T(\br)\equiv\exp(i\br\cdot \hat{\bf P}) \; ,
\ee
we find
$$
\psi(\br)=\hat T^+(\br)\psi(0)\hat T(\br) \; .
$$
This is similar to the dependence of the field operator on time, as
expressed through the evolution operator in Eq. (49). That is, the 
translation operator (62) describes the spatial variation in a manner
similar to the description of the time evolution by means of the 
evolution operator $\hat U(t)$. 

The translation operator (62) enjoys the properties
$$
\hat T(0)=\hat 1 \; , \quad \hat T^+(\br)\hat T(\br) = \hat 1 \; ,
\quad
\hat T^{-1}(\br)=\hat T(-\br)=\exp(-i\br\cdot\hat{\bf P}) \; ,
\quad
\hat T(\br)\hat T(\br') = \hat T(\br+\br') \; .
$$
Hence the family $\{ \hat T(\br)\}$ forms a unitary translation group.

\subsection{Coherent states}

Coherent states are the eigenstates of the annihilation operator. A
coherent state $|\eta>\in{\cal F}$ is a solution of the eigenproblem
\be
\label{63} 
\psi(x)|\; \eta> \; =\eta(x)|\; \eta> \; ,
\ee
where $\psi(x)\equiv\psi(x,0)$ and $\eta(x)$ is called {\it coherent field}.
Being a vector of the Fock space, the coherent state has the form

\begin{eqnarray}
|\eta>  = \left [ \begin{array}{l}
\nonumber
\eta_0 \\
\eta_1(x_1) \\
\eta_2(x_1,x_2) \\
. \\
. \\
. \\
\eta_n(x_1,\ldots,x_n) \\
. \\
.\\
. \\
\end{array} \right ] \; .
\end{eqnarray}

For short, we shall denote this coherent state as
$$
|\eta>=[\eta_n(x_1,\ldots,x_n)] \; .
$$
A conjugate to $|\eta >$ is $|\eta>^+\equiv<\eta|$. The scalar product 
of two coherent states is 
$$
<\eta|\xi>=\sum_{n=0}^\infty (\eta_n,\xi_n) \; ,
$$
where the scalar product of the components is
$$
(\eta_n,\xi_n) = \int \eta_n^*(x_1,\ldots,x_n)\xi_n(x_1,\ldots,x_n)\;
dx_1\ldots dx_n \; .
$$

By the definition of the annihilation operator, we have
$$
\psi(x)|\; \eta>\; = \left [ \sqrt{n+1}\;
\eta_{n+1}(x_1,\ldots,x_n,x)\right ] \; .
$$
Taking into account the definition of the coherent state (63), we get
the recursion relation 
$$
\sqrt{n+1}\; \eta_{n+1}(x_1,\ldots,x_n,x) = \eta(x)\eta_n(x_1,\ldots,x_n) \; .
$$
The iterative solution of the latter recursion relation results in the form 
\be
\label{64}
\eta_n(x_1,\ldots,x_n) = \frac{\eta_0}{\sqrt{n!}} \; \prod_{i=1}^n
\eta(x_i) \; .
\ee
The scalar product of the components (64) gives 
$$
(\eta_n,\xi_n) = \frac{\eta_0^*\xi_0}{n!}\; (\eta,\xi)^n \; .
$$

The scalar product of two coherent states reads as 
$$
<\eta|\xi>\; = \sum_{n=0}^\infty \; \frac{\eta_0^*\xi_0}{n!} \;
(\eta,\xi)^n = \eta_0^*\xi_0 e^{(\eta,\xi)}  \; ,
$$
where
$$
(\eta,\xi) \equiv \int \eta^*(x)\xi(x)\; dx \; .
$$
The coherent state can be normalized as 
$$
||\; |\eta>||^2 =<\eta|\eta>=1 \; ,
$$
which defines
$$
|\eta_0| =\exp\left\{ -\; \frac{1}{2}\; (\eta,\eta)\right \} \; .
$$

In this way, the scalar product of two normalized coherent states becomes
$$
<\eta|\xi>\; = \exp\left\{ -\; \frac{1}{2}\; (\eta,\eta) + (\eta,\xi) - \;
\frac{1}{2}\; (\xi,\xi)\right \} \; .
$$
Since $0 < | <\eta|\xi>| \leq 1$, it follows that the coherent states are
not orthogonal with each other. The set $\{ |\eta>\}$ forms an overcomplete
basis for which the resolution of unity holds true:
$$
\int |\eta><\eta|\; {\cal D}\eta = \hat 1 \; .
$$

Integration over the coherent fields can be introduced as follows. Let 
us take an orthonormal complete basis $\{\chi_k(x)\}$ and let us expand 
the coherent field as
$$
\eta(x) =\sum_k c_k\chi_k(x) \; , \quad c_k\equiv|c_k|e^{i\al_k} \; ,
$$
with $\al_k$ real-valued and $0\leq|c_k|\leq 1$. Then the integration over
the complex-valued coherent field is defined with the help of the 
differential measure 
$$
{\cal D}\eta \equiv \prod_k d|c_k|^2 \; \frac{d\al_k}{2\pi} \; .
$$

A coherent state $|\eta>$ is an eigenstate of the annihilation operator, 
but it is not an eigenstate of the creation operator $\psi^\dgr(x)$. This
is seen from the action 
$$
\psi^\dgr(x)|\; \eta>\; = \left [ \sqrt{n}\;  \hat S
\eta_{n-1}(x_1,\ldots,x_{n-1})\; \dlt(x_n-x)\right ] \; .
$$
However, we have the equation
$$
<\eta |\; \psi^\dgr(x) = \; <\eta|\; \eta^*(x) \; .
$$

In view of Eq. (64), the coherent state can be represented as
$$
|\eta> \; = \sum_{n=0}^\infty\; \frac{\eta_0}{n!} \left (
\int \eta(x)\psi^\dgr(x)\; dx\right )^n |\; 0> \; .
$$
In turn, the latter takes the form
\be
\label{65}
|\eta>\; = \eta_0\exp\left\{ \int \eta(x)\psi^\dgr(x)\; dx\right\}| \; 0> \; .
\ee

Since the coherent state pertains to the ideal Fock space, its components
possess the symmetry property  
$$
\hat P_{ij}\eta_n(x_1,\ldots,x_n) =\pm\eta_n(x_1,\ldots,x_n) \; 
$$
with respect to the permutation of the variables $x_i$ and $x_j$, for $i\neq j$.
Therefore the coherent fields satisfy the commutation relations 
$$
\eta(x_i)\eta(x_j) - \eta(x_j)\eta(x_i) = 0 \qquad ({\rm Bose})
$$
in the case of bosons and the anticommutation relations
$$
\eta(x_i)\eta(x_j) + \eta(x_j)\eta(x_i) = 0 \qquad ({\rm Fermi})
$$
in the case of fermions. For bosons, the coherent field is a complex-valued
function.  While the fermionic coherent fields are Grassmann variables.  

Time-dependent coherent states can be defined in the same way as it is done 
for other states in the Fock space:
\be
\label{66}
|\eta(t)>\; =\hat U(t)|\eta> = [\eta_n(x_1,\ldots,x_n,t)] \; .
\ee
The invariance of the normalization is assumed, so that
$$
<\eta(t)|\eta(t)>\; = \; <\eta|\eta> = 1 \; .
$$
By treating the coherent state as an eigenstate of the annihilation 
operator satisfying the eigenproblem
\be
\label{67}
\psi(x)|\eta(t)>\; = \eta(x,t)|\eta(t)> \; ,
\ee
we get the definition of a time-dependent coherent field $\eta(x,t)$.  
The latter equation is equivalent to 
$$
\psi(x)\hat U(t)|\eta>\; = \eta(x,t)\hat U(t)|\eta> \; ,
$$
from where
$$
\hat U^+(t)\psi(x)\hat U(t)|\eta>\; =\eta(x,t)|\eta> \; .
$$
With $\psi(x,t)=\hat U^+(t)\psi(x)\hat U(t)$, we obtain another form of 
Eq. (67) as
\be
\label{68}
\psi(x,t)|\eta>\; = \; \eta(x,t)|\eta> \; .
\ee

Using the time-dependent coherent states, it is possible to pass to the 
coherent-state representation for the operators of observables. For instance,
the number-of-particle operator yields
$$
<\eta|\hat N(t)|\eta>\; = \int |\eta(x,t)|^2\; dx \; .
$$
And for the Hamiltonian, we find
$$
<\eta|\hat H(t)|\eta>\; = \int \eta^*(x,t)\left [ - \;
\frac{\nabla^2}{2m} + U(x)\right ]\eta(x,t)\; dx +
$$
$$
+ \frac{1}{2} \int |\eta(x,t)|^2 \Phi(x,x') |\eta(x',t)|^2\; dx dx' \; .
$$
The temporal derivative gives
$$
<\eta\; |\; i\; \frac{\prt}{\prt t}\; \psi(x,t)\; |\; \eta>\; =
i\; \frac{\prt}{\prt t}\; \eta(x,t) \; .
$$

The evolution equation (55) transforms into the equation
\be
\label{69}
i\; \frac{\prt}{\prt t}\; \eta(x,t) =
\frac{\dlt<\eta|\hat H(t)|\eta>}{\dlt\eta^*(x,t)} \; .
\ee
Explicitly, we obtain the {\it coherent-field equation}
\be
\label{70}
i\; \frac{\prt}{\prt t} \; \eta(x,t) = \left [ -\;
\frac{\nabla^2}{2m} + U(x) + \int \Phi(x,x')|\eta(x',t)|^2 \; dx'
\right ]\; \eta(x,t) \; .
\ee
This is the nonlinear Schr\"odinger equation.

Presenting the solution in the form 
$$
\eta(x,t) =\eta(x) e^{-iEt} \; ,
$$
we get the stationary coherent-field equation 
\be
\label{71}
\left [ -\; \frac{\nabla^2}{2m} + U(x) + \int \Phi(x,x')|\eta(x')|^2 \;
dx' \right ] \eta(x) = E\eta(x) \; .
\ee
It is important to keep in mind that, as is seen, the above equation has
a meaning, provided that the interaction potential is integrable, such 
that
$$
\left | \int \Phi(x,x')\; dx'\right | < \infty \; .
$$
The coherent-field is defined by this equation up to an arbitrary global 
phase, that is, the general solution is $\eta_\al(x,t) = e^{i\al}\eta(x,t)$,
with $\al$ being an arbitrary real-valued quantity. 

More information on coherent states can be found in the books 
\cite{Klauder_45,Yukalov_46}.

\subsection{Density matrices}

Statistical averages of the products of field operators define the reduced 
density matrices \cite{Coleman_25}. An $n$-order density matrix is defined as
\be
\label{72}
\rho_n(x_1,\ldots,x_n,x_1',\ldots,x_n',t) \equiv {\rm Tr}\;
\psi(x_1)\ldots\psi(x_n)\hat\rho(t)\psi^\dgr(x_n')\ldots\psi^\dgr(x_1') =
\ee
$$
= {\rm Tr}\; \psi(x_1,t)\ldots \psi(x_n,t) \hat\rho\psi^\dgr(x_n',t)
\ldots\psi^\dgr(x_1',t) =
$$
$$
=\;  < \psi^\dgr(x_n',t)\ldots \psi^\dgr(x_1',t)
\psi(x_1,t)\ldots\psi(x_n,t) > \; .
$$
They are symmetric with respect to the simultaneous permutation of $x_i$ 
with $x_j$ and $x_i'$ with $x_j'$, 
$$
\hat P_{ij}\rho_n(x_1,\ldots,x_n,x_1',\ldots,x_n',t) =
\rho_n(x_1,\ldots,x_n,x_1',\ldots,x_n',t) \; .
$$
Also, they enjoy the following properties:
$$
\rho^*_n(x_1,\ldots,x_n,x_1',\ldots,x_n',t) =
\rho_n(x_n',\ldots,x_1',x_n,\ldots,x_1,t) \; ,
$$
$$
\rho_n(x_1,\ldots,x_n,x_1',\ldots,x_n',t) =
\rho_n(x_n,\ldots,x_1, x_n',\ldots,x_1',t) \; ,
$$
$$
\rho^*_n(x_1,\ldots, x_n,x_1',\ldots, x_n',t) =
\rho_n(x_1',\ldots, x_n', x_1,\ldots, x_n,t) \; .
$$

The matrices are called {\it reduced} because their partial integration 
reduces them to lower-order matrices. Suppose we consider the restricted 
number-conserving space $\cF_N$, for which the number of particles is fixed, 
so that for $\vp_N\in\cF_N$, one has $\hat N \vp_N = N \vp_N$. Then, it 
follows that $<\hat N \hat A>= N<\hat A>$. For the second-order density 
matrix, we have
$$
\rho_2(x,x'',x',x'',t) \; = \;
< \hat n(x'',t)\psi^\dgr(x',t) \psi(x,t)> - \; 
\dlt(x'-x'') \rho_1(x,x'',t) \; ,
$$
where $\hat n(x,t) \equiv \psi^\dgr(x,t)\psi(x,t)$ is the number-of-particle 
density. Integrating over $x''$ yields
$$
\int \rho_2(x,x'',x',x'',t)\; dx'' \; = \; 
< \hat N \psi^\dgr(x',t)\psi(x,t)> - \rho_1(x,x',t) \; .
$$
On the restricted Fock space $\cF_N$,
$$
< \hat N \psi^\dgr(x',t)\psi(x,t)> \; = \; N \rho_1(x,x',t) \; .
$$
Therefore,
$$
\int \rho_2(x,x'',x',x'',t)\; dx'' =  (N-1) \rho_1(x,x',t) \; .
$$

The name of the matrices contains the term {\it density}, since the 
diagonal element of the first-order matrix  
$$
\rho_1(x,x',t) =\; <\psi^\dgr(x',t)\psi(x,t)> 
$$
defines the particle density
$$
\rho_1(x,x,t)=<\psi^\dgr(x,t)\psi(x,t)>\equiv\rho(x,t) \; .
$$

The name {\it matrices} means that these objects can be treated as matrices.
Thus, the matrix 
$$
\hat\rho_n=[\rho_n(x_1,\ldots, x_n,x_1',\ldots, x_n',t)]  
$$
is a matrix with respect to the $x$-variables. This matrix is evidently 
self-adjoint, $\hat\rho_n^+=\hat\rho_n$, and 
semipositive, since
$$
\frac{(f_n,\hat\rho_nf_n)}{||f_n||^2} \geq 0 \; .
$$
Here, the scalar product 
$$
(f_n,\hat\rho_n f_n) \equiv \int f_n^*(x_1,\ldots, x_n)
\rho(x_1,\ldots, x_n,x_1',\ldots, x_n',t) f_n(x_1',\ldots, x_n')\;
dx_1\ldots dx_n\; dx_1'\ldots dx_n'
$$
and the vector norm 
$$
||f_n||^2 =(f_n,f_n) = \int f_n^*(x_1,\ldots,x_n) f_x(x_1,\ldots,x_n)\;
dx_1 \ldots dx_n
$$
are used. 

The norm of an $n$-order density matrix is given by the expression
$$
||\hat\rho_n|| \equiv \sup_{||f_n||=1}\; \left |\left (
f_n,\hat\rho_nf_n\right )\right |  \; .
$$

The eigenproblem  
$$
\hat\rho_n f_{nj} =\lbd_{nj}\; f_{nj} \; ,
$$
implies the equation
$$
\int \rho_n(x_1,\ldots, x_n,x_1',\ldots, x_n',t)
f_{nj}(x_1',\ldots,x_n')\;dx_1' \ldots dx_n' =
\lbd_{nj} f_{nj}(x_1,\ldots, x_n) \; .
$$
Being self-adjoint and semipositive, the density matrix enjoys real 
non-negative eigenvalues. The largest of the eigenvalues defining the 
norm 
$$
||\hat\rho_n|| =\sup_j \lbd_{nj} \equiv \lbd_n
$$
is of special importance characterizing the level of ordering in the 
system. 

The normalization for the density matrices can be summarized as follows:
$$
{\rm Tr}\hat\rho_1 \equiv \int \rho_1(x,x,t)\; dx = N \; ,
$$
$$
{\rm Tr}\hat\rho_2 \equiv \int \rho_2(x_1,x_2,x_1,x_2,t)\; dx_1 dx_2 =
N(N-1) \; ,
$$
$$
{\rm Tr}\hat\rho_n \equiv \int \rho_n(x_1,\ldots,x_n,x_1,\ldots,x_n,t)\;
dx_1\ldots dx_n
= \frac{N!}{(N-n)!} \; .
$$

The level of ordering in the system can be described 
\cite{Coleman_25,Yukalov_26} by the {\it order indices}
\be
\label{73}
\om(\hat\rho_n) \equiv \frac{\log||\hat\rho_n||}{\log{\rm Tr}\hat\rho_n} \; .
\ee
Keeping in mind that $N\gg n\geq 1$, we may use the Stirling formula
$z! \simeq \sqrt{2\pi}\; e^{-z} \; z^{z+1/2}$, in which $|z|\ra\infty$. 
Then, we have
$$
\frac{N!}{(N-n)!} \simeq \left ( \frac{N}{e}\right )^n \; , \quad
\log\; \frac{N!}{(N-n)!} \simeq n\log N \; .
$$
As a result, the order index (73) reads as 
\be
\label{74}
\om(\hat\rho_n) = \frac{\log\lbd_n}{n\log N} \; ,
\ee
for $N\gg n\geq 1$. 

The order indices allow us to distinguish between long-range order,
mid-range order, and short-range order \cite{Coleman_25,Yukalov_26}.
This is contrary to the order parameters that distinguish only the 
long-range order from the absence of such an order. 

By means of the reduced density matrices, it is possible to express the 
statistical averages of operators of observables. In the general case,
we have  
$$
<\hat A(t)> \; = \sum_{n=0}^\infty \frac{1}{n!}\;
\int \lim_{\{x_i'\ra x_i\} } \; A_n(x_1,\ldots, x_n)
\rho_n(x_1,\ldots,x_n,x_1',\ldots,x_n',t)\; dx_1\ldots dx_n \; .
$$
In particular cases, for the average of the number-of-particle operator,
we get
$$
<\hat N>\; = \int \rho_1(x,x,t)\; dx = N \; .
$$
For the kinetic-energy operator,
$$
<\hat K>\; = \int \lim_{x'\ra x}\left ( -\;
\frac{\nabla^2}{2m}\right )\rho_1(x,x',t)\; dx \; .
$$
And for the potential-energy operator
$$
<\hat W>\; = \int U(x)\rho_1(x,x,t)\; dx +
\frac{1}{2}\; \int \Phi(x_1,x_2)\rho_2(x_1,x_2,x_1,x_2,t)\; dx_1 dx_2 \; . 
$$
The total internal energy is $E\equiv\; <\hat H>\; =\; <\hat K> + <\hat W>$.

The equations of motion for the reduced density matrices follow from their
definition and Eq. (54) for the field operator. The first-order density 
matrix satisfies the equation 
\be
\label{75}
i\; \frac{\prt}{\prt t} \; \rho_1(x_1,x_2,t) = \left [ -\;
\frac{\nabla_1^2}{2m} + U(x_1) + \frac{\nabla_2^2}{2m}\; -\; U(x_2)
\right ]\; \rho_1(x_1,x_2,t) +
\ee
$$
+ \int \left [ \Phi(x_1,x_3) -\Phi(x_2,x_3)\right ]\;
\rho_2(x_1,x_3,x_2,x_3,t)\; dx_3 \; ,
$$
involving the second-order matrix. The equation for the second-order matrix
involves the third-order matrix, and so on. The sequence of these equations  
is called the {\it Bogolubov - Born - Green - Kirkwood - Yvon hierarchy}
\cite{Bogolubov_44,Kirkwood_47,Bogolubov_48,Bogolubov_49}.

Particle correlations are characterized by the {\it pair correlation function}
$$
g(x,x',t) \equiv \;
\frac{<\psi^\dgr(x,t)\psi^\dgr(x',t)\psi(x',t)\psi(x,t)>}{\rho(x,t)\rho(x',t)} \; ,
$$
which is expressed through the second-order density matrix:
$$
g(x,x',t) = \frac{\rho_2(x,x',x,x',t)}{\rho(x,t)\rho(x',t)} \; .
$$
The pair correlation function is real, $g^*(x,x',t) = g(x,x',t)$ and
symmetric, $g(x',x,t) = g(x,x',t)$. It satisfies the asymptotic condition
$$
g(x,x',t)\ra 1 \quad (|x-x'|\ra\infty),
$$
following from the clustering property (42).

\section{Statistical states}

In this section, different types of statistical states are treated and their
properties are discussed. Several important mathematical facts are presented,
which are widely used for describing quantum systems.

\subsection{Gibbs entropy}

Gibbs entropy is the central notion for characterizing statistical states.
It is also called Shannon entropy or information entropy, since it quantifies
the amount of the unavailable information on the considered system. For a 
statistical operator $\hat\rho$, the Gibbs entropy is
\be
\label{76}
S(t)\equiv -{\rm Tr}\hat\rho(t)\ln\hat\rho(t) \; .
\ee
Since $\hat\rho(t)$ is semipositive, bounded, and normalized to unity, $S(t)$
is semipositive, $S(t)\geq 0$. 

Here and in what follows, the functions of operators are defined by the rule
$$
f(\hat A)\equiv \sum_{n=0}^\infty \; \frac{f^{(n)}(a)}{n!}\; \left
( \hat A - a\right )^n \; , \qquad f^{(n)}(a)\equiv \frac{d^n f(a)}{da^n} \; ,
$$
where $a$ is a convenient nonoperator quantity. Below, several important facts, 
related to the Gibbs entropy are proved. 

\vskip 2mm

{\bf Liouville theorem}. Gibbs entropy does not change with time:
\be
\label{77}
S(t) = S(0) \; .
\ee

\vskip 2mm

{\it Proof}. The time evolution of the statistical operator is given by the
relation
$$
\hat\rho(t) =\hat U(t)\;\hat\rho\; \hat U^+(t) \; ,
$$
where $\hat\rho \equiv\hat\rho(0)$. By the definition of the function of an 
operator, we have
$$
\ln\hat\rho(t) = \hat U(t)\;\ln\hat\rho\;\hat U^+(t) \; .
$$
Therefore
$$
S(t) = -{\rm Tr}\; \hat\rho(0)\ln\hat\rho(0) = S(0) \; , 
$$
which proves equality (77). $\square$

\vskip 2mm

{\bf Gibbs inequality}. Let $\hat A$ and $\hat B$ be semi-positive operators, 
possessing traces, then
\be
\label{78}
{\rm Tr} \hat A \ln \hat A - {\rm Tr} \hat A \ln \hat B \geq
{\rm Tr} \left ( \hat A - \hat B \right ) \; .
\ee

\vskip 2mm

{\it Proof}. For a semi-positive operator $\hat A$, one has
$$
\ln \hat A \geq 1 - \hat A^{-1} \; .
$$
Similarly,
$$
\ln\hat A - \ln\hat B = \ln(\hat A \hat B^{-1}) \geq 1 - \hat A^{-1} \hat B \; .
$$
From here, 
$$
\hat A \ln(\hat A \hat B^{-1}) \geq \hat A - \hat B \; .
$$
Then inequality (78) follows. $\square$  

\vskip 2mm

Let us introduce the {\it entropy operator}
\be
\label{79}
\hat S(t)\equiv -\ln\hat\rho(t) \; .
\ee
And let us define the {\it average entropy}
\be
\label{80}
\overline S(t)\equiv{\rm Tr}\hat\rho(0)\hat S(t) \; .
\ee
Notice that the average entropy does not equal ${\rm Tr}\hat\rho(t)\hat S(0)$.

\vskip 2mm

{\bf Nondecrease of average entropy}. The average entropy (80) does
not decrease with time:
\be
\label{81}
\overline S(t)\geq \overline S(0) \; .
\ee

\vskip 2mm

{\it Proof}. By the definition of the average entropy, one has 
$$
\overline S(t) - \overline S(0) ={\rm Tr}\hat\rho\left [
\ln\hat\rho - \ln\hat\rho(t)\right ] \; .
$$
The proof follows the same arguments as in the proof of the Gibbs 
inequality. We have
$$
\ln\hat\rho -\ln\hat\rho(t) =\ln\hat\rho \hat\rho^{-1}(t) \; .
$$
Since
$$
\ln\hat\rho \hat\rho^{-1}(t) \geq 1 - \hat\rho^{-1}\hat\rho(t) \; ,
$$
we get
$$
\overline S(t) - \overline S(0) \geq {\rm Tr}\hat\rho \left [ 1 -
\hat\rho^{-1}\hat\rho(t) \right ] \; .
$$
Then
$$
{\rm Tr}\hat\rho \left [ 1 -\hat\rho^{-1}\hat\rho(t) \right ] =
{\rm Tr}\hat\rho - {\rm Tr}\hat\rho(t) \; .
$$
As far as ${\rm Tr}\hat\rho = {\rm Tr}\hat\rho(t) = 1$, we come to the 
inequality
$$
\overline S(t) - \overline S(0) \geq 0 \; ,
$$
which yields Eq. (81). $\square$

\subsection{Information functionals}

The Gibbs entropy serves as an efficient tool for defining the form of the 
statistical operators. This is done employing the {\it principle of minimal 
information}, which implies a conditional maximum of entropy under given 
additional constraints. The idea of this principle goes back to Gibbs
\cite{Gibbs_34,Gibbs_35,Gibbs_50} and has also been considered by Shannon 
\cite{Shannon_51} and Janes \cite{Janes_52}. General formulation and 
different applications have been studied in Refs. \cite{Yukalov_33,Yukalov_46}.

Suppose that some operators of local observables $\hat A_i\in{\cal A}$ at 
$t=0$ are assumed to give the known observable quantities
\be
\label{82}
A_i\equiv\; <\hat A_i>\; = {\rm Tr}\hat\rho \hat A_i \; ,
\ee
enumerated with $i=0,1,2,\ldots$. One of the quantities is always $\hat A_0\equiv\hat 1$, 
which guarantees the normalization condition 
$$
<\hat 1>={\rm Tr}\hat\rho=1 \; .
$$
The operators entering constraint (82) are termed the {\it constraint 
operators}. The information functional, or simply information, is
\be
\label{83}
I[\hat\rho] \equiv - S + \sum_i \lbd_i \left ( 
{\rm Tr}\hat\rho \hat A_i - A_i \right ) \; ,
\ee
where $\lbd_i$ are Lagrange multipliers, controlling the validity of 
conditions (82) through the variation $\dlt I[\hat\rho]/\dlt\lbd_i = 0$. 
The statistical operator is supposed to be such that to provide the minimum
of information (83). Minimization with respect to $\hat\rho$ implies
$$
\frac{\dlt I[\hat\rho]}{\dlt\hat\rho} = 0\; , \quad
\frac{\dlt^2 I[\hat\rho]}{\dlt\hat\rho^2} > 0 \; .
$$
Using the derivatives
$$
\frac{\dlt}{\dlt\hat\rho}\; {\rm Tr}\hat\rho \hat A_i = \hat A_i\; ,
\quad \frac{\dlt}{\dlt\hat\rho}\; \ln\hat\rho = \hat\rho^{-1} 
$$
leads to the {\it Gibbs statistical operator}
\be
\label{84}
\hat\rho = \frac{1}{Z}\; \exp\left ( -\sum_{i>0} \lbd_i\hat A_i\right )\; ,
\ee
where 
$$
Z= {\rm Tr} \exp\left ( - \sum_{i>0} \lbd_i\hat A_i \right ) 
$$
is called the {\it partition function}. The second derivative is positive, 
since
$$
\frac{\dlt^2 I[\hat\rho]}{\dlt\hat\rho^2} = \hat\rho^{-1} > 0 \; .
$$

In the same way, it is possible to find the statistical operator, when the
local operator densities at $t=0$ yield the known densities of observables
\be
\label{85}
A_i(x) \equiv \; <\hat A_i(x)>\; = {\rm Tr}\hat\rho \hat A_i(x) \; .
\ee
Then one defines the local information functional
\be
\label{86}
I_{loc}[\hat\rho] \equiv - S + \sum_i \int \lbd_i(x) \left [ 
{\rm Tr}\hat\rho \hat A_i(x) - A_i(x) \right ]\; dx \; .
\ee
The minimization of information (86) results in the statistical operator
\be
\label{87}
\hat\rho = \frac{1}{Z}\; \exp\left\{ - \sum_{i>0} \;
\int \lbd_i(x)\hat A_i(x)\; dx \right \}\; , 
\ee
with the partition function
$$
Z ={\rm Tr}\exp\left\{ - \sum_{i>0}\; \int \lbd_i(x)
\hat A_i(x)\; dx\right \} \; .
$$
The statistical operator (87) is called the {\it local-equilibrium}
or {\it quasi-equilibrium} statistical operator.

\subsection{Equilibrium states}

A {\it statistical state} is the set $\{<\hat A(t)>\}$ of the averages of
all operators of local observables. By this definition, a statistical state
possesses several general properties that are studied in the algebraic 
approach \cite{Emch_53}. 

An {\it equilibrium state} does not depend on time, so that 
$<\hat A(t)>=<\hat A(0)>$ for all averages from the state $\{ <\hat A(t)>\}$.
This implies that $\hat\rho(t)=\hat\rho(0)=\hat\rho$, because of which
$[\hat H,\; \hat\rho]=0$. The latter means that the statistical operator
$\hat\rho$ is an integral of motion. Hence it has to be a function of 
$\hat H$ and other integrals of motion. An {\it integral of motion} $\hat C$
is an operator for which
$$
[\hat C,\; \hat H]=0\; , \quad  i\; \frac{d}{dt}\;\hat C =
[\hat C,\; \hat H]=0 \; .
$$
The Hamiltonian $\hat H$ is, clearly, also an integral of motion.

Replacing in the information functionals $\hat A_i$ by $\hat C_i$, and 
minimizing the information, we get the {\it Gibbs statistical operator}
$$
\hat\rho = \frac{1}{Z}\; \exp\left ( - \sum_i \lbd_i \hat C_i\right ) \; ,
$$
with the partition function
$$
Z = {\rm Tr}\exp \left ( - \sum_i \lbd_i \hat C_i\right ) \; .
$$
The statistical state $\{<\hat A>\}$, composed of the averages 
$<\hat A>\; \equiv {\rm Tr}\hat\rho\hat A$, with the Gibbs statistical 
operator, is termed the {\it Gibbs state}. 

One says that $\hat\rho$ defines a statistical ensemble, which is a collection 
of possible dynamical states of a physical system. More rigorously, a 
statistical ensemble is a pair $\{\cF,\hat\rho\}$ of the Fock space of 
microscopic states $\cF$ and a statistical operator $\hat\rho$. Depending on 
the choice of the constraint operators, there are different statistical 
ensembles represented by different statistical operators.    

If, in addition to the normalization condition, one considers the sole 
constraint operator that is the Hamiltonian energy operator $\hat H$, then
the principle of minimal information yields the {\it Gibbs canonical operator}
\be
\label{88}
\hat\rho =\frac{1}{Z} \; e^{-\bt\hat H}\; , \quad
Z ={\rm Tr} e^{-\bt\hat H} \; ,
\ee
where $\beta$ is inverse temperature, $\bt T\equiv 1$. The statistical operator 
$\hat\rho$ describes the {\it Gibbs canonical ensemble}. In this ensemble, the 
role of a thermodynamic potential is played by the {\it free energy} 
$F=F(T,V,N,{\bf M})$, which can be defined by the equality
$$
{\rm Tr}  e^{-\bt\hat H} \equiv  e^{-\bt F}\; . 
$$
Therefore
\be
\label{89}
F\equiv -T\ln Z = -T\ln {\rm Tr} e^{-\bt\hat H} \; .
\ee
Other thermodynamic quantities can be found through the free energy, for instance
$$
S = -\left (\frac{\prt F}{\prt T}\right )_{VN} = \bt\left (<\hat H>\; - F\right )\; , 
\quad E\equiv \; <\hat H> = F + TS \; .
$$
The Gibbs entropy (information entropy) coincides with the thermodynamic
entropy,
$$
S = - {\rm Tr}\hat\rho\ln\hat\rho = \bt(E - F) \; .
$$

When, in addition to the normalization condition, as the constraint operators 
one takes the Hamiltonian energy $\hat H$ and the number-of-particle operator 
$\hat N$, then one gets the {\it grand canonical operator}
\be
\label{90}
\hat\rho = \frac{1}{Z}\; e^{-\bt(\hat H -\mu\hat N)}\; , \quad
Z= {\rm Tr} e^{-\bt(\hat H -\mu\hat N)} \; ,
\ee
where $\mu$ is chemical potential. This statistical operator describes the 
{\it grand canonical ensemble}. The corresponding thermodynamic potential is 
the grand potential $\Om=\Om(T,V,\mu,{\bf M})$ defined by the equality
$$
{\rm Tr} e^{-\bt(\hat H -\mu\hat N)} \equiv e^{-\bt\Om}\; .
$$
This means that
\be
\label{91}
\Om = - T\ln {\rm Tr} e^{-\bt(\hat H -\mu\hat N)}\; .
\ee
The grand potential $\Om = F -\mu N$ defines other thermodynamic quantities, 
for example, the entropy
$$
S =-\left ( \frac{\prt\Om}{\prt T}\right )_{V\mu} = \bt(E -\mu N - \Om) \; .
$$
The Gibbs and thermodynamic entropies coincide, so that the above expression
of the entropy equals $S = -{\rm Tr}\hat\rho\ln\hat\rho$.

Generally, different statistical ensembles can describe different physical
situations. If the considered systems are physically different, then one
of the ensembles may be appropriate for characterizing one of the system, 
while another ensemble may be not appropriate for the same system, but 
appropriate for the other system. In that sense, the statistical ensembles 
do not need to be completely equivalent. Though in the majority of the cases,
the ensembles are equivalent, so that any of them can be used.    

It is possible to distinguish the ensemble equivalence in two senses.
One meaning is the {\it thermodynamic equivalence of ensembles}, when the
thermodynamic characteristics, calculated for different ensembles, coincide,
after being expressed through the same variables, so that
$$
S  = -\left ( \frac{\prt F}{\prt T}\right )_{VN} =
-\left ( \frac{\prt\Om}{\prt T}\right )_{V\mu} \; , \quad
P = -\left ( \frac{\prt F}{\prt V}\right )_{TN} =
-\left ( \frac{\prt\Om}{\prt V}\right )_{T\mu} \; .
$$

Another meaning is the {\it statistical equivalence of ensembles}, when the
statistical states coincide, so that the observable quantities calculated in
one ensemble, say in the canonical ensemble giving $<\hat A>_N$, coincide 
with the same observable calculated in another ensemble, say in the grand 
canonical ensemble giving $<\hat A>_\mu$, after taking into account the 
relation between the chemical potential and the number of particles.

\subsection{Representative ensembles}

In order to correctly characterize a system, one has to chose a 
{\it representative statistical ensemble}, which is an ensemble that uniquely 
describes the considered system \cite{Yukalov_33,Gibbs_34,Ter_54,Yukalov_55}.
Actually, the problem of ensemble equivalence is meaningful only for 
representative ensembles \cite{Yukalov_55}. 

An {\it equilibrium statistical ensemble} is a pair $\{\cF,\hat\rho\}$ of
the Fock space and a statistical operator. To uniquely describe the system, 
the ensemble must be representative, which requires that the statistical 
operator has to be defined taking into account all constraints that are 
necessary for the unique system description. For this, one has to consider
the necessary constraint operators, $\hat C_i^+=\hat C_i$ that are assumed 
to be self-adjoint, but not necessarily integrals of motion. These constraint
operators define the corresponding {\it statistical conditions}
\be
\label{92}
C_i \; = \; < \hat C_i>\; = \; {\rm Tr}\hat\rho \hat C_i \; ,
\ee
with the trace over $\cF$. 

The first standard constraint is the normalization condition for the 
statistical operator,
$$
1 \; = \; < \hat 1_\cF> \; = \; {\rm Tr}\hat\rho \; ,
$$
where $\hat 1_\cF$ is a unity operator in $\cF$. Another usual condition 
is the definition of the internal energy
$$
E \; = \; < \hat H > \; = \; {\rm Tr} \hat\rho \hat H \; .
$$

Keeping in mind the normalization condition, the definition of the internal 
energy, and other constraints (92), required for the unique description of 
the system, we construct the information functional 
$$
I[\hat\rho] = {\rm Tr}\hat\rho \ln \hat\rho + \lbd_0 
( {\rm Tr}\hat\rho - 1 ) + \bt ( {\rm Tr}\hat\rho \hat H - E)
+\bt \sum_i \nu_i ( {\rm Tr}\hat\rho \hat C_i - C_i ) \; ,
$$
in which $\lbd_0$, $\bt$, and $\bt\nu_i$ are the Lagrange multipliers.

The minimization of the information functional gives the statistical operator
\be
\label{93}
\hat\rho = \frac{1}{Z} \; e^{-\bt H} \; ,
\ee
in which the {\it grand Hamiltonian}
\be
\label{94}
H = \hat H + \sum_i \nu_i \hat C_i 
\ee
is introduced. The partition function is $Z=\exp(1+\lbd_0) = {\rm Tr} \exp{-\bt H}$.
Such a statistical operator defines the representative statistical ensemble $\{\cF,\hat\rho\}$.

\subsection{Action functional}

A nonequilibrium statistical ensemble is a pair $\{\cF,\hat\rho(t)$, where
the statistical operator is a function of time characterized by an evolution 
operator and an initial condition $\hat\rho\equiv \hat\rho(0)$.

In quantum field theory, the system evolution is defined by an action 
functional \cite{Kleinert_56}. The temporal energy operator is written in the 
form
\be
\label{95}
\hat E \equiv \int \psi^\dgr(x,t) \; i \; \frac{\prt}{\prt t}\;
\psi(x,t)\; dx \; .
\ee
The system dynamical properties are described by the Lagrangian 
$ \hat L \equiv \hat E - \hat H$ and the {\it action functional}
\be
\label{96}
A[\psi] \equiv \int \left (\hat L - \nu_i \hat C_i\right ) dt = 
\int \left ( \hat E - H \right ) dt \; ,
\ee
taking into account the constraints that uniquely define the system. Here
$H$ is the grand Hamiltonian (94).

The equations of motion are given by the extremum of the action functional,
\be
\label{97}
\frac{\dlt A[\psi]}{\dlt\psi^\dgr(x,t)} = 0 \; , \qquad
\frac{\dlt A[\psi]}{\dlt\psi(x,t)} = 0 \; .
\ee
From the form of the action functional (96), it follows that the extremization 
of the action functional is identical to the variational equation
\be
\label{98}
i\; \frac{\prt}{\prt t}\; \psi(x,t) = 
\frac{\dlt H}{\dlt\psi^\dgr(x,t)} \; .
\ee
And the latter, as has been proved \cite{Yukalov_43}, is equivalent to 
the Heisenberg equation of motion
\be
\label{99}
i\; \frac{\prt}{\prt t}\; \psi(x,t) = [ \psi(x,t), H] \; .
\ee

A representative ensemble takes into account all statistical conditions, 
necessary for the unique description of the considered statistical system.
By defining a representative ensemble, one comes to the notion of the 
grand Hamiltonian. And, as is shown above, the grand Hamiltonian governs both 
the statistical properties of equilibrium systems as well as the temporal 
evolution of nonequilibrium systems. This means that the Liouville equation 
for the statistical operator must also contain the grand Hamiltonian.

\subsection{Fluctuations and stability}

In section 2.2, it is explained that in equilibrium systems, susceptibilities
are to be positive and finite. Here we show how the susceptibilities are 
connected with fluctuations of observable quantities. 

Observable quantities are represented by self-adjoint operators. Fluctuations 
of an observable, corresponding to an operator $\hat A$, are characterized 
by the {\it variance} or {\it dispersion}
\be
\label{100}
{\rm var}(\hat A) \equiv \; <\hat A^+ \hat A>\; - \; |<\hat A>|^2 \; .
\ee
Generally, variance (100) can be defined for any $\hat A$. For a self-adjoint 
$\hat A^+ =\hat A$, it reduces to
\be
\label{101}
{\rm var}(\hat A) \equiv \; <\hat A^2>\; - <\hat A>^2 \; .
\ee
By this definition, ${\rm var}(\hat A) \geq 0$, since
$$
{\rm var}(\hat A) = \; < (\hat A -\; <\hat A>)^2> \; .
$$
That is why susceptibilities are to be non-negative. 

Let us consider {\it specific heat}
$$
C_V = \frac{1}{N} \left ( \frac{\prt E}{\prt T}\right )_{VN} =
\frac{T}{N}\left ( \frac{\prt S}{\prt T}\right )_{VN} \; ,
$$
which can also be expressed through the second derivatives
$$
C_V = -\; \frac{T}{N} \left ( \frac{\prt^2 F}{\prt T^2}\right )_{VN}=
-\; \frac{T}{N} \left ( \frac{\prt^2\Om}{\prt T^2}\right )_{V\mu}\; .
$$
Using this definition, for the specific heat we have
\be
\label{102}
C_V =\frac{{\rm var}(\hat H)}{NT^2} \; ,
\ee
which shows that it characterizes energy fluctuations. 

Consider now the {\it isothermal compressibility}
$$
\kappa_T = -\; \frac{1}{V} \left ( \frac{\prt V}{\prt P}\right )_{TN} =
\frac{1}{N\rho} \left ( \frac{\prt N}{\prt\mu}\right )_{TV} \; ,
$$
which can also be written as
$$
\kappa_T = -\; \frac{1}{N\rho}\left (\frac{\prt^2\Om}{\prt\mu^2}\right )_{TV}
= \frac{1}{\rho^2} \left ( \frac{\prt^2 P}{\prt\mu^2}\right )_{TV}\; .
$$
Therefore for the compressibility, we get
\be
\label{103}
\kappa_T = \frac{{\rm var}(\hat N)}{N\rho T} \; ,
\ee
which means that it describes the number-of-particle fluctuations.

The number-of-particle variance can be written as
$$
{\rm var}(\hat N) =\int [ <\hat n(x)\hat n(x')>\; - \rho(x)\rho(x')]\; dx dx' \; ,
$$
being expressed through the density-density correlation function
$$
<\hat n(x)\hat n(x')>\; = \rho_2(x,x',x,x') + \rho(x)\dlt(x-x') \; .
$$
Therefore, it can be written as 
\be
\label{104}
{\rm var}(\hat N) = N + \int \rho(x)\rho(x') [ g(x,x') -1] \; dx dx' \; .
\ee

Considering the {\it magnetic susceptibility}, we take into account that 
an external magnetic field $\bB$ enters the Hamiltonian through the 
expression $ - \sum_{i=1}^N \hat\bM_i \cdot \bB$, where $\hat\bM_i$ is
the magnetic-moment operator of an $i$-particle. The magnetic susceptibility 
tensor is defined as
$$
\chi_{\al\bt} = \frac{1}{N}\; \frac{\prt M^\al}{\prt B^\bt} \; , 
$$
where the components of the average magnetization are  
$$
M^\al = -\; \frac{\prt G}{\prt B^\al} =\; <\hat M^\al> \; , \quad
\hat\bM \equiv \sum_{i=1}^N \hat\bM_i \; .
$$
Then the susceptibility tensor is
$$
\chi_{\al\bt} = -
\;\frac{1}{N}\; \frac{\prt^2 G}{\prt B^\al\prt B^\bt} = 
\frac{<\hat M^\al\hat M^\bt>-<\hat M^\al><\hat M^\bt>}{NT}\; .
$$
The diagonal elements of the tensor give the magnetic susceptibilities
\be
\label{105}
\chi_{\al\al} = \frac{{\rm var}(\hat M^\al)}{NT} \; ,
\ee
which shows that they describe magnetic-moment fluctuations. 

Being proportional to the variances of the related observables, all these
susceptibilities are non-negative:
$$
C_V\geq 0 \; , \quad \kappa_T\geq 0 \; , \quad \chi_{\al\al} \geq 0 \; ,
$$
which is a part of the stability conditions. For the stability of an
equilibrium system it is also required that the susceptibilities be finite:
$$
C_V < \infty\; , \quad \kappa_T <\infty\; , \quad \chi_{\al\al} <\infty \; .
$$
This means that the stability conditions, in terms of the variance of 
observables, are
\be
\label{106}
0 \leq \frac{{\rm var}(\hat A)}{N} < \infty \; .
\ee
This has to be valid for any $N$, including $N\ra\infty$.  

When ${\rm var}(\hat A)\propto N$, this is called the {\it thermodynamically 
normal} dispersion corresponding to thermodynamically normal fluctuations.
And if ${\rm var}(\hat A)\propto N^\gm$, with $\gm >1$, this is called the 
{\it thermodynamically anomalous} dispersion describing thermodynamically 
anomalous fluctuations. Thermodynamically anomalous fluctuations imply the
system instability.

Summarizing, the susceptibilities
$$
C_V = \frac{{\rm var}(\hat H)}{NT^2}\; , \quad \kappa_T =
\frac{{\rm var}(\hat N)}{N\rho T} \; , \quad \chi_{\al\al} =
\frac{{\rm var}(\hat M^\al)}{NT}
$$
describe the fluctuations of the corresponding observables, they are expressed 
through the variance of the related operators and have to be non-negative and 
finite for any equilibrium system.

\subsection{Gibbs-Bogolubov inequality}

A very important result, widely used in statistical physics is the inequality,
first, proved by Gibbs for classical systems and then extended by Bogolubov
to quantum systems, because of which it is named the Gibbs-Bogolubov 
inequality. 

Let us consider two self-adjoint operators $\hat A$ and $\hat B$. And let us
define the functionals
$$
F(\hat A) \equiv -T \ln{\rm Tr} e^{-\bt\hat A}\; , \quad
F(\hat B) \equiv -T \ln{\rm Tr} e^{-\bt\hat B} \; .
$$
We introduce the averages
$$
<\hat A>_A \; \equiv{\rm Tr}\hat\rho_A\hat A \; , \quad 
<\hat A>_B \; \equiv{\rm Tr}\hat\rho_B\hat A \; ,
$$
in which 
$$
\hat\rho_A \equiv \frac{e^{-\bt\hat A}}{{\rm Tr}e^{-\bt\hat A}} \; ,
\quad \hat\rho_B \equiv \frac{e^{-\bt\hat B}}{{\rm Tr}e^{-\bt\hat B}} \; .
$$
Then the following theorem is valid.

\vskip 2mm

{\bf Gibbs-Bogolubov inequality}. For the quantities $F(\hat A)$ and 
$F(\hat B)$, one has
\be
\label{107}
F(\hat A) \leq F(\hat B) \; + \; <\hat A - \hat B>_B
\ee
and 
\be
\label{108}
<\hat A - \hat B>_A \; + \; F(\hat B) \leq F(\hat A) \; .
\ee

\vskip 2mm

{\it Proof}. According to the above definitions, we have
$$
{\rm Tr}\hat\rho_B\ln \hat\rho_B = \bt\left [ F(\hat B) -\;
<\hat B>_B\right ] \; ,
$$
$$
{\rm Tr}\hat\rho_B\ln \hat\rho_A = \bt\left [ F(\hat A) -\;
<\hat A>_B\right ] \; .
$$
By the Gibbs inequality (78), 
$$
{\rm Tr}\hat\rho_B \ln \hat\rho_B \geq {\rm Tr}\hat\rho_B \ln \hat\rho_A \; .
$$
From here, we get inequality (107). Interchanging $\hat A$ and $\hat B$, we
come to inequality (108). $\square$

\vskip 2mm

The most often used application of the Gibbs-Bogolubov inequality is for 
finding the best variational approximation for a complicated problem. Suppose 
we need to calculate the free energy  
$$
F(\hat H) = -T\ln {\rm Tr} e^{-\bt\hat H} \; , 
$$
which is impossible to calculate exactly. Then we can take an approximate
free energy
$$
F(\hat H_{app}) = -T\ln {\rm Tr} e^{-\bt\hat H_{app}} \; ,
$$
defined through an approximate Hamiltonian $H_{app}$. We use the notation
$$
<\hat A>\; = {\rm Tr}\;\hat\rho\hat A\; , \quad \hat\rho =
\frac{\exp(-\bt\hat H)}{{\rm Tr}\exp(-\bt\hat H)} \; ,
$$
$$
<\hat A>_{app} = {\rm Tr}\;\hat\rho_{app}\hat A\; , \quad \hat\rho_{app} =
\frac{\exp(-\bt\hat H_{app})}{{\rm Tr}\exp(-\bt\hat H_{app})} \; .
$$
Then, by the Gibbs-Bogolubov inequality,
\be
\label{109}
F(\hat H)\leq F(\hat H_{app}) \; + \; <\hat H - \hat H_{app}>_{app} \; .
\ee
The best approximation is obtained by minimizing the right-hand side of 
inequality (109), that is, by choosing the approximate Hamiltonian that
provides such a minimization. 

A particular form of the Gibbs-Bogolubov inequality is derived for zero 
temperature, when $T\ra 0$, hence $\bt\ra\infty$. By the definition of 
the trace,
$$
{\rm Tr} e^{-\bt\hat H} = \sum_k \vp_k^+  e^{-\bt\hat H} \vp_k \; ,
$$
where $\vp_k$ can be taken as the eigenfunctions of the Hamiltonian,
$$
\hat H\vp_k = E_k\vp_k \; .
$$
The {\it ground-state energy} is 
$$
E_0 \equiv \min_k E_k = (\vp_0,\hat H\vp_0) \; .
$$
Assuming a non-degenerate spectrum, the above trace can be rewritten as 
$$
\sum_k\vp_k^+  e^{-\bt\hat H} \vp_k =  e^{-\bt E_0} \left [ 1 +
\sum_{k\neq 0} e^{-\bt (E_k -E_0)} \right ] \; .
$$
Since $E_k - E_0 > 0$ for $k \neq 0$, we have
$$
{\rm Tr} e^{-\bt\hat H} \simeq e^{-\bt E_0} \quad (\bt\ra \infty) \; .
$$
Therefore
$$
F(\hat H) \simeq E_0 \quad (T\ra 0) \; .
$$
Respectively, when temperature tends to zero, we get 
$$
<\hat A>\; \simeq (\vp_0,\hat A\vp_0) \quad (T\ra 0) \; .
$$
For the approximate Hamiltonian, one has
$$
\hat H_{app}\vp_k^{app} = E_k^{app}\vp_k^{app} \; , \quad
E_0^{app} =\left (\vp_0^{app},\hat H_{app}\vp_0^{app}\right ) \; .
$$
Similarly to the case of the exact Hamiltonian,
$$
\hat F(\hat H_{app}) \simeq E_0^{app} \quad (T\ra 0) 
$$
and
$$
<\hat A>_{app} \simeq \left (\vp_0^{app},\hat A\vp_0^{app}\right )
\quad (T\ra 0) \; . 
$$
In that way, the Gibbs-Bogolubov inequality at $T=0$ reduces to the 
inequality
\be
\label{110}
E_0 \leq \left (\vp_0^{app},\hat H \vp_0^{app}\right ) \; ,
\ee
which is called the {\it Rayleigh-Ritz inequality} or the variational principle.

\subsection{Bogolubov inequalities}

Bogolubov has proved several more useful inequalities 
\cite{Bogolubov_30,Bogolubov_48,Bogolubov_49}. One of them concerns the 
statistical average of the product of two operators $<\hat A\hat B>$, with 
a Hamiltonian $H$. Generally, this average $<\hat A\hat B>$ can be treated
as a bilinear form, that is a form linear in both $\hat A$ and $\hat B$.
The required conditions are $(<\hat A^+\hat A>)\; \geq 0$ and 
$<\hat A\hat B>^*=<\hat B^+\hat A^+>$. For statistical averages, these 
conditions are always valid. Note that $\hat A^+\hat A$ is a semi-positive 
operator.

\vskip 2mm

{\bf Bogolubov inequality for products}. The bilinear form $<\hat A\hat B>$,
satisfies the inequality 
\be
\label{111}
\left | < \hat A\hat B>\right |^2 \leq \;\; < \hat A\hat A^+>
<\hat B^+\hat B> \; .
\ee

\vskip 2mm

{\it Proof}. Let us consider the form 
$$
< \left (x \hat B^+ + y\hat A\right ) 
\left ( x^* \hat B + y^* \hat A^+\right ) > \;\; \geq \; 0 \; ,
$$
which is non-negative for arbitrary complex numbers $x$ and $y$. This can 
be rewritten as
$$
|x|^2 < \hat B^+ \hat B> + xy^* < \hat B^+ \hat A^+ > +
x^*y < \hat A \hat B> + |y|^2 < \hat A \hat A^+ > \;\; \geq 0 \; .
$$
Setting
$$
x = - < \hat A \hat B>\; , \qquad y = \; < \hat B^+ \hat B> \; 
$$
yields
$$
< \hat A \hat A^+ > <  \hat B^+ \hat B >^2 - 
\left | < \hat A \hat B> \right |^2 < \hat B^+ \hat B > \; \;
\geq \; 0 \; .
$$
If $<\hat B^+\hat B>\neq 0$, then the Bogolubov inequality follows.

And if $<\hat B^+\hat B>=0$, then we set $x=z<\hat A\hat B>$ and $y=1$, 
with $z$ being a real-valued number. This gives
$$
2z \left | < \hat A \hat B > \right |^2 + < \hat A \hat A^+> \;\;
\geq 0 \; .
$$
Let us send $z\ra -\infty$. Then the left-hand side tends to $-\infty$, 
if $<\hat A\hat B>$ is nonzero. This is impossible. Hence 
$< \hat A \hat B > \; = \; 0$ when $< \hat B^+ \hat B > \; = \; 0$.
The Bogolubov inequality again holds true. $\square$

\vskip 2mm

Another inequality has been proved by Bogolubov for the average of the 
commutators of two operators $\hat A$ and $\hat B$.

\vskip 2mm

{\bf Bogolubov inequality for commutators}. For any two operators
$\hat A$ and $\hat B$ the inequality holds:
\be
\label{112}
\left | < [ \hat A, \hat B ] > \right |^2 \leq \frac{\bt}{2}
< \hat A^+ \hat A + \hat A \hat A^+ > \left | < [ \hat B^+,
[H,\hat B] ] > \right | \; .
\ee

\vskip 2mm

The proof of this inequality is a bit cumbersome, because of which it is 
omitted here \cite{Bogolubov_30}.

\subsection{Nernst law}

Nernst formulated the law as an empirical fact valid for any equilibrium 
system. But in quantum statistical mechanics, this law can be proved.

Let $\zeta_0 \geq 1$ be the number of quantum states $\vp_0$ possessing the 
same energy $E_0 \equiv (\vp_0,\hat H\vp_0)$ at $T=0$. This parameter 
$\zeta_0$ is called the {\it degeneracy factor}. It is independent from 
thermodynamic variables, being defined by the quantum properties of the 
system.

\vskip 2mm

{\bf Nernst law}. The entropy $S = - {\rm Tr}\; \hat\rho \ln\hat\rho$ for 
an equilibrium system at zero-temperature is
\be
\label{113}
\lim_{T\ra 0} S = \ln\zeta_0 \geq 0 \; .
\ee

\vskip 2mm

{\it Proof}. The statistical operator of an equilibrium system has the form 
$\hat\rho = e^{-\bt\hat H}/{\rm Tr}e^{-\bt\hat H}$. When temperature tends 
to zero, we have 
$$
{\rm Tr}\; e^{-\bt\hat H} \simeq \zeta_0 e^{-\bt E_0} \quad (T\ra 0) \; .
$$
Then limit (113) follows. $\square$

\vskip 2mm

The quantity $S_0 \equiv \ln\zeta_0$ has the meaning of the Boltzmann entropy.
Since the entropy $S$ is an extensive quantity, one has 
$\lim_{N\ra\infty} S/N < \infty$. Assuming that the degeneracy is finite, 
$\zeta_0 < \infty$, the quantity $S_0$ can be neglected in the thermodynamic 
limit when
$$
\lim_{N\ra\infty} \; \frac{S_0}{N} = \lim_{N\ra\infty} \; 
\frac{\ln\zeta_0}{N} = 0 \; .
$$

But $S_0$ cannot be neglected if $\zeta_0 \propto e^{\al N}$, with $\al > 0$. 

From the Nernst law it follows that the specific heat at zero temperature 
is zero.

\vskip 2mm

{\bf Zero-temperature specific heat}. The limit of the specific heat at zero 
temperature is zero: 
\be
\label{114}
\lim_{T\ra 0} \; C_V = 0 \; .
\ee

\vskip 2mm

{\it Proof}. At temperature tending to zero, one has
$$
S_0 = \lim_{T\ra 0} S = \lim_{T\ra 0} \; \frac{TS}{T} =
\lim_{T\ra 0} \; \frac{\prt(TS)}{\prt T} \; .
$$
Using the expression of the specific heat
$$
C_V = \frac{T}{N} \left ( \frac{\prt S}{\prt T}\right )_V = 
\frac{1}{N} \left ( \frac{\prt E}{\prt T}\right )_V \; ,
$$
we come to the equality
$$
S_0 = S_0 + N \lim_{T\ra 0} C_V \; .
$$
From here, limit (114) follows. $\square$

\vskip 2mm

It is important to emphasize that, for any degeneracy factor $\zeta_0$,
the limits
$$
\lim_{T\ra 0} F = E_0 \; , \quad \lim_{T\ra 0} <\hat A>\; = \;
(\vp_0,\hat A\vp_0) 
$$
are valid, since $\zeta_0$ does not depend on $T$, and 
$$
\lim_{T\ra 0} \; T\ln\zeta_0 = 0 \; .
$$

Knowing the behavior of the free energy at low temperature makes it possible 
to derive an equation for the free energy at arbitrary temperatures. To this 
end, because of the relations
$$
F = E - TS \; , \quad S = -\; \frac{\prt F}{\prt T} \; , 
$$
we have the limits 
$$
\lim_{T\ra 0} F = E_0 \; , \quad
\lim_{T\ra 0} \; \frac{\prt F}{\prt T} = - S_0 \; ,
$$
serving as boundary conditions for the equation
$$
T \; \frac{\prt F}{\prt T}=  F -  E \; .
$$
Integrating the latter equation over temperature between zero and $T$, we 
obtain the expression
\be
\label{115}
F = E_0 - TS_0 - T \int_0^T \; \frac{E-E_0}{T^2}\; dT \; .
\ee
If $S_0$ can be neglected in the thermodynamic limit, then the expression for 
the free energy at large $N \gg 1$ becomes
\be
\label{116}
F = E_0 - T \int_0^T \; \frac{E-E_0}{T^2}\; dT \; .
\ee
The integral over $T$ is well defined because the energy $E$, as $T \ra 0$, 
tends to $E_0$ not slower than as $T^2$.

\subsection{Parameter variation}

It often happens that the Hamiltonian $\hat H = \hat H(\lbd)$ depends 
on a parameter $\lbd$, and one needs to find the variation of an observable 
quantity $<\hat A> \; = \; {\rm Tr} \;\hat\rho \hat A$ that is the average 
of an operator $\hat A = \hat A(\lbd)$, which can also depend on this 
parameter. When the Hamiltonian is a function of a parameter, the statistical 
operator is the function
$$
\hat\rho = \frac{e^{-\bt\hat H}}{{\rm Tr} e^{-\bt\hat H}} = \hat\rho(\lbd) 
$$
of this parameter as well.
 
Before considering the variation of the parameter $\lbd$, let us define the
{\it covariance} of two operators $\hat A$ and $\hat B$:
$$
{\rm cov}(\hat A,\hat B) \equiv \frac{1}{2} < [\hat A,\; \hat B]_+> -
<\hat A> <\hat B> \; ,
$$
where the notation for the anticommutator 
$[\hat A,\; \hat B]_+ \equiv \hat A \hat B + \hat B \hat A$ is used. The 
covariance is symmetric,
$$
{\rm cov}(\hat A,\hat B) = {\rm cov}(\hat B,\hat A) \; ,
$$
and additive,
$$
{\rm cov}(\hat A +\hat B,\hat C) = {\rm cov}(\hat A,\hat C) +
{\rm cov}(\hat B,\hat C) \; .
$$
The diagonal covariance is the variance, i.e., the dispersion that for 
a self-adjoint operator is
$$
{\rm cov}(\hat A,\hat A) = \; <\hat A^2> - <\hat A>^2 \; = \;
{\rm var}(\hat A) \; .
$$

For a function of an operator $f=f(\hat A)$, the derivative with respect
to the parameter is defined as the symmetrized derivative
$$
\frac{\prt}{\prt\lbd}\; f(\hat A) \equiv \frac{1}{2}\; \left [
\frac{\prt f}{\prt\hat A}\; , \; \frac{\prt\hat A}{\prt\lbd}
\right ]_+ \; .
$$

Direct differentiation yields 
\be
\label{117}
\frac{\prt}{\prt\lbd}<\hat A> \; = \; < \frac{\prt\hat A}{\prt\lbd}> -
\bt\; {\rm cov}\left (\hat A, \; \frac{\prt\hat H}{\prt\lbd}\right ) \; .
\ee

Considering the zero-temperature limit, we have the following properties.

\vskip 2mm

{\bf Proposition 1}. Let $\hat A$ be a self-adjoint integral of motion,
then
\be
\label{118}
\lim_{T\ra 0} \; \bt\; {\rm cov}
\left (\hat A,\; \frac{\prt\hat H}{\prt\lbd}\right ) = 0 \; .
\ee

\vskip 2mm

{\it Proof}. A self-adjoint operator is such for which 
$(\vp_k,\hat A\vp_p) = (\hat A\vp_k,\vp_p)$ for any $\vp_k$, $\vp_p$ from 
the Fock space. An operator is the integral of motion when 
$[\hat A,\; \hat H]= 0$. In that case, it enjoys a common set of 
eigenfunctions with the Hamiltonian, so that $\hat A\vp_k = A_k\vp_k$ and
$\hat H\vp_k = E_k\vp_k$. For the normalized functions, such that 
$(\vp_k,\vp_k) = 1$, one has $\prt(\vp_k,\vp_k)/\prt\lbd = 0$.
Therefore
$$
\frac{\prt}{\prt\lbd} \left (\vp_k,\hat A\vp_k\right ) =
\left (\vp_k,\; \frac{\prt\hat A}{\prt\lbd}\;\vp_k\right ) \; .
$$
When $T\ra 0$, then $\lim_{T\ra 0} <\hat A> \; = \left (\vp_0,\hat A\vp_0\right )$, 
with $\vp_0$ being the ground state. In that way,
$$
\lim_{T\ra 0} \; \frac{\prt}{\prt\lbd}<\hat A> \; = 
\lim_{T\ra 0} \; < \frac{\prt\hat A}{\prt\lbd}> \; .
$$
Then, in view of equality (117), limit (118) follows. $\square$

\vskip 2mm

{\bf Proposition 2}. For the derivative of the entropy $S$ with respect to a
parameter $\lbd$, the zero-temperature limit
\be
\label{119}
\lim_{T\ra 0} T\; \frac{\prt S}{\prt\lbd} = 0 
\ee
is valid. 

\vskip 2mm

{\it Proof}. Let us take in Eq. (117) $\hat A = \hat H$. Then for 
$E\equiv<\hat H>$, at any $T\geq 0$, we have
$$
\frac{\prt E}{\prt\lbd} = \; < \frac{\prt\hat H}{\prt\lbd} > -
\bt\; {\rm cov}\left (\hat H,\; \frac{\prt\hat H}{\prt\lbd}\right ) \; .
$$
For the normalized functions,
$$
\frac{\prt}{\prt\lbd} \left (\vp_k,\hat H\vp_k\right ) = \left (
\vp_k,\; \frac{\prt\hat H}{\prt\lbd}\;\vp_k\right ) \; .
$$
According to property (118), one has
$$
\lim_{T\ra 0} \; \bt\; {\rm cov} \left (
\hat H,\; \frac{\prt\hat H}{\prt\lbd}\right ) = 0 \; .
$$
It is straightforward to check that 
$$
< \frac{\prt\hat H}{\prt\lbd} > \; = \; \frac{\prt F}{\prt\lbd}\; ,
\quad F = - T\ln \;{\rm Tr} e^{-\bt\hat H} \; .
$$
This leads to the relation
$$
\frac{\prt E}{\prt\lbd} = \frac{\prt F}{\prt\lbd} \; - \;
\bt\; {\rm cov} \left (\hat H,\; \frac{\prt\hat H}{\prt\lbd}\right )\; .
$$
Since $F = E - TS$, we get
$$
\frac{\prt F}{\prt\lbd} = \frac{\prt E}{\prt\lbd} \; - \; T \;
\frac{\prt S}{\prt\lbd}  \; , \quad
\frac{\prt E}{\prt\lbd} = \; < \frac{\prt\hat H}{\prt\lbd} > + \;
T \; \frac{\prt S}{\prt\lbd} \; .
$$
Comparing the above formulas, we obtain
$$
\frac{\prt S}{\prt\lbd}  = -\bt^2\; {\rm cov} \left (
\hat H, \; \frac{\prt\hat H}{\prt\lbd} \right ) \; ,
$$
which, according to Eq. (118), gives limit (119).

Note that the same result can be derived from the expression 
$S\equiv -{\rm Tr}\hat\rho\ln\hat\rho$, from which
$$
\frac{\prt S}{\prt\lbd}  = \bt\; {\rm cov} \left (
\ln\hat\rho, \; \frac{\prt\hat H}{\prt\lbd} \right ) \; .
$$
Substituting here $\ln\hat\rho = \bt(F-\hat H)$ and taking into account 
Eq. (118) yields limit (119). $\square$

\vskip 2mm

{\bf Proposition 3}. Let the Hamiltonian be the sum $\hat H = \hat K + \hat W$
of the kinetic-energy operator and potential-energy operator, respectively.
Then the variation of the internal energy $E$ with respect to mass $m$, at 
zero temperature, gives
\be
\label{120}
\frac{\prt E}{\prt m} = -\; \frac{1}{m} < \hat K > +
<\frac{\prt\hat W}{\prt m}>  \quad (T=0) \; .
\ee

\vskip 2mm

{\it Proof}. Let us set $\lbd=m$. From the form of the kinetic-energy operator, 
we have
$$
\frac{\prt\hat K}{\prt m} = -\; \frac{1}{m}\; \hat K \; ,
\quad \frac{\prt\hat H}{\prt m} = -\; \frac{1}{m}\; \hat K +
\frac{\prt\hat W}{\prt m} \; .
$$
This yields
$$
\frac{\prt E}{\prt m} =  -\; \frac{1}{m}< \hat K > +
<\frac{\prt\hat W}{\prt m}> + \; T\; \frac{\prt S}{\prt m}\; ,
$$
which is equivalent to
$$
\frac{\prt F}{\prt m} = -\; \frac{1}{m} <\hat K> +
<\frac{\prt\hat W}{\prt m} > \; .
$$
According to Proposition 2, we have
$$
\frac{\prt S}{\prt m} = - \bt^2\; {\rm cov}\left ( \hat H,\;
\frac{\prt\hat H}{\prt m}\right ) \; , \quad
\lim_{T\ra 0} T\; \frac{\prt S}{\prt m} = 0 \; .
$$
Thus we come to Eq. (120). $\square$

\vskip 2mm

Equation (120) shows that the Landau-Lifshitz \cite{Landau_28} relation
$<\hat K> \; = -m\; \prt E/\prt m$, at $T=0$, is valid only when    
$\prt\hat W/\prt m = 0$. However the latter is not always true. For 
example, in the case of a dilute system of cold atoms, one has
$\prt\hat W/\prt m = - \hat W/m$. Then
$$
\frac{\prt\hat H}{\prt m} =  -\; \frac{1}{m} \; \hat H \; , \quad
\frac{\prt S}{\prt m} = \frac{\bt^2}{m}\; {\rm var}(\hat H) \; .
$$
In view of the expression for the specific heat
$$
C_V \equiv \frac{1}{N} \left ( \frac{\prt E}{\prt T}\right )_V =
\frac{{\rm var}(\hat H)}{NT^2} \; ,
$$
we get
$$
\frac{\prt S}{\prt m} = \frac{N}{m}\; C_V \; , \quad
\lim_{T\ra 0} TC_V = 0 \; .
$$
Therefore, instead of the Landau-Lifshitz relation, we have
\be
\label{121}
\frac{\prt E}{\prt m} = -\; \frac{E}{m} \quad (T=0) \; .
\ee
This shows that one has to be cautious when invoking equations presumed 
to be well known, not forgetting to check the region of applicability of 
these equations.

\subsection{Kubo formula}

When the system is subject to an external perturbation, its observable 
quantities change. Such a change can be described by the Kubo formula
\cite{Kubo_57,Bonch_58}. 

Let the system, at time $t=0$, be characterized by a Hamiltonian $H$, and
a statistical operator $\hat\rho(t)$. Generally, $H$ is a grand Hamiltonian.
In the case when there are no external perturbations, the statistical operator 
satisfies the Liouville equation 
$$
i\; \frac{d}{dt}\; \hat\rho(t) = [ H,\hat\rho(t)] \; . 
$$
The statistical operator evolution can be represented as  
$\hat\rho(t) = \hat U(t) \hat\rho(t_0)\hat U^+(t)$ by means of the evolution
operator satisfying the Schr\"{o}dinger equation
$$
i\; \frac{d}{dt} \; \hat U(t) = H \hat U(t) \; ,
$$
with the initial condition $\hat U(t_0) = 1$. The observable quantity, 
corresponding to an operator $\hat A(t)=\hat U^+(t) \hat A(t_0)\hat U(t)$,
when there is no perturbation, is given by the average
$$
< \hat A(t)> \; \equiv \; {\rm Tr} \hat\rho(t) \hat A(t_0) =
{\rm Tr} \hat\rho(t_0) \hat A(t) \; .
$$

But let us assume that at the moment $t\geq t_0$, the Hamiltonian suddenly 
changes to $H_{out}$. Hence now the statistical operator becomes 
$\hat \rho_{out}(t)$ that satisfies the Liouville equation 
$$
i\; \frac{d}{dt} \; \hat\rho_{out}(t) = 
[ H_{out},\hat\rho_{out}(t) ] \; ,
$$
with the initial condition $\hat\rho_{out}(t_0)=\hat\rho(t_0)$. 
Now, the observable quantity, corresponding to the operator $\hat A$, 
reads as
$$
< \hat A(t) >_{out} \; \equiv \; {\rm Tr}\hat\rho_{out}(t)\hat A(t_0) \; .
$$

The change of the observable quantity, called {\it increment}, is defined as
\be
\label{122}
\Dlt < \hat A(t)> \; \equiv \; < \hat A(t)>_{out} - < \hat A(t)> \; . 
\ee

Let the new Hamiltonian have the form $H_{out} = H + \Dlt H$. Then the 
Liouville equation becomes 
$$
i\; \frac{d}{dt} \; \hat\rho_{out}(t) = [ H,\hat\rho_{out}(t)] +
[ \Dlt H,\hat\rho_{out}(t)] \; .
$$
Looking for the solution of the Liouville equation in the form 
$$
\hat\rho_{out}(t)=\hat U(t)\hat D(t)\hat U^+(t)
$$
and introducing the notation
$$
\Dlt H(t)\equiv\hat U^+(t)\Dlt H\hat U(t) \; ,
$$
we obtain the equation
$$
i\; \frac{d}{dt}\; \hat D(t) = [\Dlt H(t),\hat D(t)]\; , 
$$
with the initial condition $\hat D(t_0)=\hat\rho_{out}(t_0)=\hat\rho(t_0)$.
The above equation can be rewritten as the integral equation
\be
\label{123}
\hat D(t) = \hat\rho(t_0) - 
i \int_{t_0}^t \; [\Dlt H(t_1),\hat D(t_1)] \; dt_1 \; .
\ee
Using this, the new statistical operator can be represented as
\be
\label{124}
\hat\rho_{out}(t) = \hat\rho(t) - 
i \hat U(t) \int_{t_0}^t \; [\Dlt H(t_1),\hat D(t_1)] \; dt_1 \; \hat U^+(t)\; .
\ee
Then increment (122) takes the form
\be
\label{125}
\Dlt<\hat A(t) >\;  = \; - 
i \int_{t_0}^t \; {\rm Tr}[\hat A(t),\Dlt H(t_1)]
\hat D(t_1) \; dt_1 \; .
\ee
This expression is exact for any $\Dlt H$.

By constructing an iterative solution for $\hat D(t)$, we start with the
zero approximation $\hat D_0(t) =\hat\rho(t_0)$. Then the first 
approximation is
$$
\hat D_1(t) = \hat D_0(t) - i \int_{t_0}^t \; 
[\Dlt H(t_1),\hat\rho(t_0)] \; dt_1 \; . 
$$
The second iteration gives
$$
\hat D_2(t) = \hat D_1(t) + (-i)^2 \; \int_{t_0}^t \;
\int_{t_0}^{t_1} \; \left [ \Dlt H(t_1), \left [ 
\Dlt H(t_2),\hat\rho(t_0)\right ] \right ]\; dt_1 \; dt_2 \; .
$$
Continuing the iteration procedure, we come to the formal solution
$$
\hat D(t) = \hat\rho(t_0) + \sum_{n=1}^\infty (-i)^n \;
\int_{t_0}^t dt_1 \; \int_{t_0}^{t_1} dt_2 \ldots \;
\int_{t_0}^{t_{n-1}} dt_n \times
$$
$$
\times \left [ \Dlt H(t_1), \left [ \Dlt H(t_2), \ldots
\left [ \Dlt H(t_n),\hat\rho(t_0)\right ] \right ] \ldots
\right ] \; .
$$
Then increment (125) becomes
\be
\label{126}
\Dlt < \hat A(t) > \; = \; -i \int_{t_0}^t < \left [ 
\hat A(t),\Dlt H(t_1)\right ] > dt_1 \; + \; 
\sum_{n=1}^\infty (-i)^{n+1} \int_{t_0}^t dt_1 \; 
\int_{t_0}^{t_1} dt_2 \ldots \;
\int_{t_0}^{t_n} dt_{n+1} \times
\ee
$$
\times \; < \left [ \left [ \ldots \left [ 
\hat A(t),\Dlt H(t_1)\right ], \Dlt H(t_2)\right ], \ldots , 
\Dlt H(t_{n+1})\right ] > \; .
$$ 

If the Hamiltonian perturbation $\Dlt H\equiv\dlt H$ is small, such 
that $\dlt H\ra 0$, then one can limit himself/herself by the first-order
approximation for increment (126). Then $\Dlt<\hat A>\equiv\dlt<\hat A>$
is termed the {\it linear response} or the {\it linear increment} that is
\be
\label{127}
\dlt < \hat A(t)>\; = \; -i \int_{t_0}^t < \left [ 
\hat A(t), \dlt H(t_1)\right ] > \; dt_1 \; ,
\ee
which is the {\it Kubo formula}. 

A particular important case is when $t_0\ra-\infty$, which is called the 
{\it adiabatic switching}. In that case, the Kubo formula (127) can be
represented as 
\be
\label{128}
\dlt<\hat A(t)>\; = \; - i \int_{-\infty}^\infty \;
\Theta(t-t_1) < \left [ \hat A(t),\dlt H(t_1)
\right ]> \; dt_1 \; ,
\ee
where $\Theta(t)$ is a unit step function
\begin{eqnarray}
\nonumber
\Theta(t) = \left\{ \begin{array}{ll}
1, & t \geq 0 \\
0, & t < 0          \; .       
\end{array} \right.
\end{eqnarray}

Let us introduce the {\it commutator Green function}
\be
\label{129}
C\left [ \hat A(t),\hat B(t') \right ] \equiv - i\Theta(t-t')
< \left [ \hat A(t),\hat B(t')\right ] > \; ,
\ee
which sometimes is named the Bogolubov-Tyablikov Green function. Then
the linear response (127) reads as 
\be
\label{130}
\dlt < \hat A(t) > \; = \; \int_{-\infty}^\infty
C\left [ \hat A(t),\dlt H(t_1)\right ] dt_1 \; .
\ee

If one assumes that the system at $t\leq t_0$ has been in equilibrium, 
with a Hamiltonian $H$ and a statistical operator $\hat\rho$, then the
commutator Green function (129) depends only on the difference of times
$t-t'$. This makes it possible to define the Fourier transform
$$
C\left [ \hat A(t),\hat B(t')\right ] = \int_{-\infty}^\infty
\chi\left (\hat A,\hat B,\om\right ) e^{-i\om(t-t')} \;
\frac{d\om}{2\pi} \; .
$$
Respectively, the inverse Fourier transform, called the {\it admittance}, 
reads as 
\be
\label{131}
\chi\left (\hat A,\hat B,\om\right ) = \int_{-\infty}^\infty
C\left [ \hat A(t), \hat B(0)\right ] e^{i\om t}\; dt \; ,
\ee
where the equality
$$
\int_{-\infty}^\infty \; e^{i\om t}\; dt = 2\pi\dlt(\om)  
$$
is used. 

Finally, we come to the expression for the linear response
\be
\label{132} 
\dlt<\hat A(t)>\;= \lim_{\om\ra 0} \chi(\hat A,\dlt H,\om) \; .
\ee
Here all averages are calculated for an equilibrium system.

\subsection{Quasi-equilibrium states}

In section 4.2, it has been mentioned that, using the local information 
functional, one can define a quasi-equilibrium, or locally equilibrium,
statistical operator. Here, we shall specify the general properties of
the corresponding quasi-equilibrium system \cite{Yukalov_46}.

For any system, the energy Hamiltonian and the number-of-particle operator
can be written as integrals of the related densities,  
$$
\hat H = \int \hat H(x)\; dx\; , \qquad \hat N = \int \hat n(x)\; dx\; ,
$$
where $\hat n(x) \equiv \psi^\dgr(x)\psi(x)$ is the density-of-particle 
operator. One can introduce the local energy density $E(x) = <\hat H(x)>$
and the local particle density $\rho(x) = <\hat n(x)>$. Their integrals 
define the internal energy and the number of particles,
$$
E=\int E(x)\; dx\; , \quad N = \int\rho(x)\; dx \; .
$$

The local information functional can be written in the form
$$
I_{loc}[\hat\rho] = {\rm Tr}\hat\rho\ln\hat\rho + 
\lbd_0\left ( {\rm Tr}\hat\rho -1
\right ) + \int \lbd_1(x)\left [ {\rm Tr}\hat\rho \hat H(x) - E(x)
\right ]\; dx +
$$
$$
+ \int \lbd_2(x)\left [ {\rm Tr}\hat\rho \hat n(x) -\rho(x)\right ]\; dx \;,
$$
with the Lagrange multipliers guaranteeing the normalization condition and
the definitions of the local energy density and local particle density. These
multipliers can be rewritten as $\lbd_0\equiv\ln Z -1$, $\lbd_1(x)\equiv\bt(x)$, 
and $\lbd_2(x)\equiv-\bt(x)\mu(x)$, where $\bt(x)$ is the local inverse 
temperature, such that $\bt(x)T(x)=1$, with $T(x)$ being the local temperature, 
and $\mu(x)$ is the local chemical potential. 

From the principle of minimal information, we find the local statistical 
operator
\be
\label{133}
\hat\rho = \frac{1}{Z}\; \exp\left\{ - \int \bt(x)\left [ \hat H(x)
- \mu(x)\hat n(x)\right ]\; dx \right \} \; ,
\ee
with the local partition function
$$
Z ={\rm Tr}\exp\left\{ -\int \bt(x)\left [ \hat H(x) -
\mu(x)\hat n(x)\right ]\; dx \right \} \; .
$$

The {\it quasi-equilibrium potential} $Q\equiv -\ln Z$ can be represented
through the {\it local grand potential} $\Om(x)$, so that 
\be
\label{134}
Q = \int \bt(x)\Om(x)\; dx \; , \quad 
Z\equiv \exp\left\{ - \int \bt(x)\Om(x)\; dx\right \} \; .
\ee

From $Q=-\ln Z$, we have
$$
\frac{\dlt Q}{\dlt\bt(x)} = E(x) - \mu(x)\rho(x) \; .
$$
While from the first relation with $\Om(x)$ in Eq. (134), we get
$$
\frac{\dlt Q}{\dlt\bt(x)} = \Om(x) +\bt(x)\; \frac{\dlt\Om(x)}{\dlt\bt(x)} \; .
$$
Comparing the above formulas, we find
$$
\frac{\dlt\Om(x)}{\dlt\bt(x)} = T(x) \left [ E(x) - \mu(x)\rho(x) -
\Om(x) \right ] \; .
$$
Because of the equality 
$$
\frac{\dlt\Om(x)}{\dlt\bt(x)} = -T^2(x) \; \frac{\dlt\Om(x)}{\dlt T(x)} \; ,
$$
we obtain
$$
\frac{\dlt\Om(x)}{\dlt T(x)} = \bt(x)\left [ \Om(x) +\mu(x)\rho(x) -
E(x)\right ] \; .
$$

The {\it local entropy} is introduced by the definition 
$$
S(x) \equiv  -\;  \frac{\dlt\Om(x)}{\dlt T(x)} \; , 
$$
which results in the expression
\be
\label{135}
S(x) =\bt(x)\left [ E(x) - \mu(x)\rho(x) -\Om(x)\right ] \; .
\ee
From here, we have the relations
$$
\bt(x) =  \frac{\dlt S(x)}{\dlt E(x)}\; , \quad
\mu(x) = -T(x)\;  \frac{\dlt S(x)}{\dlt\rho(x)} \; .
$$

The Gibbs entropy (information entropy) can be represented as an integral 
of the local entropy:
$$
S = -{\rm Tr}\hat\rho\ln\hat\rho = \int S(x)\; dx \; .
$$
Here the local entropy is as defined in Eq. (135), since
$$
S =\int \bt(x)\left [ E(x) -\mu(x)\rho(x)\right ]\; dx - Q \; .
$$

Taking into account the equalities
$$
\frac{\dlt Q}{\dlt\mu(x)} = -\bt(x)\rho(x) \; , \quad
\frac{\dlt Q}{\dlt\mu(x)} =\bt(x) \; \frac{\dlt \Om(x)}{\dlt\mu(x)}\; ,
\quad \rho(x) = -\; \frac{\dlt\Om(x)}{\dlt\mu(x)} \; ,
$$
we come to the {\it local Gibbs-Duhem relation} 
\be
\label{136}
\dlt\Om(x)=-S(x)\dlt T(x)-\rho(x)\dlt\mu(x) \; .
\ee
Introducing the local pressure $P(x)\equiv -\Om(x)$ yields 
\be
\label{137}
\dlt P(x)=S(x)\dlt T(x) +\rho(x)\dlt\mu(x) \; .
\ee

The local quantities naturally appear in nonuniform systems, such as 
the systems of trapped atoms.

\subsection{Gauge symmetry}

One of the most important properties of systems is gauge symmetry. When
the system is gauge symmetric, there can be neither Bose-Einstein 
condensation nor superconductivity.   

The typical energy Hamiltonian is
\be
\label{138}
\hat H = \int \psi^\dgr(x)\left [ -\; \frac{\nabla^2}{2m} +
U(x)\right ]\psi(x)\; dx +
$$
$$
+ \frac{1}{2} \int \psi^\dgr(x)\psi^\dgr(x') \Phi(x,x')\psi(x')\psi(x)\;
dx dx' \; ,
\ee
where $U(x)$ is an external potential and $\Phi(x,x')$ is a pair 
interaction potential. 

The number-of-particle operator is an integral of motion,
$$
[\hat N,\; \hat H]=0 \; , \quad \hat N = \int \psi^\dgr(x)\psi(x)\; dx \; .
$$
This is termed the number-of-particle conservation. 

The {\it global gauge transformation} is accomplished by the action of 
the operator 
\be
\label{139}
\hat U_\vp \equiv e^{i\vp\hat N} \; ,
\ee
where $\vp$ is the global phase, being any real number. 

Since $[\hat N,\; \hat H]=0$, Hamiltonian (138) is gauge symmetric:
$$
\hat U^+_\vp\hat H\hat U_\vp =\hat H \; .
$$
In other words, $\hat H$ is invariant under the field-operator transformation
$$
\psi(x)\ra \hat U^+_\vp\psi(x)\hat U_\vp \; .
$$

The gauge transformation $\hat U_\vp$ enjoys the following properties:
$$
\hat U_0 =\hat 1\; , \quad \hat U_\vp^{-1} = e^{-i\vp\hat N} \; ,
\quad \hat U^+_\vp \hat U_\vp = \hat 1 \; .
$$
It also possesses the property of additivity: 
$\hat U_\vp\hat U_{\vp'}=\hat U_{\vp+\vp'}$. Therefore the family 
$\{\hat U_\vp\}$ forms a group, called the {\it gauge group}. 

Among various averages of operator products, one distinguishes two 
principally different types of the averages. One type, called 
{\it normal averages} consists of the products containing equal numbers 
of creation and destruction field operators, e.g., as in the average
$<\psi^\dgr(x)\psi(x')>$. The other type of the averages, termed 
{\it anomalous averages} consists of the products with unequal numbers
of the creation and destruction operators, for instance, as in the 
averages $<\psi(x)>$ or $<\psi(x)\psi(x')>$.

\vskip 2mm

{\bf Statement 1}. If the number-of-particle operator is an integral of 
motion, then all anomalous averages are identically zero. 

\vskip 2mm

{\it Proof}. The particle-number operator has the following properties
of commutation with the field operators: 
$$
[\psi(x),\; \hat N] =\psi(x) \; , \quad [\psi^\dgr(x),\; \hat N]
=-\psi^\dgr(x) \; .
$$
From here, one gets 
$$
\psi(x)\hat N = (1+\hat N)\psi(x) \; , \quad 
\psi(x)\hat N^n = (1 + \hat N)^n\psi(x) \; . 
$$
The gauge-transformation operator (139) can be represented as 
$$
\hat U_\vp = \sum_{n=-}^\infty \frac{(i\vp)^n}{n!}\; \hat N^n \; .
$$
Using this, we come to the relations
$$
\psi(x)\hat U_\vp = e^{i\vp}\hat U_\vp\psi(x) \; , \quad
\hat U^+_\vp\psi(x)\hat U_\vp = e^{i\vp} \psi(x) \; , \quad
\hat U^+_\vp\psi^\dgr (x)\hat U_\vp = e^{-i\vp} \psi^\dgr(x) \; .
$$

In equilibrium, $[\hat \rho, \hat H] = 0$. Since $\hat N$ is an integral 
of motion, it follows that 
$$
[\hat N,\; \hat\rho]=0\; , \quad [\hat U_\vp,\; \hat\rho]=0 \; .
$$

Because of the above equalities, we obtain 
$$
<\psi(x)\psi(x')>\; = \; {\rm Tr}\hat\rho\psi(x)\psi(x') =
{\rm Tr} \hat U_\vp\hat\rho\hat U_\vp^+\psi(x)\psi(x') =
$$
$$
={\rm Tr}\hat\rho\hat U^+_\vp\psi(x)\psi(x')\hat U_\vp =
e^{2i\vp}{\rm Tr}\hat\rho\psi(x)\psi(x') \; .
$$
Therefore the equality
$$
<\psi(x)\psi(x')>\left ( 1 - e^{2i\vp}\right ) = 0
$$
is valid for any $\vp$. This implies that 
\be
\label{140}
<\psi(x)\psi(x')>\; = 0\; , 
\ee
while $<\psi^\dgr(x)\psi(x')>\; \neq 0$. In the same way, it is 
straightforward to show that
$$
<\psi(x)> \left ( 1 - e^{i \vp} \right ) = 0 
$$
foe any $\vp$, hence $<\psi(x)> = 0$. Generally, we find the equality
$$
<\prod_{i=1}^n \psi^\dgr(x_i)\; \prod_{j=1}^m \psi(x_j)>\;
\left [ 1 - e^{i\vp(m-n)}\right ] = 0 \; ,
$$
telling us that it is zero for $m\neq n$. Because this is zero for an 
arbitrary $\vp$, we come to the conclusion that
\be
\label{141}
<\prod_{i=1}^n \psi^\dgr(x_i) \prod_{j=1}^m \psi(x_j)>\; = 0 \quad
(m\neq n) \; .
\ee
Thus, all anomalous averages are identically zero. $\square$

\vskip 2mm

Contrary to this, the normal averages are not zero,
$$
< \prod_{i=1}^n \psi^\dgr(x_i)\psi(x_i')>\; \neq 0 \; .
$$

Since the appearance of the anomalous averages necessarily accompanies
Bose-Einstein condensation or superconducting transition 
\cite{Lieb_1,Yukalov_12}, this means that these transitions cannot occur
in a gauge-symmetric system, but the symmetry breaking must happen.

\subsection{Translation invariance}

When the operator of momentum is an integral of motion, there exist
constraints on the averages, from which it follows that such a system has 
to be uniform. In this section, we understand the variable $x$ as a spatial 
variable, $x\ra\br$. 

Let us assume that there is no external potential, $U(\br)=0$. Then the 
operator of momentum is an integral of motion:
$$
[\hat\bP,\; \hat H]=0 \; , \quad 
\hat\bP = \int \psi^\dgr(\br)(-i\nabla)\psi(\br)\; d\br \; .
$$
This is called the {\it momentum conservation}. The family of the 
{\it translation transformations} 
\be
\label{142}
\hat T(\br) =\exp(i\hat\bP\cdot\br)
\ee
forms the translation group $\{\hat T(\br)\}$. Recall that, according to 
Sec. 3.8, one has $\psi(\br) = \hat T^+(\br)\psi(0)\hat T(\br)$. When 
there is no external potential, the momentum operator commutes with 
Hamiltonian (138), so that
$$
[\hat T(\br),\; \hat H]=0\; , \quad [\hat T(\br),\; \hat\rho]=0 \; ,
$$
for an equilibrium $\hat\rho$. 

\vskip 2mm

{\bf Statement 2}. When the operator of momentum is an integral of motion, 
then 
\be
\label{143}
<\prod_{i=1}^n \psi^\dgr(\br_i+\br)\; \prod_{j=1}^m \psi(\br_j +\br)>\; =
\; < \prod_{i=1}^n \psi^\dgr(\br_i) \; \prod_{j=1}^m \psi(\br_j)> \; ,
\ee
for any integers $m,n$.
 
\vskip 2mm

{\it Proof}. Starting with the case of two field operators, we have
$$
<\psi^\dgr(\br_1+\br)\psi(\br_2+\br)>\; = {\rm Tr}\hat\rho
\hat T^+(\br)\psi^+(\br_1)\psi(\br_2)\hat T(\br) =
$$
$$
={\rm Tr}\hat T^+(\br)\hat\rho\hat\psi^\dgr(\br_1)\psi(\br_2)\hat T(\br) =
{\rm Tr}\hat\rho\psi^\dgr(\br_1)\psi(\br_2) \; .
$$
That is,
$$
<\psi^\dgr(\br_1+\br)\psi(\br_2+\br)>\; =
\; <\psi^\dgr(\br_1)\psi(\br_2)> \; .
$$
Similarly, for the product of any number of field operators, we obtain
Eq. (143). $\square$

\vskip 2mm

In particular, we have
$$
<\psi^\dgr(\br_1)\psi(\br_2)>\; = \; <\psi^\dgr(\br_1-\br_2)\psi(0)> \; ,
$$
which leads to the uniform particle density
$$
<\psi^\dgr(\br)\psi(\br)>\; = \; <\psi^\dgr(0)\psi(0)> \; .
$$

For a uniform system of $N$ atoms in volume $V$, it is convenient to use
the Fourier basis $\{\vp_k(\br)\}$ of plane waves
$$
\vp_k(\br) = \frac{1}{\sqrt{V}}\; e^{i\bk\cdot\br} \; .
$$
The basis is orthonormalized, $(\vp_k,\vp_p)=\dlt_{kp}$, and complete,
$$
\sum_k \vp_k(\br)\vp_k^*(\br') = \dlt(\br-\br') \; .
$$

Imposing the periodic boundary conditions $\vp_k(\br+{\bf L})=\vp_k(\br)$,
where ${\bf L} = \{ L^\al\}$ and $V = \prod_\al L^\al$, on has 
$\bk\cdot{\bf L} = 2\pi$, hence
$$
\bk=\{ k^\al\}\; , \quad
k^\al = \frac{2\pi}{L^\al}\; n^\al \quad (n^\al=0,\pm 1,\pm 2,\ldots) \; . 
$$
In the thermodynamic limit, when $N\ra \infty, V\ra\infty, L^\al\ra \infty$,
the difference between the neighbor wave numbers $\Dlt k^\al = 2\pi/L^\al$,
for which $\Dlt n^\al = 1$, tends to zero, $\Dlt k^\al\ra 0$. In that case,  
the summation over the wave numbers $k^\al$, which is equivalent to the 
summation over $n^\al$, so that $\sum_k \equiv \sum_{ \{ n^\al\} }$, can be
replaced by the integration 
$$
\sum_{ \{ n^\al\} } \ra \int d{\bf n} \; \quad ({\bf n}=\{ n^\al\}) \; .
$$
Since
$$
dn^\al = L^\al\; \frac{dk^\al}{2\pi} \; , \quad
d{\bf n} = V\; \frac{d\bk}{(2\pi)^3} \; , 
$$
the replacement takes the form
$$
\sum_k \ra V \int \frac{d\bk}{(2\pi)^3} \; .
$$

Passing to the momentum representation, one expands the field operators 
over the Fourier basis,
$$
\psi(\br) = \sum_k a_k \vp_k(\br) \; , 
$$
where
$$
a_k=(\vp_k,\psi) = \int \vp_k^*(\br)\psi(\br)\; d\br \; .
$$
From the commutation relations $[\psi(\br,\; \psi(\br')]_\mp = 0$ and
$[\psi(\br,\; \psi^\dgr(\br')]_\mp = \dlt(\br-\br')$, one gets the 
commutation relations 
$$
[a_k,\; a_p]_\mp = 0\; , \quad [a_k,\; a^\dgr_p]_\mp =\dlt_{kp} \; .
$$

Because of the momentum conservation, the system is translation invariant, 
such that 
$$
\frac{\prt}{\prt\br}\; <\psi^\dgr(\br_1+\br)\psi(\br_2+\br)>\; = 0 \; .
$$
Passing to the momentum representation in the average
$$
<\psi^\dgr(\br_1+\br)\psi(\br_2+\br)>\; =\; \sum_{kp}
<a_k^\dgr a_p>\; \vp_k^*(\br_1+\br)\vp_p(\br_2+\br)
$$
yields
$$
\frac{\prt}{\prt\br}\; <\psi^\dgr(\br_1+\br)\psi(\br_2+\br)>\; =
- i\sum_{kp} <a_k^\dgr a_p> (\bk-{\bf p})\vp_k^*(\br_1+\br)
\vp_p(\br_2+\br) \; = 0 \; .
$$
From here, it is seen that for a uniform system, one has the properties
\be
\label{144}
<a_k^\dgr a_p>\; =\dlt_{kp} <a_k^\dgr a_k>\; , \quad
<a_k a_p>\; = \dlt_{-kp} <a_{-k} a_k> \; . 
\ee
In general, the average
\be
\label{145}
< \prod_{i=1}^n a^\dgr_{k_i}\; \prod_{j=1}^m a_{p_j}>\; \neq 0
\ee
is not zero only when 
$\bk_1 + \ldots + \bk_n = {\bf p}_1 + \ldots + {\bf p}_m$. 

In this way, if the operator of momentum is an integral of motion, the 
system is uniform and cannot characterize nonuniform systems, such as 
crystalline matter. To be able to describe the latter, translation 
symmetry breaking is necessary.

\subsection{Symmetry breaking}

As is stressed in the previous two sections, in order to describe some 
thermodynamic phases, it is necessary to consider systems with broken 
symmetries. For instance, if the Hamiltonian of a magnetic system is 
invariant under spin inversion or rotation, that is, exhibits the related
symmetry with respect to the magnetic moment ${\bf M}$, then the latter is
identically zero \cite{Bogolubov_49}.

Or, when the operator of momentum is an integral of motion, 
$[\hat\bP,\; \hat H]=0$, the system enjoys translation symmetry, as is shown
in Sec. 4.14, because of which the average 
$\langle \psi^\dgr(\br)\psi(\br) \rangle = \langle \psi^\dgr(0)\psi(0) \rangle$ 
does not depend on the spatial variable. But the average
\be
\label{146}
\rho(\br)=<\psi^\dgr(\br)\psi(\br)>
\ee
represents the spatial distribution of particles, that is, the particle 
density. If $\rho(\br)=\rho(0)=const$, then there can be no crystalline matter. 

Similarly, as is discussed in Sec. 4.13, when the system is gauge-symmetric,
all anomalous averages are identically zero, hence there can be neither 
Bose-Einstein condensation nor superconductivity.  

The above examples demonstrate that, for characterizing some thermodynamic 
phases, one has to look for states with a broken symmetry. The general idea
of how the symmetry, prescribed by the Hamiltonian, can be broken is 
straightforward. One has to impose some additional conditions, boundary 
conditions, initial conditions, or like that, for selecting from the total 
Fock space the microscopic states possessing the required symmetry properties. 
This is analogous to the selection of required solutions of nonlinear 
equations possessing multiple solutions. 

The direct way of selecting the required states is by the 
{\it method of restricted trace}. One defines a subspace of the Fock space
${\cal F}_\nu\in{\cal F}$, whose states enjoy the necessary symmetry properties.
And, calculating the observable quantities, one takes the trace only over the 
states of the selected subspace,
\be
\label{147}
<\hat A>_\nu \equiv {\rm Tr}_{{\cal F}_\nu}\; \hat\rho\hat A \; .
\ee

A general method of symmetry breaking has been developed by Bogolubov
\cite{Bogolubov_49} and is called the {\it method of Bogolubov quasiaverages}.
The idea of the method is as follows. Let the system Hamiltonian $\hat H$
possess some symmetry related to an integral of motion $\hat C$, so that
$[\hat C,\; \hat H]=0$. Suppose we wish to break the corresponding symmetry.
For this purpose, we introduce the Hamiltonian
\be
\label{148}
\hat H_\nu(\ep)\equiv \hat H +\ep\hat\Gm_\nu \; ,
\ee
adding to $\hat H$ a term $\hat\Gm_\nu$, called {\it source}, that breaks the 
symmetry because it does not commute with the integral of motion $\hat C$,  
$$
[\hat C,\; \hat\Gm_\nu] \neq 0\; , \quad
[\hat C,\; \hat H_\nu(\ep)]\neq 0 \; .
$$
The average of the source $<\Gm_\nu>$ is assumed to be an extensive quantity 
proportional to $N$ for large $N \gg 1$. The index $\nu$ enumerates the 
admissible phases.

The {\it quasiaverage} of an operator $\hat A(\bf r)$ is
\be
\label{149}
<\hat A(\br)>_\nu \equiv \lim_{\ep\ra 0}\; \lim_{N\ra\infty}\;
<\hat A(\br)>_{H_\nu(\ep)} \; ,
\ee
where
$$
<\hat A>_H\; \equiv 
\frac{{\rm Tr}\hat A e^{-\bt\hat H}}{{\rm Tr}e^{-\bt\hat H}} \; .
$$
Here $\lim_{N\ra\infty}$ implies the thermodynamic limit, when $N\ra\infty$, 
$V\ra\infty$ and $N/V \ra const$. It is principal that the limits $\ep\ra 0$ 
and $N\ra\infty$ do not commute. 

Instead of the double limit, as in the Bogolubov method, it is possible to 
employ a single limit, as in the {\it method of thermodynamic quasiaverages}
\cite{Yukalov_59}, by introducing the Hamiltonian
\be
\label{150}
\hat H_\nu \equiv H + \frac{1}{N^\al} \; \hat\Gm_\nu \quad (0 < \al < 1) \; ,
\ee
where the restriction $0 < \al < 1$ guarantees that the factor $1/N^\al$, in
the thermodynamic limit, goes to zero slower than $1/N$, which is necessary 
for the efficient selection of the states with broken symmetry. The average 
of the source $<\Gm_\nu>$ is again assumed to be an extensive quantity 
proportional to $N$ for large $N \gg 1$. Then the {\it thermodynamic 
quasiaverage} is defined by the sole thermodynamic limit 
\be
\label{151}
<\hat A(\br)>_\nu \equiv \lim_{N\ra\infty}\; <\hat A(\br)>_{H_\nu} \; .
\ee

For example, to break the spin rotational symmetry, it is sufficient to add
to the Hamiltonian the source representing the action of an external magnetic 
field. To break the translation symmetry, which is necessary for describing 
crystalline matter, it is sufficient to add the source
\be
\label{152}
\hat\Gm_\nu  = \int \psi^\dgr(\br) U(\br)\psi(\br)\; d\br\; ,
\ee
in which the external potential is periodic over the lattice vectors $\{\ba\}$,
such that $U(\br+\ba) = U(\br)$ and $[\hat\bP,\; \hat\Gm_\nu]\neq 0$. Then 
the particle density, defined as the Bogolubov quasiaverage
\be
\label{153}
\rho_\nu(\br) \equiv\lim_{\ep\ra 0}\; \lim_{N\ra\infty}\;
<\psi^\dgr(\br)\psi(\br)>_{H_\nu(\ep)} \; ,
\ee
where the index $\nu$ corresponds to the chosen crystalline lattice vectors 
$\{\ba\}$, is periodic over the lattice vectors,
\be
\label{154}
\rho_\nu (\br+\ba)=\rho_\nu (\br) \; .
\ee
Or one can use the thermodynamic quasiaverages (151) defining the single limit
\be
\label{155}
\rho_\nu(\br) \equiv \lim_{N\ra\infty}\; <\psi^\dgr(\br)\psi(\br)>_{H_\nu} \; .
\ee

The method of breaking the translation symmetry by adding a source with an 
external potential that is periodic over the considered crystalline lattice
was, first, advanced by Kirkwood \cite{Kirkwood_47}. Bogolubov 
\cite{Bogolubov_49} extended this method to the case of an arbitrary symmetry 
and gave it a precise mathematical meaning. The replacement of the double limit 
in the Bogolubov method by a single thermodynamic limit was suggested in 
Ref. \cite{Yukalov_59}. Some other methods of symmetry breaking are reviewed 
in Ref. \cite{Yukalov_33}.

Generally, there can exist several equilibrium states of a system, under the
same thermodynamic variables. However, not all those states are stable. The 
system always chooses a state that is the most stable. When it happens that 
a state with a broken symmetry is more stable than the symmetric state, the 
system passes to that more stable state with broken symmetry. This is termed 
{\it spontaneous symmetry breaking}.

\section{Quadratic Hamiltonians}

Averages can be calculated exactly, when the system Hamiltonian is quadratic
with respect to field operators. However, such a trivial situation happens
rather rarely, corresponding to ideal gases. Interacting systems, as a rule,
do not allow for exact calculations. But approximate calculations are possible 
by reducing the exact Hamiltonian to an approximate quadratic form.

\subsection{Occupation numbers}

In many cases, it is convenient to invoke the representation of occupation 
numbers. This representation is introduced as follows. Let us, first, pass
to the quantum-number representation by expanding the field operator over
an orthonormal basis,
\be
\label{156}
\psi(x) = \sum_k a_k\vp_k(x) \; ,
\ee
where the operators $a_k$ and $a_k^\dgr$ generate the Fock space ${\cal F}(a_k)$,
with $k$ being a set of quantum numbers (multi-index). Depending on the particle
statistics, the commutation relations are 
$$
[a_k,a_p]_\mp=0\; , \quad [a_k,a_p^\dgr]_\mp=\dlt_{kp} \; ,
$$
the upper sign corresponding to Bose statistics, while the lower, to 
Fermi statistics. 

The {\it occupation-number operator} is 
\be
\label{157}
\hat n_k \equiv a_k^\dgr a_k \; . 
\ee
This is a self-adjoint operator, $\hat n_k^\dgr=\hat n_k$, and it is
semipositive, since, for any $f\in{\cal F}(a_k)$, one has
$$
(f,\hat n_kf)=(f,a_k^\dgr a_kf)=(a_kf,a_kf)=||a_k f||^2\geq 0 \; .
$$
The number-of-particle operator can be expressed as the sum
$$
\hat N = \int \psi^\dgr(x)\psi(x)\; dx  = \sum_k a_k^\dgr a_k = \sum_k \hat n_k  
$$
of the occupation-number operators.

For any type of statistics, the commutation relations are valid:
$$
[\hat n_k,\; \hat n_p]=0 \; , \quad [ a_k,\;\hat n_p ] = \dlt_{kp} a_k \; ,
\quad [ a_k^\dgr,\;\hat n_p ] = -\dlt_{kp} a_k^\dgr \; ,
$$
where all square brackets represent commutators.

The eigenvalues $\nu_k$ of $\hat n_k$, called the {\it occupation numbers},
are given by the eigenproblem   
\be
\label{158}
\hat n_k|k>=\nu_k|k> \; . 
\ee
They are non-negative, $\nu_k\geq 0$, since $\hat n_k$ is semipositive. 

For {\it Fermi statistics}, the eigenvalues are easily found. As far as 
$a_k a_k = - a_k a_k$, one has $a_k^2 = 0$. Then
$$
\hat n_k^2=a_k^\dgr a_k a_k^\dgr a_k = a_k^\dgr(1-a_k^\dgr a_k) a_k =
a_k^\dgr a_k =\hat n_k \; .
$$
This gives $\nu_k^2=\nu_k$, since $\hat n_k^2|k>=\nu_k^2|k>$. From here
it follows that $\nu_k=0,1$, for Fermi statistics. 

But in the case of {\it Bose statistics}, the situation is a bit more involved.
First, we notice that if $|k>$ is an eigenvector of $\hat n_k$, then $(a_k)^m|k>$ 
and $(a_k^\dgr)^m|k>$ are also its eigenvectors. To prove this, let us consider
the equalities 
$$
\hat n_k a_k|k>\; = \left ( a_k\hat n_k +[\hat n_k,a_k]\right )|k>\; =
(\nu_k-1)a_k|k> \; ,
$$
$$
\hat n_k a_k^\dgr|k>\; = \left ( a_k^\dgr \hat n_k +[\hat n_k,a_k^\dgr]
\right ) |k>\; = (\nu_k+1)a_k^\dgr|k> \; .
$$
Continuing further, we find
$$
\hat n_k (a_k)^m|k>\; = (\nu_k - m)(a_k)^m|k> \; , \quad
\hat n_k (a_k^\dgr )^m|k>\; = (\nu_k + m)(a_k^\dgr)^m|k> \; .
$$
The eigenvalues of $\hat n_k$ cannot be negative, which means that, after 
some $m$, the null vector $|0>$ must exist, such that 
$$
(a_k)^m|k>\; = |0> \; , \quad a_k|0>=0 \; .
$$
For the null vector, the eigenvalue of $\hat n_k$ is zero. Hence, the 
eigenvalues $\nu_k$ of $\hat n_k$ are non-negative integers 
$\nu_k=0,1,2,\ldots$ ranging between $0$ and $m$. The latter can be set to be
arbitrarily large. 

The occupation-number representation is convenient for the use in the case
of quadratic Hamiltonians, such as in the case of ideal gases, or when the
Hamiltonian can be approximately reduced to the quadratic form. Then any 
function $f(\hat n_k)$ of the occupation-number operator, acting on the
eigenvectors of the latter, yields $f(\hat n_k)|k>=f(n_k)|k>$.

\subsection{Noninteracting particles}

The simplest case of a Hamiltonian, quadratic in the field operators, is 
that characterizing the ideal gases of noninteracting particles. Let us 
consider the Gibbs grand ensemble, with the grand Hamiltonian 
$H\equiv \hat H - \mu\hat N$. For noninteracting particles, one has
\be
\label{159}
H = \int \psi^\dgr(x)\left [ -\; \frac{\nabla^2}{2m} + U(x) - \mu\right ]
\psi(x)\; dx \; ,  
\ee
where $U(x)$ is an external potential. The field operators can be expanded, 
as in Eq. (156), over the orthonormal and complete basis $\{\vp_k(x)\}$ of 
the solutions to the Schr\"{o}dinger equation
$$
\left [ -\; \frac{\nabla^2}{2m} + U(x) - \mu\right ] \vp_k(x) =
\ep_k\; \vp_k(x) \; .
$$
As a result, Hamiltonian (159) reduces to the simple form
\be
\label{160}
H = \sum_k (\ep_k - \mu) a_k^\dgr a_k = \sum_k \om_k a_k^\dgr a_k \; ,
\ee
with the single-particle spectrum $\om_k \equiv \ep_k -\mu$.  

This Hamiltonian is evidently invariant under the local gauge transformations
$$
a_k \ra a_k e^{i\al_k} \; ,
$$
where $\al_k$ is a real-valued number. Therefore all anomalous averages are 
identically zero, such as $<a_k a_p>\; = 0$, while for the normal averages, 
we have $<a_k^\dgr a_p>\; = \dlt_{kp}<a_k^\dgr a_k>$. The average of the 
occupation-number operator $\hat n_k \equiv a_k^\dgr a_k$ gives the
{\it occupation-number distribution}
\be
\label{161}
n_k\equiv\; <\hat n_k> \; . 
\ee

To calculate the distribution $n_k$, we need to find the partition function 
$Z = {\rm Tr} e^{-\bt H}$. To this end, we notice that
$$
{\rm Tr}\exp\left ( - \bt\sum_k \om_k a_k^\dgr a_k\right ) =
{\rm Tr} \prod_k \exp\left ( - \bt \om_k\hat n_k \right ) =
\prod_k \sum_{\nu_k} \exp ( -\bt\om_k\nu_k) \; ,
$$
with $\nu_k=0,1,2,\ldots$ for Bose statistics and $\nu_k=0,1$ for Fermi 
statistics. Hence, for Bose statistics,
$$
\sum_{\nu_k} e^{-\bt\om_k\nu_k}  = 1 + e^{-\bt\om_k} +
e^{-2\bt\om_k} + \ldots  = \frac{1}{1-e^{-\bt\om_k} } \; .
$$
Consequently,
$$
Z = \prod_k \left ( 1 - e^{-\bt\om_k} \right )^{-1} \quad ({\rm Bose}) \; .
$$
For Fermi statistics,
$$
\sum_{\nu_k} e^{-\bt\om_k\nu_k}  = 1 + e^{-\bt\om_k} \; .
$$
Therefore,
$$
Z = \prod_k \left ( 1 + e^{-\bt\om_k} \right ) \quad ({\rm Fermi}) \; .
$$
Combining both these cases, we get
\be
\label{162}
Z = \prod_k \left ( 1 \mp e^{-\bt\om_k} \right )^{\mp 1} \; .
\ee

The grand potential is
$$
\Om = -T \ln Z = - T\ln{\rm Tr} e^{-\bt H} = 
\pm T \sum_k \ln\left ( 1 \mp e^{-\bt\om_k} \right ) \; .
$$
From here, we find the occupation-number distribution
\be
\label{163}
n_k =\; <\hat n_k>\; = \frac{\dlt\Om}{\dlt\om_k} = \frac{1}{e^{\bt\om_k}\mp 1} \; .
\ee
Using the identity $e^{\bt\om_k} =(1 \pm n_k)/n_k$, we get the grand potential  
\be
\label{164}
\Om =\mp T \sum_k \ln(1\pm n_k) \; .
\ee

Other thermodynamic quantities can also be calculated exactly using the 
derivatives
$$
\frac{\prt n_k}{\prt T} = \bt^2 \om_k n_k(1\pm n_k) \; , \quad
\frac{\prt n_k}{\prt\mu} = \bt n_k(1 \pm n_k) \; .
$$
Thus the internal energy is
$$
E =\; <H> + \mu = \sum_k \ep_k n_k \; . 
$$
For the entropy, we have 
$$
S= -\left ( \frac{\prt\Om}{\prt T} \right )_{V\mu} = 
\sum_k \left [ \bt\om_k n_k \pm \ln(1 \pm n_k)\right ] \; .
$$
The number of particles is 
$$
N = -\; \frac{\prt\Om}{\prt\mu} =\sum_k n_k \; . 
$$
Since $\Om = -PV$, the pressure becomes
$$
P =\pm\; \frac{T}{V}\; \sum_k \ln(1 \pm n_k) \; .
$$
The free energy is $F = E - TS = \Om + \mu N$. 

It is worth recalling that $\hat n_k$ is a semipositive operator, hence
the distribution $n_k = <\hat n_k>$ has also to be semipositive. 
This implies that $e^{\bt\om_k} \geq \pm 1$. For Fermi statistics, 
$n_k\geq 0$ in any case. But for Bose statistics, it should be 
$e^{\bt\om_k} \geq 1$, which means that $\om_k\geq 0$, so that 
$\mu\leq\ep_k$. 

The specific heat reads as
\be
\label{165}
C_V = \frac{T}{N}\left ( \frac{\prt S}{\prt T}\right )_{V\mu} = 
\frac{\bt^2}{N}\; \sum_k \om_k^2 n_k (1\pm n_k) \; ,
\ee
where the derivative
$$
\frac{\prt S}{\prt T} = \bt^3 \sum_k \om_k^2 n_k (1\pm n_k)
$$
is used. For a stable system, it should be: $0 \leq C_V < \infty$.

For the isothermal compressibility, using the derivative
$$
\frac{\prt N}{\prt\mu} = \bt\sum_k n_k (1 \pm n_k) \; , 
$$
we have
\be
\label{166}
\kappa_T = \frac{1}{\rho N}\left ( \frac{\prt N}{\prt\mu} \right )_{TV} =
\frac{{\rm var}(\hat N)}{\rho NT} \; ,
\ee
with the number-operator dispersion
\be
\label{167}
{\rm var}(\hat N) = \sum_k n_k (1\pm n_k) \; .
\ee
The system is stable provided that $0\leq \kappa_T <\infty$, which implies
that ${\rm var}(\hat N)\sim N$. 

Similar expressions arise for interacting systems treated in a mean-field 
approximation, with the difference that $\om_k$ becomes an effective spectrum
defined by the corresponding self-consistency conditions.

\subsection{Correlation functions}

In the case of the quadratic grand Hamiltonian (160), it is straightforward 
to calculate various correlation functions. Let $b_i$ be either $a_{k_i}$ 
or $a_{k_i}^\dgr$. And let us consider the correlation functions of the type
$<b_1b_2\ldots b_{2n}>$, containing even numbers of the field operators. Such
correlators can be expressed through the combinations of the lower-order 
correlators, as we shall illustrate for the fourth-order correlation function. 

\vskip 2mm

{\bf Wick theorem}. The correlation function $<b_1b_2b_3b_4>$, where the 
averaging is defined using the equilibrium statistical operator  
$\hat\rho = e^{-\bt H}/ Z$, with Hamiltonian (160), reads as
\be
\label{168}
<b_1b_2b_3b_4>\; = \; <b_1b_2><b_3b_4> \; \pm \;
<b_1b_3><b_2b_4>\; + \; <b_1b_4><b_2b_3> \; .
\ee

\vskip 2mm

{\it Proof}. With  Hamiltonian (160), we have
$$
a_k H = (H+\om_k) a_k\; , \quad a_k H^n = (H+\om_k)^n a_k \; ,
$$
from where
$$
a_k e^{-\bt H} = e^{-\bt H} a_k e^{-\bt\om_k} \; , \quad
a_k^\dgr e^{-\bt H} = e^{-\bt H} a_k^\dgr e^{\bt\om_k} \; .
$$
Therefore
$$
(1\pm n_k) a_k\hat\rho = n_k \hat\rho a_k \; , \quad
n_k a_k^\dgr\hat\rho = \hat\rho a_k^\dgr (1\pm n_k) \; .
$$
Using this, for any operator $\hat A$, we get 
$$
(1\pm n_k)<\hat A a_k> \; = n_k<a_k\hat A>\; , \quad
n_k<\hat A a_k^\dgr>\; = (1\pm n_k)<a_k^\dgr\hat A> \; ,
$$
$$
<a_k \hat A>\; = (1\pm n_k)<[a_k,\; \hat A]_\mp >\; , \quad
<a_k^\dgr \hat A>\; = \mp n_k <[a_k^\dgr,\; \hat A]_\mp > \; .
$$

Recall the commutation relations
$$
[a_k,\; a_p]_\mp = 0 \; , \quad [a_k^\dgr,\; a_p^\dgr]_\mp = 0 \; ,
$$
$$
[a_k,\; a_p^\dgr]_\mp = \dlt_{kp} \; , \quad
[a_k^\dgr,\; a_p]_\mp = \mp\dlt_{kp} 
$$
and the properties
$$
<a_k^\dgr a_p>\; = \dlt_{kp} n_k \; , \quad <a_k a_p^\dgr>\; =
\dlt_{kp}(1\pm  n_k) \; .
$$
Excluding here $\dlt_{kp}$, we find
$$
(1\pm n_k) [a_k,\; b_i]_\mp = \; <a_k b_i>\; , \quad
\mp n_k  [a_k^\dgr,\; b_i]_\mp =\mp\; <a_k^\dgr b_i> \; .
$$
Using the equality  
$$
[A,\; BCD]_\mp = [A,\; B]_\mp CD \pm B[A,\; C]_\mp D +
BC[A,\; D]_\mp \; ,
$$
valid for any operators $A,B,C,D$, we come to the expression
$$
<a_kb_2b_3b_4>\; = \; <a_kb_2><b_3b_4>\; \pm \; <a_kb_3><b_2b_4>\; +
\; <a_kb_4><b_2b_3>\; ,
$$
and to the same expression for $a_k^\dgr$. This proves the theorem. $\square$

\vskip 2mm

Because of relation (156), the equivalent representation is valid for
the field operators $\psi(x)$. Let $\psi_i$ be either $\psi(x_i)$ or 
$\psi^\dgr(x_i)$. Then
$$
<\psi_1\psi_2\psi_3\psi_4>\; = \; <\psi_1\psi_2><\psi_3\psi_4>\; \pm \;
<\psi_1\psi_3><\psi_2\psi_4>\; + \; <\psi_1\psi_4><\psi_2\psi_3> \; .
$$

Generally, any correlator $<b_1 b_2\ldots b_{2n}>$, calculated with
the quadratic Hamiltonian (160), results in the combination
$$
<b_1b_2b_3b_4\ldots b_{2n}>\; = \; <b_1b_2><b_3b_4> \ldots <b_{2n-1}b_{2n}>\;
\pm
$$
$$
\pm \; <b_1b_3><b_2b_4>\ldots <b_{2n-1}b_{2n}>\; + \;
<b_1b_4><b_2b_3>\ldots <b_{2n-1}b_{2n}> \; + \ldots \; .
$$
In the same way, the correlator $<\psi_1\psi_2\ldots\psi_{2n}>$ can be 
represented as the combination of the products of all possible pair
{\it contractions} $<\psi_i \psi_j>$.  

However, one has to be cautious when the quadratic Hamiltonian is not exact,
but is a result of an approximation. Then calculating the correlators of the
orders higher than second can lead to unreasonable conclusions. This is the 
simple consequence of the approximation theory prescribing that, if the main
Hamiltonian is of second order, then only the correlators of second order are 
reliable. In brief, accepting an approximation of some order, one should not
go outside of the region of applicability of that approximation, but must 
restrain calculations to the same order.

\subsection{Canonical transformations}

The general form of a quadratic Hamiltonian is 
\be
\label{169}
H = \int \left [\psi^\dgr(x) A(x,x') \psi(x') + \frac{1}{2}\; B(x,x')
\psi^\dgr(x)\psi^\dgr(x') + \frac{1}{2}\; B^*(x,x')\psi(x)\psi(x')
\right ] \; dx dx' \; .
\ee
Such a quadratic form can be diagonalized by means of canonical 
transformations \cite{Bogolubov_30}.

The Hamiltonian can be rewritten in the quantum-number representation, 
for which one expands the field operators over an orthonormal complete basis,
labeled by quantum multi-indices $k$,
$$
\psi(x) = \sum_k a_k \vp_k(x) \; , \quad
a_k = \int \vp_k^*(x)\psi(x)\; dx \; .
$$
Introducing the notation
$$
A(x,x') \equiv \sum_{kp} A_{kp}\vp_k(x)\vp_p^*(x') \; , \quad
B(x,x') \equiv \sum_{kp} B_{kp} \vp_k(x)\vp_p(x') \; ,
$$
one gets the Hamiltonian
\be
\label{170}
H =\sum_{kp} \left ( A_{kp} a_k^\dgr a_p + \frac{1}{2}\;
B_{kp}\; a_k^\dgr a_p^\dgr + \frac{1}{2}\; B_{kp}^* a_p a_k \right ) \; .
\ee
In general, the Hamiltonian can contain nonoperator terms, which are not 
included here.

The Hamiltonian is self-adjoint, $H^+ = H$, which requires that 
$A_{kp}^*=A_{pk}$. Also, it should be that $B_{pk} = \pm B_{kp}$, since
$$
\sum_{kp} B_{kp}a_k^\dgr a_p^\dgr = \pm \sum_{kp} B_{kp} a_p^\dgr a_k^\dgr
= \pm \sum_{kp} B_{pk} a_k^\dgr a_p^\dgr \; . 
$$

The general form of the {\it Bogolubov canonical transformations} is
\be
\label{171}
a_k = \sum_p \left ( u_{kp} b_p + v_{kp}^* b_p^\dgr \right )\; , \quad
a_k^\dgr = \sum_p \left ( u_{kp}^* b_p^\dgr + v_{kp} b_p \right )\; .
\ee
It is termed canonical because it does not change the particle statistics,
so that the commutation relations for the operators $a_k$ are the same as 
these for the operators $b_k$:
$$
[a_k,\; a_p]_\mp = 0 \; , \quad [a_k,\; a_p^\dgr]_\mp = \dlt_{kp} \; ,
$$
$$
[b_k,\; b_p]_\mp = 0 \; , \quad [b_k,\; b_p^\dgr]_\mp =  \dlt_{kp} \; .
$$
This imposes the restrictions
$$
\sum_q \left ( u^*_{kq}v_{pq} \mp v_{kq}u^*_{pq}\right ) = 0\; ,
\quad \sum_q \left ( u^*_{kq}u_{pq} \mp v_{kq}v^*_{pq}\right ) =\dlt_{kp} \; .
$$

One requires that the Hamiltonian would acquire the diagonal form called the 
{\it Bogolubov Hamiltonian}
\be
\label{172}
H =\sum_k \om_k b_k^\dgr b_k + E_0 \; . 
\ee
The diagonalization condition gives 
$$
E_0 = E_0^* = -\sum_{kp} \om_k |v_{pk}|^2  \; ,
$$
while the spectrum $\om_k = \om_k^*$ and the expansion functions are defined
by the {\it Bogolubov equations} 
\be
\label{173}
\sum_q \left [ ( A_{kq} - \dlt_{kq}\om_p) u_{qp} + B_{kq} v_{qp} \right ]= 0 \; , 
\quad \sum_q \left [ ( A_{kq}^* + \dlt_{kq}\om_p) v_{qp} + B_{kq}^*
u_{qp} \right ] = 0 \; .
\ee

The operators $a_k$ describe particles, while the operators $b_k$ correspond
to quasiparticles, that is, to excitations. The distribution of particles is 
$$
n_k\equiv<a_k^\dgr a_k> = \sum_p \left [ \left ( |u_{kp}|^2 \pm |v_{kp}|^2\right )\pi_k
+ |v_{kp}|^2\right ] \; , 
$$
while the distribution of excitations is
$$
\pi_k \equiv \; < b_k^\dgr b_k>\; = \frac{1}{e^{\bt\om_k}\mp 1} \; .
$$
The particle distribution is summed to the total number of particles  
$N = \sum_k n_k$. But the sum over the excitation distribution is not fixed, 
being dependent on thermodynamic variables. 

Because of the diagonal form of the Bogolubov Hamiltonian (172), one has 
$$
<b_k> \; =0 \; , \quad <b_k b_p>\; = 0 \; , 
\quad  <b_k^\dgr b_p>\; = \dlt_{kp}\pi_k \; ,
\quad  <b_k b_p^\dgr >\; = \dlt_{kp}(1\pm \pi_k) \; .
$$

As follows from transformation (171), $<a_k^\dgr>=0$, that is, there is 
no generation of single particles. However, there can exist the anomalous 
average
$$
<a_k a_p>\; = \sum_q \left [ \left ( v_{kq}^* u_{pq} \pm
u_{kq} v_{pq}^*\right ) \pi_q + u_{kq} v_{pq}^* \right ] \; .
$$
The nonzero $<a_k a_p>\neq 0$ implies the breaking of gauge symmetry for 
pairs, which can be created or annihilated together. Respectively, the 
anomalous average 
$$
<\psi(x)\psi(x')>\; = \sum_{kp} <a_k a_p> \vp_k(x)\vp_p(x') 
$$
is also nonzero.

\subsection{Linear terms}

It may happen that the Hamiltonian contains the terms linear in the field 
operators, as in the expression
\be
\label{174}
H = \sum_{kp} \left ( A_{kp} a_k^\dgr a_p + \frac{1}{2}\;
B_{kp} a_k^\dgr a_p^\dgr + \frac{1}{2}\; B_{kp}^* a_p a_k \right ) +
\sum_k \left ( C_k a_k^\dgr + C_k^* a_k \right ) \; .
\ee
This Hamiltonian is not gauge invariant, because of which $<a_k>$ is,
generally, nonzero, meaning that these particles can be created and 
annihilated. Since the particle spin is conserved, such a situation can 
occur only for particles of spin $0$, that is, for Bose particles. 
Therefore, the operators $a_k$ here satisfy the Bose commutation relations.

To reduce Hamiltonian (174) to a diagonal form, one, first, diagonalizes the 
quadratic part of the Hamiltonian as earlier in the previous section, which 
yields
$$
H = \sum_k \left ( \om_k b_k^\dgr b_k + D_k b_k^\dgr + D_k^* b_k\right ) + E_0 \; , 
$$
with
$$
D_k \equiv \sum_p \left ( C_p u_{pk}^* + C_p^* v_{pk}\right ) \; .
$$
Then one invokes another canonical transformation 
\be
\label{175}
\hat b_k = b_k +\frac{D_k}{\om_k} \; ,
\ee
where the operators $\hat b_k$ satisfy the same Bose commutation relations
as the operators $a_k$, 
$$
[\hat b_k,\; \hat b_p] = 0 \; , \quad 
[\hat b_k,\; \hat b_p^\dgr] = \dlt_{kp} \; .
$$
As a result, we come to the diagonal form
\be
\label{176}
H =\sum_k \om_k \hat b_k^\dgr \hat b_k + \tilde E_0\; , 
\ee
in which
$$
\tilde E_0 = E_0  - \sum_k \frac{|D_k|^2}{\om_k} \; .
$$

The diagonal form of Hamiltonian (176) tells us that 
$$
<\hat b_k>\; = 0\; , \quad <\hat b_k \hat b_p>\; = 0 \; , 
$$
and the distribution of excitations is given by the equation
$$
<\hat b_k^\dgr \hat b_p>\; = \frac{\dlt_{kp}}{e^{\bt\om_k}\mp 1} \; .
$$

However the quasiparticles, characterized by the operators $b_k$, are 
not conserved and can be created or destructed as single objects as well
as pairs:
$$
<b_k>\; = -\; \frac{D_k}{\om_k} \; , \quad
<b_kb_p>\; = \frac{D_kD_p}{\om_k\om_p}\; , \quad
<b_k^\dgr b_k>\; = \; <\hat b_k^\dgr\hat b_k>\; + \; \frac{|D_k|^2}{\om_k^2} \; .
$$
The same concerns the original particles described by the operators $a_k$,
which can be created or annihilated by pairs as well as by single particles. 
For example,
$$
<a_k> = - \sum_p \left ( \frac{D_p}{\om_p} u_{kp} + 
\frac{D_p^*}{\om_p} v_{kp}^* \right ) \; .
$$
This shows that, if the Hamiltonian contains the terms linear in the field 
operators, as in Eq. (174), hence the gauge symmetry is broken, then $<a_k>$
is not zero.

\subsection{Hartree approximation}

The Hamiltonian of interacting particles is not quadratic and, generally, 
is given by the expression  
\be
\label{177}
H = \int \psi^\dgr(x) \left [ -\; \frac{\nabla^2}{2m} + U(x) - \mu\right ]
\psi(x)\; dx +
\frac{1}{2}\int \psi^\dgr(x)\psi^\dgr(x')\Phi(x,x')\psi(x')\psi(x)\; dx dx' \; .
\ee
To reduce it to a quadratic form, one needs to resort to some approximations.  
The simplest such an approximation is the {\it Hartree approximation} that,
for averages, corresponds to the decoupling
\be
\label{178}
<\psi^\dgr(x_1)\psi^\dgr(x_2)\psi(x_3)\psi(x_4)>\; = \;
<\psi^\dgr(x_1)\psi(x_4)><\psi^\dgr(x_2)\psi(x_3)> \; .
\ee
In terms of the field operators, this is equivalent to the equality
$$
\psi^\dgr(x_1)\psi^\dgr(x_2)\psi(x_3)\psi(x_4) =
\psi^\dgr(x_1)\psi(x_4) <\psi^\dgr(x_2)\psi(x_3)> +
$$
\be
\label{179}
+ <\psi^\dgr(x_1)\psi(x_4)>\psi^\dgr(x_2)\psi(x_3) -
<\psi^\dgr(x_1)\psi(x_4)><\psi^\dgr(x_2)\psi(x_3)> \; .
\ee

Introducing the {\it Hartree potential}
$$
U_H(x) \equiv \int \Phi(x,x')\rho(x')\; dx'  
$$
and the average density
$$
\rho(x) =\; <\hat n(x)>\; , \quad \hat n(x) \equiv \psi^\dgr(x)\psi(x) 
$$
makes it possible to reduce the initial Hamiltonian (177) to the Hartree 
approximation 
\be
\label{180}
H= \int\psi^\dgr(x)\left [ -\; \frac{\nabla^2}{2m} + U(x) + U_H(x) -\mu
\right ]\psi(x)\; dx + E_0 \; ,
\ee
where
$$
E_0 = -\; \frac{1}{2} \int \Phi(x,x')\rho(x)\rho(x')\; dx dx' \; .
$$

If we choose the orthonormal basis $\{\vp_k(x)\}$ as the set of solutions
to the equation 
$$
\left [ -\; \frac{\nabla^2}{2m} + U(x) + U_H(x)\right ]\vp_k(x) =
\ep_k\vp_k(x) \; ,
$$
then, expanding the field operators $\psi(x)$ over this basis, we come to
the Hamiltonian
\be
\label{181} 
H = \sum_k \om_k a_k^\dgr a_k + E_0 \;  \quad (\om_k = \ep_k -\mu) \; .
\ee
For the average density, we have
$$
\rho(x) = \sum_k n_k |\vp_k(x)|^2 \; , \quad
n_k =\; < a_k^\dgr a_k>\; = \left ( e^{\bt\om_k}\mp 1 \right )^{-1} \; .
$$

As an example, let us consider a uniform system, when there is no external 
potential, $U(\br)=0$, and let $x=\br$. Also, we take into account that
the interaction potential usually depends on the difference of the spatial 
variables, $\Phi(\br,\br')=\Phi(\br-\br')$. Then the expansion basis is 
composed of the plane waves $\vp_k(\br) = e^{i\bk\cdot\br}/\sqrt{V}$ and the 
average density is
$$ 
\rho(\br) = \frac{1}{V}\; \sum_k n_k = \rho \; ,
$$
which is a constant defining the chemical potential $\mu=\mu(T,\rho)$. 
Consequently,
$$
U_H(\br)=\rho\Phi_0\; , \quad \Phi_0 \equiv \int \Phi(\br)\; d\br \; ,
\quad \ep_k = \frac{k^2}{2m} +\rho\Phi_0 \; .
$$
The internal energy of the system is 
$$
E=\; <H> + \mu N = E_0 +\sum_k n_k\ep_k\; , \quad
E_0 = -\; \frac{1}{2}\; N\rho\Phi_0 \; ,   
$$
with $E_0$ being the ground-state energy. 

As is seen, for a uniform system, the Hartree approximation is applicable
only if the interaction potential is integrable:
$$
\left | \int \Phi(\br)\; d\br \right | <\infty \; .
$$

Another example is a periodic system with a crystalline lattice defined 
by the crystalline vectors $\{\ba_i\}$, with $i = 1, 2, \ldots, N$ 
enumerating the lattice sites. Then the expansion basis can be formed by 
well localized functions $\vp_i(\br)=\vp(\br-\ba_i)$, called localized 
orbitals. Good localization means that the {\it overlap integral}
$$
(\vp_i,\vp_j) \equiv \int \vp^*(\br-\ba_i)\vp(\br-\ba_j)\; d\br
$$
is small. By the Cauchy-Schwarz inequality,
$$
|(\vp_i,\vp_j)|^2 \leq |(\vp_i,\vp_i)(\vp_j,\vp_j)| \; .
$$
Since the orbitals are normalized, $(\vp_i,\vp_i)=1$, one has 
$|(\vp_i,\vp_j)|\leq 1$. And the orbitals are well localized, when
$|(\vp_i,\vp_j)|\ll 1$ $(i\neq j)$. If the orbitals are so well localized 
that they do not intersect in space, then the requirement for the 
integrability of the interaction potential is not compulsory.

\subsection{Hartree-Fock approximation}

The Hartree approximation can work well when the particle exchange is 
negligible, for instance, as in the case of well localized orbitals. When 
the particle exchange is important, one has to resort to the Hartree-Fock 
approximation. This approximation implies for the averages the decoupling
$$
<\psi^\dgr(x_1)\psi^\dgr(x_2)\psi(x_3)\psi(x_4)>\; = \;
<\psi^\dgr(x_1)\psi(x_4)><\psi^\dgr(x_2)\psi(x_3)>\pm \; .
$$
\be
\label{182}
\pm <\psi^\dgr(x_1)\psi(x_3)><\psi^\dgr(x_2)\psi(x_4)> \; .
\ee
And in terms of the operators, this is equivalent to the construction 
$$
\psi^\dgr(x_1)\psi^\dgr(x_2)\psi(x_3)\psi(x_4) =\psi^\dgr(x_1)\psi(x_4)
<\psi^\dgr(x_2)\psi(x_3)> +
$$
$$
+ <\psi^\dgr(x_1)\psi(x_4)>\psi^\dgr(x_2)\psi(x_3) -
<\psi^\dgr(x_1)\psi(x_4)><\psi^\dgr(x_2)\psi(x_3)> \pm
$$
$$
\pm \psi^\dgr(x_1)\psi(x_3)<\psi^\dgr(x_2)\psi(x_4)> \pm
<\psi^\dgr(x_1)\psi(x_3)>\psi^\dgr(x_2)\psi(x_4) \mp
$$
\be
\label{183}
\mp <\psi^\dgr(x_1)\psi(x_3)><\psi^\dgr(x_2)\psi(x_4)> \; .
\ee

Let us define the Hartree potential $U_H(x)$, as in the previous section.  
And let us introduce the {\it Hartree-Fock operator potential}
$$
\hat U_{HF}(x) \equiv U_H(x) + \hat U_{ex}(x) \; ,
$$
where $\hat U_{ex}(x)$ is the {\it exchange operator potential} defined 
by the action
$$
\hat U_{ex}(x)\psi(x) \equiv \pm
\int \Phi(x,x')\rho_1(x,x')\psi(x')\; dx' \; .
$$
Then the Hartree-Fock approximation for Hamiltonian (177) leads to
\be
\label{184}
H = \int \psi^\dgr(x)\left [ -\; \frac{\nabla^2}{2m} + U(x) +
\hat U_{HF}(x) - \mu\right ] \psi(x) \; dx + E_0^{HF} \; ,
\ee
where
$$
E_0^{HF} = -\; \frac{1}{2} \int \Phi(x,x') \left [ \rho(x)\rho(x') \pm
|\rho_1(x,x')|^2 \right ] \; dx dx' 
$$
and the first-order density matrix is
$$
\rho_1(x,x')=\; <\psi^\dgr(x')\psi(x)>  = \sum_k n_k\vp_k(x)\vp_k^*(x') \; . 
$$

One can choose the expansion basis made of the functions that are the 
solutions to the equation
$$
\left [ -\; \frac{\nabla^2}{2m} + U(x) +
\hat U_{HF}(x) \right ] \vp_k(x) = \ep_k\vp_k(x) \; .
$$
The difference with the Hartree approximation is that now $\hat U_{HF}(x)$ 
is a nonlocal potential.

For a uniform system, when $x \ra\br$ and $U(\br) = 0$, the expansion basis
is composed of the plane waves. The exchange operator potential acts on the
latter according to the rule 
$$
\hat U_{ex}(\br)\vp_k(\br) =
\pm \int \Phi(\br-\br') \rho_1(\br,\br')\vp_k(\br') \; d\br' \; .
$$
Assuming that the interaction potential is integrable, one invokes the Fourier 
transform
$$
\Phi(\br) = \frac{1}{V} \; \sum_k \Phi_k e^{i\bk\cdot\br} \; , \quad
\Phi_k = \int \Phi(\br) e^{- i\bk\cdot\br} \; d\br \; .
$$

Hamiltonian (184) becomes
\be
\label{185} 
H = \sum_k (\ep_k -\mu) a_k^\dgr a_k + E_0^{HF} \; ,
\ee
with the spectrum
$$
\ep_k = \frac{k^2}{2m} +\rho\Phi_0 \pm \frac{1}{V}\;\sum_p n_p\Phi_{k-p}  
$$
and the nonoperator term 
$$
E_0^{HF} = -\; \frac{1}{2}\left ( N\rho\Phi_0 \pm \frac{1}{V}\;
\sum_{kp} \Phi_{k+p} n_k n_p \right ) \; .
$$
In the case of an isotropic system, one has $\Phi_{-k}=\Phi_k$ and 
$n_{-k}=n_k$.

\subsection{Hartree-Fock-Bogolubov approximation}

The Hartree-Fock approximation preserves the gauge symmetry. Therefore it 
cannot be applied for systems where gauge symmetry becomes broken. That is, 
the Hartree-Fock approximation is not applicable for systems with Bose-Einstein 
condensate as well as for superconducting systems. 

The approximation that takes into account possible gauge symmetry breaking
is the Hartree-Fock-Bogolubov approximation that for the averages reads as
the decoupling 
$$
<\psi^\dgr(x_1)\psi^\dgr(x_2)\psi(x_3)\psi(x_4)>\; = \;
<\psi^\dgr(x_1)\psi(x_4)><\psi^\dgr(x_2)\psi(x_3)> \pm
$$
\be
\label{186}
\pm <\psi^\dgr(x_1)\psi(x_3)><\psi^\dgr(x_2)\psi(x_4)> +
<\psi^\dgr(x_1)\psi^\dgr(x_2)><\psi(x_3)\psi(x_4)> \; .
\ee
This differs from the Hartree-Fock approximation by the term containing the
anomalous averages $<\psi(x)\psi(x')>$ and $<\psi^\dgr(x)\psi^\dgr(x')>$.
To express the Hartree-Fock-Bogolubov approximation in the operator terms,
we shall use the short-hand notation for the field operators 
$\psi_i\equiv\psi(x_i)$ and $\psi_i^\dgr\equiv\psi^\dgr(x_i)$. Then this 
approximation corresponds to the operator expression
$$
\psi^\dgr_1\psi^\dgr_2\psi_3\psi_4 = \psi^\dgr_1\psi_4
<\psi^\dgr_2\psi_3> + <\psi^\dgr_1\psi_4>\psi^\dgr_2\psi_3 -
$$
$$
- <\psi^\dgr_1\psi_4><\psi^\dgr_2\psi_3> \pm
\psi^\dgr_1\psi_3<\psi^\dgr_2\psi_4> \pm <\psi^\dgr_1\psi_3>
\psi^\dgr_2\psi_4 \mp
$$
\be
\label{187}
\mp <\psi^\dgr_1\psi_3><\psi^\dgr_2\psi_4>
+ \psi^\dgr_1\psi^\dgr_2<\psi_3\psi_4>
+ <\psi^\dgr_1\psi^\dgr_2>\psi_3\psi_4 -
<\psi^\dgr_1\psi^\dgr_2><\psi_3\psi_4> \; .
\ee

Hamiltonian (177) reduces to the form
$$
H = \int \psi^\dgr(x)\left [ - \; \frac{\nabla^2}{2m} + U(x) +
\hat U_{HF}(x) - \mu\right ] \psi(x)\; dx +
$$
\be
\label{188}
+ \frac{1}{2} \int \Phi(x,x')\left [
\psi^\dgr(x)\psi^\dgr(x')<\psi(x')\psi(x)> +
\psi(x')\psi(x)<\psi^\dgr(x)\psi^\dgr(x')>\right ] \; dx dx' + E_0^{HFB} \; ,
\ee
in which
$$
E_0^{HFB} = E_0^{HF} -\; \frac{1}{2} \int \Phi(x,x') |<\psi(x')\psi(x)>|^2
\; dx dx' \; .
$$

Expanding the field operators over a basis and introducing the notation
$$
A_{kp}  \equiv \int \vp_k^*(x) \left [ - \; \frac{\nabla^2}{2m} + U(x) +
\hat U_{HF}(x) - \mu\right ] \vp_p(x)\; dx \; ,
$$
$$
B_{kp} \equiv \int \vp_k^*(x) \Phi(x,x') <\psi(x')\psi(x)> \vp_p^*(x')\;
dx dx' \; ,
$$
we can transform Hamiltonian (188) to the expression
\be
\label{189}
H = \sum_{kp} \left ( A_{kp} a_k^\dgr a_p +
\frac{1}{2}\; B_{kp} a_k^\dgr a_p^\dgr + \frac{1}{2}\; B^*_{kp} a_p a_k
\right ) + E_0^{HFB} \; .
\ee
The operator part here is equivalent to that in Eq. (170), hence can be 
diagonalized in the same way as in Sec. 5.4. 

As in the previous cases, the interaction potential, generally, has to be 
integrable for the applicability of this approximation.

\subsection{Kirkwood approximation}

Quite often, the interaction potentials are not integrable, because of which
the direct application of the mean-field approximations of the previous 
sections can be inadmissible. 

An example of a nonintegrable potential is the singular power-law potential
$\Phi(\br) \propto 1/r^n$, where $r\equiv |\br|$. In the three-dimensional 
space, the integral
$$
\int \Phi(\br)\; d\br \propto \int_0^R \frac{dr}{r^{n-2}} \; ,
$$
where $R$ is the system radius, diverges at zero, if $n\geq 3$. This is 
classified as ultraviolet divergence. 

A rather popular interaction potential is the Lennard-Jones potential
$$
\Phi(\br) = 4\ep\left [ \left ( \frac{\sgm}{r}\right )^{12} -
\left ( \frac{\sgm}{r}\right )^{6} \right ] \; ,
$$
describing the interactions between many neutral particles, such as
atoms and molecules. The {\it nonintegrable} potentials are often termed 
the {\it hard-core} potentials. In the case of the Lennard-Jones potential,
$\sgm$ is the hard-core radius.

Kirkwood \cite{Kirkwood_47} suggested a method of dealing with hard-core
interaction potentials by employing the modified Hartree decoupling
\be
\label{190}
<\psi^\dgr(\br)\psi^\dgr(\br')\psi(\br')\psi(\br)> \; = g(\br,\br')
<\psi^\dgr(\br)\psi(\br)><\psi^\dgr(\br')\psi(\br')> \; ,
\ee
which differs from the Hartree approximation by the factor $g(\br,\br')$
that takes into account short-range particle correlations, tending to zero, 
where the interaction potential tends to infinity. The actual form of 
$g(\br,\br')$ should be defined from other physical assumptions.  
In the operator representation, the Kirkwood approximation reads as
$$
\psi^\dgr(\br)\psi^\dgr(\br')\psi(\br')\psi(\br) = g(\br,\br') \left [
\psi^\dgr(\br)\psi(\br)<\psi^\dgr(\br')\psi(\br')> + \right.
$$
\be
\label{191}
\left. + <\psi^\dgr(\br)\psi(\br)>\psi^\dgr(\br')\psi(\br') -
 <\psi^\dgr(\br)\psi(\br)><\psi^\dgr(\br')\psi(\br')>\right ] \; .
\ee

Substituting form (191) into the Hamiltonian, one meets the effective
{\it Kirkwood potential} 
$$
U_K(\br) \equiv \int \Phi(\br,\br') g(\br,\br') \rho(\br')\; d\br' \; ,
$$
which is similar to the Hartree potential, but with the principal 
difference that now, instead of the bare interaction potential, one has
the {\it smoothed potential}
$$
\overline\Phi(\br,\br') \equiv g(\br,\br')\Phi(\br,\br') \; , 
$$
The latter is integrable, 
$$
\left | \int \overline\Phi(\br,\br')\; d\br' \right | < \infty \; ,
$$
even if the bare interaction potential is not. 

Equation (190), as is evident, is nothing but the definition of the pair 
correlation function $g(\br,\br')$. Kirkwood suggested to extract the pair 
correlation function $g(\br,\br')$ from experiment. This could be done by 
measuring the {\it structure factor}
\be
\label{192}
S(\bk) \equiv \frac{1}{N} \int \left [
<\hat n(\br)\hat n(\br')> -\rho(\br)\rho(\br')\right ]
e^{-i\bk\cdot(\br-\br')} \; d\br d\br' \; ,
\ee
where $\hat n(\br) = \psi^\dgr(\br)\psi(\br)$ is the density operator.
The density-density correlation function can be expressed through the pair
correlation function:
\be
\label{193}
<\hat n(\br)\hat n(\br')>\; = \rho(\br)\dlt(\br-\br') +
\rho(\br)\rho(\br') g(\br,\br')\; , 
\ee
which gives
\be
\label{194}
S(\bk) = 1 +\frac{1}{N} \; \int \rho(\br)\rho(\br') [ g(\br,\br') -1]
e^{-i\bk\cdot(\br-\br')}\; d\br d\br' \; .
\ee

In the uniform case, when $\rho(\br)=\rho$ and $g(\br,\br')=g(\br-\br')$,
one has
\be
\label{195}
S(\bk) =  1 +\rho \int [g(\br)-1] e^{-i\bk\cdot\br}\; d\br \; . 
\ee
Taking the inverse Fourier transform yields
$$
g(\br) = 1 + \frac{1}{N}\; \sum_k \left [ S(\bk) - 1\right ]
e^{i\bk\cdot\br} \; .
$$
For a large system, one passes from summation to integration, obtaining 
\be
\label{196}
g(\br) = 1 + \frac{1}{\rho}\; \int \left [ S(\bk) - 1\right ]
e^{i\bk\cdot\br} \frac{d\bk}{(2\pi)^3} \; .
\ee
Measuring in experiment the structure factor $S(\bf k)$, one can calculate
the pair correlation function by means of relation (196).

\subsection{Correlated approximations}

The Kirkwood idea can be generalized as follows \cite{Yukalov_60}. 
For simplicity, we again use the notation of Sec. 5.8 denoting the field
operators as $\psi_i\equiv\psi({\bf r}_i)$. And let us employ a brief 
notation $g_{12} \equiv g({\bf r}_1, {\bf r}_2)$ for the pair correlation 
function. The general form of a correlated approximation is
\be
\label{197}
\psi^\dgr_1\psi^\dgr_2\psi_3\psi_4 = \sqrt{g_{12} g_{34}} 
\left [ \psi^\dgr_1\psi^\dgr_2\psi_3\psi_4 \right ]_{MF} \; ,
\ee 
where the term in the square brackets in the right-hand side implies some
of the mean-field approximations. For instance, using for $[\ldots]_{MF}$ 
the Hartree approximation leads to the Kirkwood approximation (190). 
Employing for $[\ldots]_{MF}$ the Hartree-Fock approximation results in the 
correlated Hartree-Fock form
$$
\psi^\dgr(\br)\psi^\dgr(\br')\psi(\br')\psi(\br) =
g(\br,\br') \left [ \psi^\dgr(\br)\psi(\br) <\psi^\dgr(\br')\psi(\br')> +
\right.
$$
$$
+ <\psi^\dgr(\br)\psi(\br)>\psi^\dgr(\br')\psi(\br') -
<\psi^\dgr(\br)\psi(\br)><\psi^\dgr(\br')\psi(\br')> \pm
$$
$$
\pm \psi^\dgr(\br)\psi(\br')<\psi^\dgr(\br')\psi(\br)> \pm
<\psi^\dgr(\br)\psi(\br')>\psi^\dgr(\br')\psi(\br) \mp
$$
\be
\label{198}
\left. \mp <\psi^\dgr(\br)\psi(\br')><\psi^\dgr(\br')\psi(\br)> \right ] \; .
\ee
In the same way, one can introduce the correlated Hartree-Fock-Bogolubov 
approximation. 

One can remember that the pair interaction potentials depend on the difference
between the spatial variables, so that $\Phi(\br,\br') = \Phi(\br-\br')$. 
Similarly, the pair correlation function also usually depends on this difference,  
$g(\br,\br') = g(\br-\br')$. Thus the smoothed interaction potential becomes
$\overline\Phi(\br) \equiv g(\br)\Phi(\br)$. Since $g(\br)\ra 0$ when 
$\Phi(\br)\ra\infty$, the smoothed potential is integrable,
$$
\left | \int \overline\Phi(\br)\; d\br \right | < \infty \; .
$$

The pair correlation function is real-valued and symmetric, such that
$g(-\br)=g(\br)$. It can be expressed through the second-order density 
matrix as in Sec. 3.10. Thus, for a uniform system, it is 
\be
\label{199}
g(\br-\br') = \frac{\rho_2(\br,\br',\br,\br')}{\rho^2} \; .
\ee
The behavior of the pair correlation function is such that $g(\br)\ra 1$,
when $\Phi(\br)\ra 0$. In view of the normalization 
$$
\int \rho_2(\br,\br',\br,\br')\; d\br d\br' = N(N-1)\; , 
$$
the property
\be
\label{200}
\frac{1}{V} \int g(\br)\; d\br = 1  -\; \frac{1}{N} 
\ee
follows.

Instead of finding an approximation for the second-order density matrix,
thus obtaining the pair correlation function (199), one often directly 
invokes some approximations for the latter. The simplest such a way is
to use the {\it cutoff approximation}
\begin{eqnarray}
\nonumber
g(\br)=\left\{ \begin{array}{cc}
0, & r\leq \sgm \\
1, & r>\sgm \; . \end{array}
\right.
\end{eqnarray}

Sometimes, one resorts to the {\it classical approximation}
$$
g(\br)=\exp\{-\bt\Phi(\br)\} \; ,
$$
which can be used only for high temperatures.

For quantum systems, it is possible to use the {\it quantum approximation},
by setting $g(\br) = |\vp(\br)|^2$, with the function $\vp(\br)$ being
defined by the Schr\"{o}dinger equation for the relative motion: 
$$
\left [ -\; \frac{\nabla^2}{2m_0} + \Phi(\br) + E \right ] \vp(\br) = 0 \; ,
$$
where
$$
m_0 \equiv \frac{m_1m_2}{m_1+m_2}\; , \quad m_1=m_2=m\; , 
\quad m_0 = \frac{m}{2} \; .
$$

If at short distance $r\ra 0$, the interaction potential behaves as 
$$
\Phi(\br) \simeq 4\ep \left ( \frac{\sgm}{r}\right )^{12} \quad (r\ra 0) \; ,
$$
then the solution to the above equation yields
$$
\vp(\br) \sim \exp\left \{ -\; \frac{\kappa}{2}
\left ( \frac{\sgm}{r}\right )^{5} \right \} \quad (\br\ra 0) \; ,
$$
with the {\it correlation parameter} $\kappa \equiv 4 \sqrt{m\ep\sgm^2}/5$. 
Therefore
$$
g(\br) \approx \exp \left \{ - \kappa \left ( \frac{\sgm}{r}\right )^{5}
\right \} \; .
$$

One also uses variational methods for constructing the pair correlation 
functions and there exist other more complicated ways of finding their
approximate expressions \cite{Yukalov_46}. When correlations and interactions 
between particles are strong, it may happen that even integrable potentials 
require to take account of the pair correlation function.

Different regimes of describing the system can be classified by considering 
{\it characteristic lengths}. The {\it interaction radius} is given by the
expression
\be
\label{201}
r_0 \equiv \sgm +
\frac{\int_{r>\sgm}|\br|\Phi(\br)d\br}{\int_{r>\sgm}\Phi(\br)d\br} \; .
\ee
The {\it scattering length} 
\be
\label{202}
a_s \equiv \frac{m}{4\pi} \int \overline\Phi(\br)\; d\br
\ee
characterizes the intensity of interactions. These lengths should be 
compared with the {\it mean interparticle distance} $a$ and the 
{\it thermal wavelength} $\lbd_T$, 
$$
a \equiv \rho^{-1/3} \; , \quad \lbd_T \equiv \sqrt{\frac{2\pi}{mT}} \; . 
$$

Particle interactions are strong, when $a_s\gg a$, while interaction 
correlations are strong, if $r_0\gg a$. For $\lbd_T\ll a$, one has the 
classical regime. And the {\it quantum regime} develops when $\lbd_T \sim a$, 
becoming essentially quantum for $\lbd_T > a$. 

The approximations of the mean-field type provide sufficiently accurate
description, when the particles in the system are strongly correlated,
so that the influence of each of them is extended onto many other particles. 
In such a case, each particle can be imagined to be subject to a mean field
created by other particles. Generally, there exist two kinds of correlations,
interaction correlations and coherence correlations. 

{\it Interaction correlations} are strong, when the interaction radius $r_0$
is much larger than the interatomic distance $a$. Then each particle 
experiences the action of many other particles of the system. Such interactions
are termed long-range.      

{\it Coherence correlations} characterize the particle correlations that
develop not merely because of particle interactions but also due to particle 
quantum statistics. The {\it coherence radius} is defined as
\be
\label{203}
 r_{coh} \equiv 
\frac{\int r | C(\br,0)| d\br }{\int | C(\br,0)| d\br } \;  ,
\ee
where the correlation function 
\be
\label{204}
C(\br,\br') \equiv \frac{\rho_1(\br,\br')}{\sqrt{\rho(\br)\rho(\br')} }
\ee
is expressed through the density matrix $\rho_1(\bf r, \bf r')$. The coherence
correlations are strong, if the coherence radius $r_{coh}$ is much larger than 
the interatomic distance $a$. 

Summarizing, particle correlations in the system are strong, provided that at 
least one of the conditions holds true, either the interaction correlations or
the coherence correlations are strong, so that either $r_0 \gg a$ or 
$r_{coh} \gg a$. That is, particle correlations are strong, when at least one 
of the following parameters is large:   
\be
\label{205}
 \frac{r_0}{a} \gg 1 \; , \qquad \frac{r_{coh}}{a} \gg 1 \;  .
\ee
In that case, it is admissible to resort to a mean-field type approximation.

\vskip 2mm

In conclusion, this Tutorial presents the basic notions of quantum statistical 
physics that are necessary for the correct description of quantum atomic systems. 
Accurately applying these notions will help to avoid mistakes that, unfortunately, 
are rather common in the current literature. The exposition in the Tutorial 
has been aimed to be, from one side, sufficiently detailed to be easily read and, 
from another side, quite brief, omitting lengthy discussions that could be found 
in the cited references. In the following parts, the material of this Tutorial 
will be applied for a more concrete description of cold atoms satisfying 
Bose-Einstein and Fermi-Dirac statistics.     

\vskip 2mm

{\bf Acknowledgement}

\vskip 2mm

The author acknowledges financial support from the Russian Foundation for
Basic Research. The help from E.P. Yukalova is appreciated.

\end{document}